\documentclass[traditabstract]{aa} % for a referee version

\usepackage{rotating}
\usepackage{multicol}
\usepackage{longtable}

\usepackage{graphicx} 
\usepackage{txfonts} 
\usepackage{natbib}  
\usepackage{multirow}
\usepackage{geometry}
\usepackage[utf8]{inputenc}

\usepackage{graphicx}
\usepackage{subcaption}
\usepackage{txfonts}
\usepackage{ulem}
\usepackage{xcolor}
\usepackage{hyperref}

\begin{document}

    \title{{\it Hubble Space Telescope} survey of Magellanic Cloud star clusters. Photometry and astrometry of 113 clusters and early results}

% \abstract{}{}{}{}{} 
% 5 {} token are mandatory
 
 %
%-------------------------------------------------------------------

%\title[] {}

%\author[ A.\,P.\,Milone et al.] 
   \author{       A.\,P.\,Milone,          \inst{1,2}
       G.\,Cordoni              \inst{1}, 
       A.\,F.\,Marino           \inst{2,3},
       F.\,D'Antona             \inst{4},
       A.\,Bellini              \inst{5},
       M.\,Di Criscienzo        \inst{4},
       E.\,Dondoglio            \inst{1},
       E.\,P.\,Lagioia\inst{1},
       N.\,Langer\inst{6},
       M.\,V.\,Legnardi\inst{1},
       M.\,Libralato\inst{5},
       H.\,Baumgardt\inst{7},
       M.\,Bettinelli\inst{1},
       Y.\,Cavecchi\inst{8},
       R.\,de Grijs\inst{9,10},
       L.\,Deng\inst{11},
       B.\,Hastings\inst{6},
       C.\,Li\inst{12},
       A.\,Mohandasan\inst{1},
       A.\,Renzini\inst{2}, 
       E.\,Vesperini\inst{13}, 
       C.\,Wang\inst{6},
       T.\,Ziliotto\inst{1}, 
       M.\,Carlos\inst{1}, 
       G.\,Costa\inst{1},
       F.\,Dell'Agli\inst{4},
       S.\,Di Stefano\inst{1},
       S.\,Jang\inst{1}, 
       M.\,Martorano\inst{1}, 
       M.\,Simioni\inst{2}, 
       M.\,Tailo\inst{14}, 
       P.\,Ventura\inst{4}} 
% %et al., \\ 
   \institute{$^1$ Dipartimento di Fisica e Astronomia ``Galileo Galilei'', Univ. di Padova, Vicolo dell'Osservatorio 3, Padova, IT-35122 \\
   $^2$ Istituto Nazionale di Astrofisica - Osservatorio Astronomico di Padova, Vicolo dell'Osservatorio 5, Padova, IT-35122\\
  $^{3}$ Istituto Nazionale di Astrofisica - Osservatorio Astrofisico di Arcetri, Largo Enrico Fermi, 5, Firenze, IT-50125 \\
 $^{4}$ INAF - Osservatorio Astronomico di Roma, Via Frascati 33, I-00040, Monte Porzio Catone, Roma, Italy \\
 $^{5}$Space Telescope Science Institute, 3800 San Martin Drive, Baltimore,  MD 21218, USA\\
 $^{6}$Argelander-Institut f\"{u}r Astronomie, Universität Bonn, Bonn, Germany; Max-Planck-Institut f\"{u}r Radioastronomie, Bonn, Germany\\
 $^{7}$ School of Mathematics and Physics, The University of Queensland, St. Lucia, QLD 4072, Australia \\
 $^{8}$ Instituto de Astronomia, Universidad Nacional Autonoma de Mexico, Circuito exterior, Ciudad de Mexico 04510, Mexico\\
 $^{9}$School of Mathematical and Physical Sciences, Macquarie University, Balaclava Road, Sydney, NSW 2109, Australia \\ %de Grijs
 $^{10}$Research Centre for Astronomy, Astrophysics and Astrophotonics, Macquarie University, Balaclava Road, Sydney, NSW 2109, Australia \\ %de griij
%%$^{11}$Key Laboratory for Optical Astronomy, National Astronomical Observatories, \\ Chinese Academy of Sciences, 20A Datun Road, Beijing 100012, People's Republic of China \\ %Deng
$^{11}$Department of Astronomy, China West Normal University, Nanchong 637002, People's Republic of China \\ % Deng
$^{12}$School of Physics and Astronomy, Sun Yat-sen University, Daxue Road, Zhuhai, 519082, People's Republic of China \\ %Li
$^{13}$ Department of Astronomy, Indiana University, Bloomington, IN 47401, USA\\
$^{14}$ Dipartimento di Fisica e Astronomia Augusto Righi, Università degli Studi di Bologna, Via Gobetti 93/2, 40129, Bologna, Italy}
%
% R.\,De Grijs
%(1) 
%(2) 
%
%
%  anna.marino@inaf.it, giacomo.cordoni@phd.unipd.it, mariavittoria.legnardi@studenti.unipd.it, sohee.jang@unipd.it, edoardo.lagioia@unipd.it, emanuele.dondoglio@studenti.unipd.it, anjmohnvp@gmail.com, marilia.gabriela.carlos@gmail.com, Marco.Martorano@ugent.be
%  franca.dantona@gmail.com, flavia.dellagli@inaf.it, paolo.ventura@inaf.it, mrctailo@gmail.com, marcella.dicriscienzo@inaf.it,
%  lichengy5@mail.sysu.edu.cn
%  matteo.simioni@inaf.it, marghebettinelli@gmail.com
% cwang@astro.uni-bonn.de, bhastings@astro.uni-bonn.de, nlanger@astro.uni-bonn.de
%} 
%\begin{document} 
%\date{Accepted 2016 March 10. Received 2016 March 8; in original form 2016 January 7} 
 
%\pagerange{\pageref{firstpage}--\pageref{lastpage}} \pubyear{2017} 
%\maketitle 
%\label{firstpage}

\titlerunning{HST survey of Magellanic Cloud clusters} 
\authorrunning{Milone et al.}

  \abstract
  {In the past years, we have undertaken an extensive investigation of star clusters and their stellar populations in the Large and Small Magellanic Clouds (LMC, SMC) based on archival images collected with the {\it Hubble Space Telescope}. 
We present photometry and astrometry of stars in 101 fields observed with the Wide Field Channel of the Advanced Camera for Surveys and the Ultraviolet and Visual Channel and the Near-Infrared Channel of the Wide Field Camera 3. 
 These fields comprise 113 star clusters.  We provide differential-reddening maps for those clusters with significant reddening variations across the field of view.
We illustrate various scientific outcomes that arise from the early inspection of the photometric catalogs. In particular, we provide new insights on the extended main-sequence turn-off (eMSTO) phenomenon: 
i) We detected eMSTOs in two clusters, KMHK\,361 and NGC\,265, which had no previous evidence of multiple populations. This finding corroborates the conclusion that the eMSTO is a widespread phenomenon among clusters younger than $\sim$2 Gyr.
ii) The homogeneous color magnitude diagrams (CMDs) of 19 LMC clusters reveal that the distribution of stars along the eMSTO  depends on cluster age.
iii) We discovered a new feature along the eMSTO of NGC\,1783, which consists of a distinct group of stars going on the red side of the eMSTO in CMDs composed of ultraviolet filters.  
Furthermore, we derived the proper motions of stars in the fields of view of clusters with multi-epoch images. Proper motions allowed us to separate the bulk of bright field stars from cluster members and investigate the internal kinematics of stellar populations in various LMC and SMC fields.
As an example, we analyze the field around NGC\,346 to disentangle the motions of its stellar populations, including  NGC\,364 and BS\,90, young and pre-MS stars in the star-forming region associated with NGC\,346, and young and old field stellar populations of the SMC.
Based on these results and the fields around five additional clusters, we find that young SMC stars exhibit elongated proper-motion distributions that point toward the LMC, thus bringing new evidence for a kinematic connection between the LMC and SMC.} 
 
   \maketitle

\section{Introduction}\label{sec:intro}

In the past years, our group has extensively used the {\it Hubble Space Telescope} ({\it HST}\,) archive to study star clusters in both Magellanic Clouds \citep[e.g.][and references therein]{milone2009a, milone2020a}. The exquisite stellar photometry and astrometry provided by {\it HST}, together with the most advanced techniques for the analysis of astronomical 
images \citep[e.g.][]{anderson2008a, sabbi2016a, bellini2017a}, has provided significant advances in understanding Magellanic-Cloud star clusters and their stellar populations.

Inspired by the discovery that the color-magnitude diagram (CMD) of the Large Magellanic Cloud (LMC) cluster NGC\,1806  is not consistent with a single isochrone \citep{mackey2007a}, we started a series of papers to investigate the so-called extended-main sequence turn off phenomenon (eMSTO), in clusters with ages from about one to 2.3\,Gyr.    
 The main results include the discovery that the eMSTO is a common feature of Magellanic Cloud clusters \citep{milone2009a}, the early discoveries of split main sequence (MS) in young Magellanic Cloud clusters \citep{milone2013a, milone2015a, milone2016a, milone2017a}, and the characterization of the multiple populations in young and intermediate-age LMC and Small Magellanic Cloud (SMC) clusters \citep{milone2018a}. We provided the first direct evidence, based on high-resolution spectra, that the blue and red MS are made up of stellar populations with different rotation rates \citep{marino2018a} and the color and magnitude of an eMSTO star depend on stellar rotation  \citep{dupree2017a, marino2018a}.    
  The high-precision photometry resulting from this project has been instrumental to shed light on the physical mechanisms that are responsible for generating multiple populations in young clusters and has been used both by our team and by other groups to constrain the effect of rotation and stellar mergers on the eMSTO and the split MS \citep[e.g.][]{bastian2009a, dantona2015a, dantona2017a, wang2022a, cordoni2022a} and the contribution of variable stars on the eMSTO \citep{salinas2018a}.
 
 Although our main purpose consisted in investigating the eMSTO phenomenon, the resulting photometric and astrometric catalogs have  been used for various investigations of stellar astrophysics, including multiple stellar populations in Magellanic Cloud globular clusters \citep[GCs,][]{lagioia2019a, lagioia2019b, milone2020a, dondoglio2021a}, photometric binaries \citep[][]{milone2009a, milone2013a}, Be stars \citep[][]{milone2018a, hastings2021a}, and extinction \citep{demarchi2020a}.
 
 Driven by these results, we decided to homogeneously analyze all archival images collected with the Ultraviolet and Visual Channel (UVIS) and the Near Infrared Channel (NIR) of Wide Field Camera 3 (WFC3) and with the Wide Field Channel of the Advanced Camera for Surveys (WFC/ACS) on board {\it HST}.    
  In this work, we present high-precision stellar positions and magnitudes for stars in 101 fields of Magellanic Clouds that include 113 star clusters.

 The paper is organized as follows. Section \ref{sec:data} describes the dataset and the methods used for homogeneously reducing the data and presents the CMDs. The methods for correcting the photometry for differential reddening and the differential-reddening maps are discussed in Section \ref{sec:reddening}, while Section \ref{sec:centri} is dedicated to the determination of the cluster centers. 
 Absolute stellar proper motions are derived in Section \ref{sec:pms}. Section \ref{sec:secrets}
 provides some scientific cases that arise from early inspection of our catalogs. Finally, we report in the appendix the serendipitous discovery of one gravitational lens and two stellar clusters.
 
\section{Data and data analysis}\label{sec:data}
The dataset used in this paper comprises images collected through the UVIS/WFC3 and NIR/WFC3, and WFC/ACS on board {\it HST}.
The images include 84 known star clusters in the  Large Magellanic Cloud (LMC) and 29 clusters in the Small Magellanic Cloud (SMC). These clusters span wide intervals of age and stellar density, from sparse star-forming regions to old and dense GCs. 
 The main properties of the available exposures are listed in Table\,\ref{tab:data}.

%\section{Data reduction}\label{sec:reduction}
Photometry and astrometry are obtained from calibrated, flat-fielded WFC3/NIR ({\rm \_flt}) exposures, while in the case of UVIS/WFC3 and WFC/ACS data we used the calibrated, flat-fielded exposures corrected for the effects of the poor charge-transfer efficiency (CTE) of the detectors \citep[{\rm \_flc,}][]{anderson2010a}.
Stars are measured by means of distinct approaches
%which are optimal for stars with different luminosities, 
 that work best in different brightness regimes, 
 as discussed in the following subsections. 

\subsection{First-pass photometry}

We accounted for spatial variations of the Point-Spread Function (PSF) by using the grids of library PSFs provided by Jay Anderson for each filter and camera. The PSFs can change from one exposure to another due to focus variations produced by the breathing of {\it HST}, small guiding inaccuracies, and residual CTE. To derive the optimal PSF we perturbed the library PSFs by using a version of the \citet{anderson2006a} computer program adapted to UVIS/WFC3 and WFC/ACS \citep[see also][]{bellini2013a}. In a nutshell, we divided each image into a grid containing $n \times n$ cells, with $n$ ranging from 1 to 5. Bright, isolated, and unsaturated stars within each cell are fitted by the library PSF model, and the residuals of the fit are iteratively used to improve the PSF model itself.
We calculated the appropriate PSF model of each star  based on its location in the detector by linearly interpolating the four nearest PSFs of the grid \citep{anderson2000a}.
The number of cells in the grid has been fixed with the aim of obtaining the best quality-fit parameters for bright stars and depends on the number of available reference stars used to constrain the PSF perturbation in the cell. 

%We iteratively measured the fluxes and positions of the selected stars by means of library PSFs and calculate the residuals between each observed stars and the corresponding PSF model. These residuals are averaged to perturb the library PSF and obtain an improved PSF. Nine iterations are used to derive the final PSF models.

These PSFs are then used to measure the magnitudes and positions of unsaturated stars in each image. Saturated stars in the UVIS/WFC3 and WFC/ACS images are measured using the methods by \citet{gilliland2004a} and \citet{gilliland2010a}. These authors noted that the total number of electrons of saturated stars in the UVIS/WFC3 and WFC/ACS detectors is conserved and this information is preserved in the ${\rm \_flt}$ images with gain=2.
Hence, we measured each saturated star in an aperture of 5-pixel radius and added the contiguous saturated pixels that had bled outside this radius \citep[see][for details]{anderson2008a}. 

All catalogs derived from each filter and camera have been tied to the same photometric zero point, corresponding to the zero point of the deepest exposure in the filter that we used as a reference frame to construct the photometric  master frame. To do this, we used the bright, unsaturated stars that are well-fitted by the PSF to calculate the  difference between the magnitudes in the master frame and in each exposure. We used the mean of these magnitude differences to transform stars measured in each exposure into this reference frame.

Stellar positions are corrected for geometric distortion by using the solutions provided by \citet{anderson2006b} for WFC/ACS and \citet{bellini2009a} and \citet{bellini2011a} for UVIS/WFC3.
%NIR ???
The coordinates of stars in all images of each cluster are transformed  into a common reference system based on Gaia Early Data Release 3 (eDR3) catalogs  \citep[][]{gaia2020a}, in such a way that the abscissa and the ordinate are aligned with the West and North direction, respectively.  
We first de-projected the right ascension and declination into the plane tangential to the center of the main cluster in the field. 
We assumed  for these coordinates a scale factor of 0.04 arcsec per pixel. 
We first used bright, unsaturated stars that are well-fitted by the PSF to derive the six-parameter linear transformations used to convert the coordinates of all stars in each exposure into this reference frame.   
Then, we derived the $3\sigma$-clipped average stellar positions to derive a new astrometric catalog, that we used as a master frame to improve the transformations.  

\subsection{Multi-pass photometry}
The main outcomes from first-pass photometry, including  PSF models, coordinate transformations, photometric zero points, stellar magnitudes, and positions, are used to simultaneously identify and measure all point-like sources in all exposures.   
To do this, we used the FORTRAN computer program KS2 developed by Jay Anderson \citep[e.g.\,][]{sabbi2016a, bellini2017a, nardiello2018a}, which is the evolution of {\it kitchen sink}, originally written to reduce WFC/ACS images  \citep{anderson2008a}. 
KS2 exploits various iterations to find and measure stars. It first identifies the brightest and most isolated stars, calculates their fluxes and positions, and subtracts them from the image. In the subsequent iterations, it finds, measures, and subtracts  stars that are gradually fainter and closer to neighbor stars. 
 We used the stellar positions and magnitudes derived from first-pass photometry to generate appropriate masks for bright stars, including saturated ones. These masks optimize the detection and measurement of faint sources that are close to bright stars. They also minimize the detection of spurious sources that are typically associated with diffraction spikes and other structures of the stellar profile.  

This program adopts three distinct methods to measure stars, each providing optimal photometry for different ranges of stellar luminosity and density.  
\begin{itemize}
    \item Method I is optimal for relatively bright stars. It provides accurate measurements of all stars that generate distinct peaks within their local 5$\times$5-pixel raster after neighbor stars are subtracted. 
    Each star is measured by using the PSF model corresponding to its position, while the sky level is estimated from the annulus between 4 and 8 pixels from the center of the star.
    \item Method II provides the best photometry for faint stars, which do have not enough flux to provide robust fits with the PSF. After subtracting neighbor stars, KS2 performs the aperture photometry of the star in the $5\times5$ pixel raster. Each pixel is properly weighted to ensure low weight to those pixels contaminated by nearby stars. The sky is calculated as in Method I.
    \item Method III provides the best photometry in very crowded regions and for faint stars when a large number of exposures are available. It works as Method II, but aperture photometry is calculated over a circle with a radius of 0.75 pixels and the local sky in the annulus between 2 and 4 pixels from the position measured during the finding stage. 
\end{itemize}
    Stellar fluxes and positions are measured in each exposure separately and then are properly averaged together to derive our best determinations of magnitudes and positions. 

\begin{figure} 
\centering
\includegraphics[width=8.5cm,trim={9.5cm 0.0cm 9.5cm 0.0cm},clip]{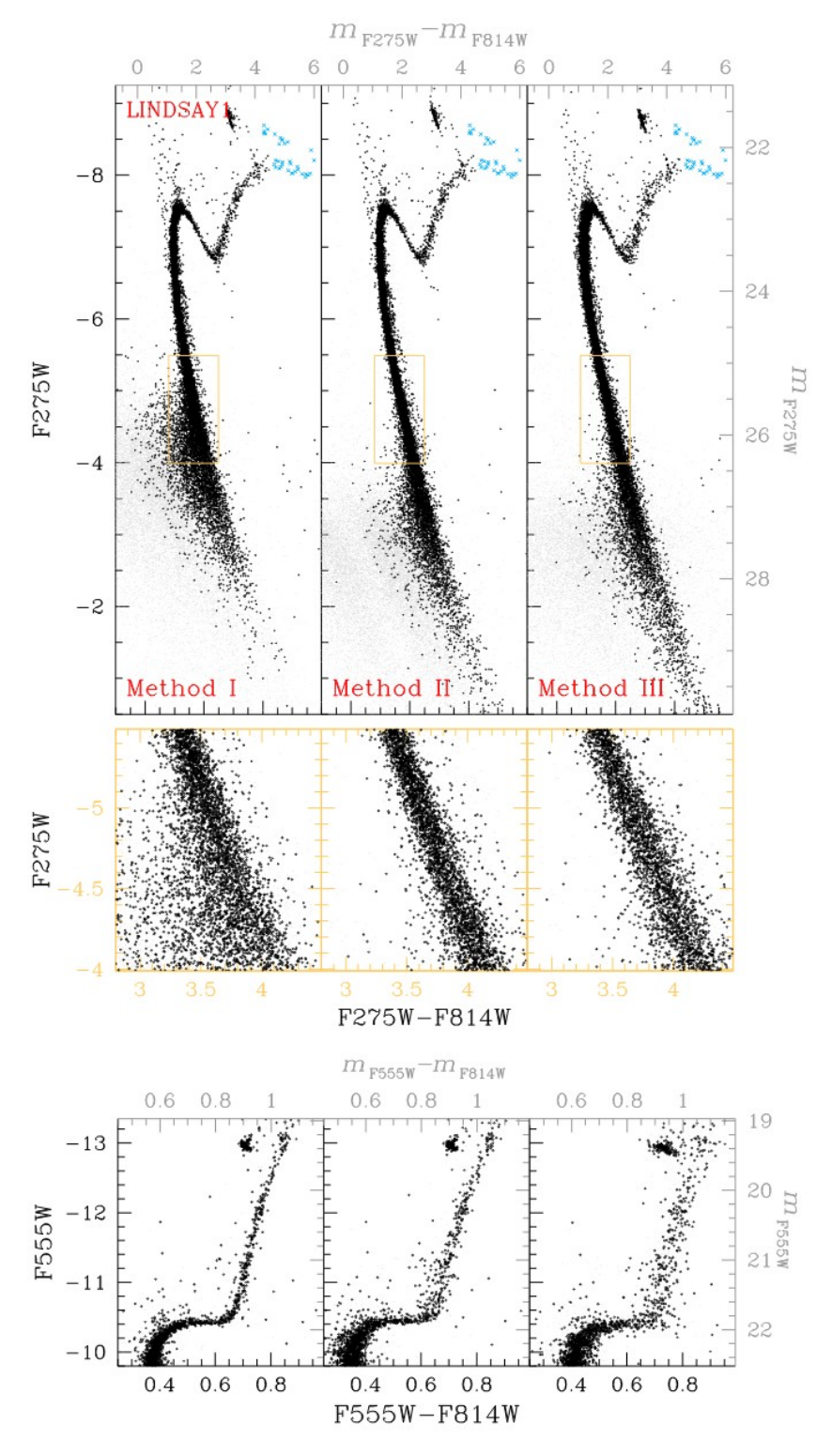}
%  \includegraphics[width=9.0cm,trim={0.5cm 7cm .5cm 4.0cm},clip]{methods.pdf}
%  \includegraphics[width=9.0cm,trim={0.5cm 17cm .5cm 4.5cm},clip]{methods2.pdf}
% \includegraphics[width=9.0cm,trim={0.5cm 15.2cm .5cm 4.5cm},clip]{methods3.pdf}
%/home/milone/NUBI/LINDSAY01/KS2
 \caption{Comparison of the instrumental F275W vs.\,F275W$-$F814W CMDs of stars in the field of view of the star cluster Lindsay\,1 as derived from Method I (top-left), Method II (top-middle) and Method III (top-right). Well-measured stars are colored black, while stars with poor photometry are plotted with light gray dots. Azure crosses mark stars where the F814W magnitude is derived from saturated images. Middle panels are zoomed-in views of the top-panel CMD around the MS. Bottom panels compare the region of the instrumental F555W vs.\,F555W$-$F814W CMDs populated by bright stars with F555W$<-9.75$ mag. Calibrated magnitudes and colors are indicated by the top and right axes.  } 
 \label{fig:methods} 
\end{figure}

    Figure\,\ref{fig:methods} compares the CMDs of stars in the field of view of Lindsay\,1 obtained from the three methods. We have chosen this GC as an example because of the deep F275W and F814W photometry available, which comprises 16 and 7 exposures in F275W and F814W, with total integration times of 27,341s and  2,206s, respectively. 
    A visual comparison of the top panels reveals that methods\,II and III are optimal for faint stars as they provide well-defined MSs. The latter method provides slightly better photometry for stars at the bottom of the MS alone, whereas Method\,II provides the best photometry for the remaining faint MS stars as highlighted in the middle panels, where we show the zoomed CMDs for MS stars with instrumental $-6<$F275W$<-4$ mag\footnote{Instrumental magnitudes are defined as the $-$2.5 log$_{10}$ of the detected photo-electrons.}. Clearly, the MS plotted in the central panel is much narrower and better defined than that shown in the left and right panels.
    On the contrary, Method I provides the best photometry for stars with bright instrumental magnitudes as demonstrated by the narrow red-giant and sub-giant branch (RGB and SGB) sequences in the bottom-left F555W vs.\,F555W$-$F814W CMDs.
 For each field, we derived three distinct catalogs from Methods\,I, II, and III. The science results shown in this paper, which are all focused on bright stars, are based on the photometry derived from Method\,I.

\subsection{Photometry calibration}
Photometry of each filter and camera has been calibrated to the Vega mag system  by computing the aperture correction to the PSF-fit-derived magnitudes and applying  to the corrected instrumental magnitude a photometric zero-point.
 To calculate the aperture corrections, we used  unsaturated and isolated stars only.
 
We measured aperture magnitudes within circular regions of $\sim$0.4 and 0.5 arcsec radius for UVIS/WFC3 and WFC/ACS, respectively. To do this, we used the drizzled  and CTE-corrected (${\rm \_drc,}$) images, which are normalized to 1s exposure time. Aperture photometry has been calibrated by adding to these instrumental magnitudes the corresponding aperture corrections and the zero points \citep{bohlin2016a, deustua2017a}.  
Finally, we calculated the 3 $\sigma$-clipped average of the difference between instrumental PSF magnitudes and calibrated aperture magnitudes for the stars in common. The resulting average values are then added to all the stars to derive calibrated magnitudes.       

\subsection{Quality parameters}\label{subsec:quality}
The computer program KS2 computes for each star various parameters that can be used as diagnostics of the photometric and astrometric quality. For each filter it provides three main quantities:
\begin{itemize}

   \item the $RADXS$ parameter is a shape parameter that indicates the amount of flux that exceeds the predictions from the best-fitting PSF \citep{bedin2008a}. It is defined as $RADXS=(\sum_{\rm i,j} pix_{\rm i,j}-PSF_{\rm i,j})/10^{-{\rm mag}/2.5}$ where the sum is calculated within  an annulus between 1.0 and 2.5 pixels from the center of the star and is normalized to the star's total flux. This quantity is negative when the object is sharper than the PSF (e.g.\,cosmic rays and PSF artifacts) and it is positive when the object is broader than the PSF (e.g.\,galaxies). The perfect PSF fit corresponds to $RADXS$=0. 

    \item the quality-fit parameter, $qfit$, which is indicative of the goodness of the PSF fit. It is defined as $qfit=\frac{\sum_{i,j} pix_{\rm i,j} PSF_{\rm i,j}}{\sqrt{\sum_{\rm i,j} pix^{2}_{\rm i,j} PSF^{2}_{\rm i,j}}}$, and it is calculated in a 5$\times$5 pixel area centered on the star, and  $pix_{\rm i,j}$ and $PSF_{\rm i,j}$ are the values of the pixel and the best-fitting PSF model, respectively, estimated in the pixel (i,j).    It ranges from unit, in the case of a perfect fit, to zero.
     
        \item the root mean scatter of the magnitude determinations, $rms$.
\end{itemize}

\begin{centering} 
\begin{figure} 
  \includegraphics[width=9.5cm,trim={0.5cm 5.5cm 1cm 4.0cm},clip]{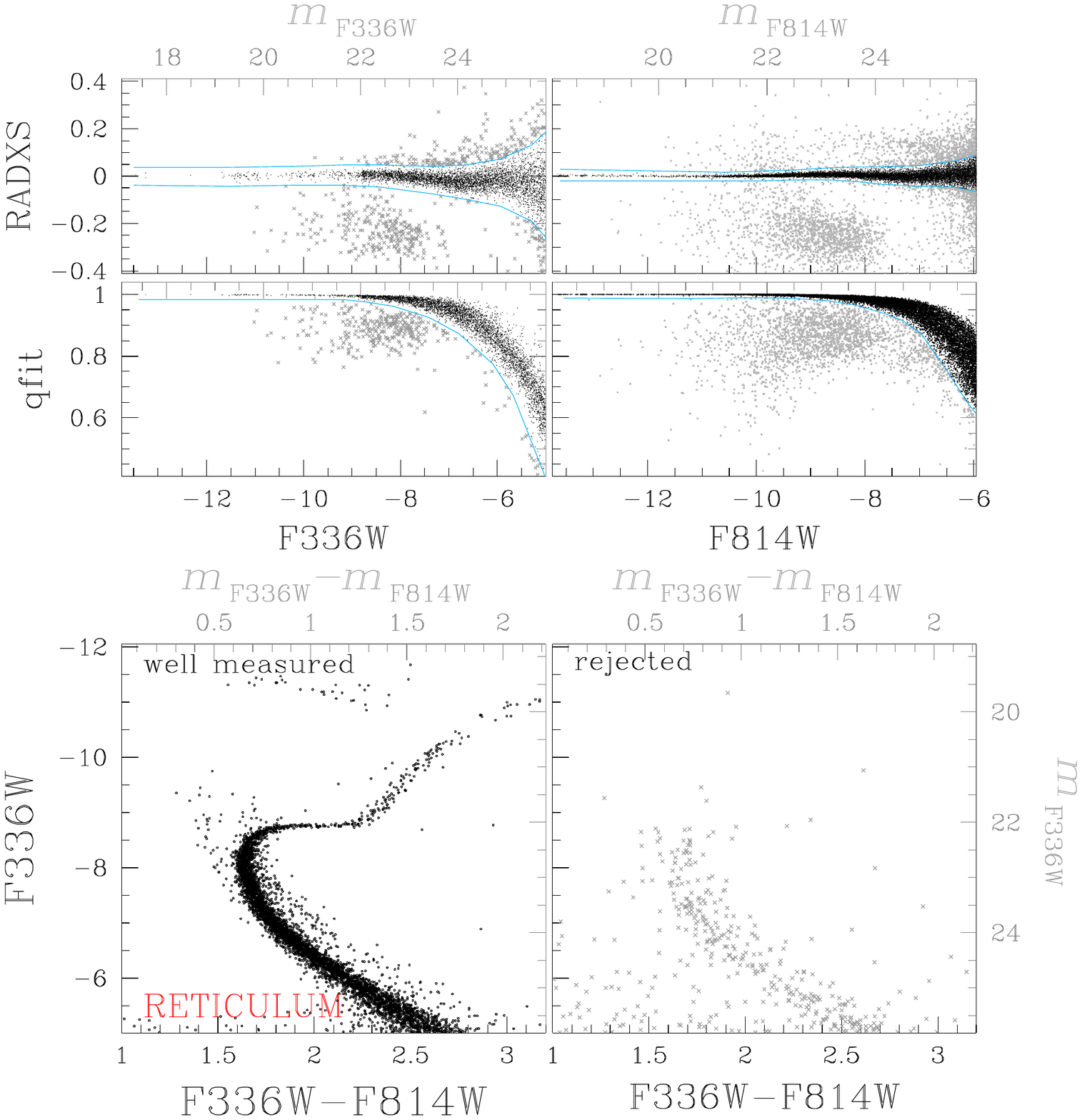}
%/home/milone/NUBI/RETICULUM/KS2/
 \caption{\textit{Top panels}. $RADXS$ and $qfit$ parameters derived from F336W  (left) and F814W (right) photometries of stars in the field of view of Reticulum. The azure lines separate well-measured stars (black dots) from poorly measured sources (gray crosses).
  The bottom panels show the instrumental F336W vs.\,F336W$-$F814W CMD for stars that pass the selection criteria in both filters (left) and for the remaining stars (right).}
 \label{fig:selezioni} 
\end{figure} 
\end{centering}

As an example, in the top-left panels of Figure \ref{fig:selezioni} we plot the $RADXS$ and $qfit$ parameters derived from F336W photometry of the star cluster Reticulum as a function of the F336W instrumental magnitude. Top-right panels show the analogous figures but for the F814W filter. The azure lines are drawn by hand with the criteria of separating the bulk of well-measured point-like sources from sources that are poorly fitted by the PSF model.
The bottom panels compare the CMD of stars that pass the selection criteria in both filters and the CMD of stars that have been rejected in at least one filter.
Although the magnitude $rms$ is another diagnostic of photometric quality, we prefer not to use it to select the sample of stars with high-quality measurements to avoid excluding variable stars. 
%%%%%%%%%%%%%%%%%%%%%%%%%%%%%%%%%%%%%%%%%%%%%%%%%%%%%%%%%%%%%%%%%%%%%%%%%%%%%

%%%%%%%%%%%%%%%%%%%%%%%%%%%%%%%%%%%%%%%%%%%%%%%%%%%%%%%%%%%%%%%%%%%%%%%%%%%%%

\subsection{The color-magnitude diagrams}\label{subsec:CMDs}
\begin{figure*} 
\centering
 \includegraphics[height=13.5cm,trim={0.0cm 0.0cm 10.0cm 1.0cm},clip]{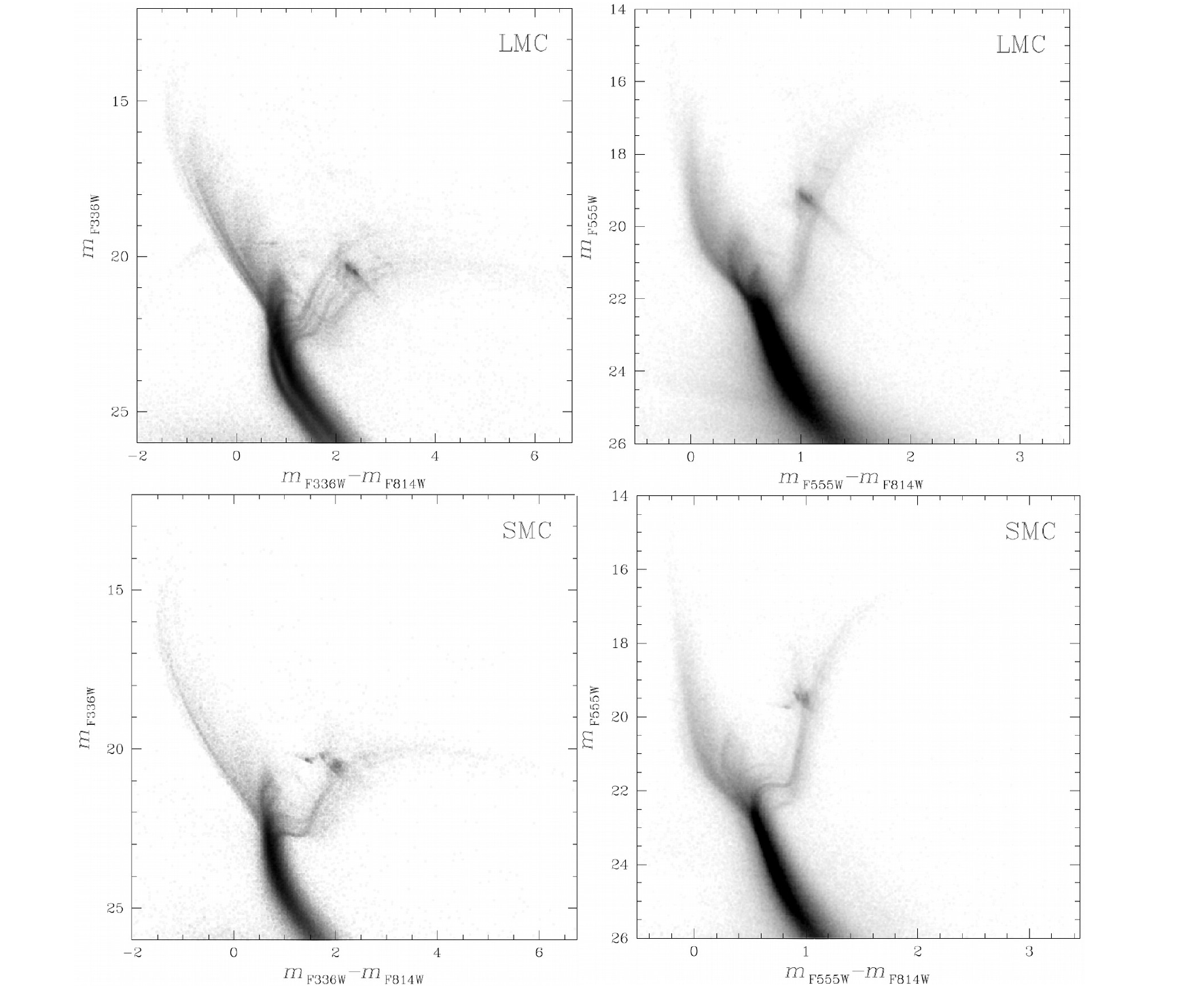}
% \includegraphics[height=8.5cm,trim={0.5cm 5.0cm 0.0cm 5.25cm},clip]{LMCf336wf814w.pdf}
%\includegraphics[height=8.5cm,trim={0.5cm 5.0cm 0.0cm 5.25cm},clip]{LMCf555wf814w.pdf}
% \includegraphics[height=8.5cm,trim={0.5cm 4.0cm 0.0cm 4.5cm},clip]{SMCf336wf814w.pdf}
%\includegraphics[height=8.5cm,trim={0.5cm 4.0cm 0.0cm 4.5cm},clip]{SMCf555wf814w.pdf}
 %/home/milone/NUBI/CATALOGHI/
 \caption{$m_{\rm F336W}$ vs.\,$m_{\rm F336W}-m_{\rm F814W}$ (left panels) and $m_{\rm F555W}$ vs.\,$m_{\rm F555W}-m_{\rm F814W}$ (right panels) Hess diagrams for all LMC (top) and SMC stars (bottom).}
 \label{fig:HESS} 
\end{figure*}

In the four panels of Figure\,\ref{fig:HESS} we show the $m_{\rm F336W}$ vs.\,$m_{\rm F336W}-m_{\rm F814W}$ (left) and the $m_{\rm F555W}$ vs.\,$m_{\rm F555W}-m_{\rm F814W}$ (right) Hess diagrams of all the observed fields in the LMC (top) and SMC (bottom).
%Some example of CMDs from our dataset are illustrated in Figure\,\ref{fig:HESS} where we show the $m_{\rm F336W}$ vs.\,$m_{\rm F336W}-m_{\rm F814W}$ and the $m_{\rm F555W}$ vs.\,$m_{\rm F555W}-m_{\rm F814W}$ Hess diagrams of all LMC (top) and SMC (bottom) stars in our dataset. 
Clearly, these diagrams reveal the complexity of stellar populations in the Magellanic Clouds, from bright and blue MSs composed of young and metal-rich stars to old and metal-poor stellar populations characterized by blue and faint MSs and faint RGBs.
%APM. Aggiungere numero di stelle.

%%%%%%%%%%%%%%%%%%%%%%%%%%%%%%%%

%%%%%%%%%%%%%%%%%%%%%%%%%%%%%%%%

\begin{figure*} 
\centering
\includegraphics[height=7.78cm,trim={0.0cm 0.0cm 0.0cm 0.0cm},clip]{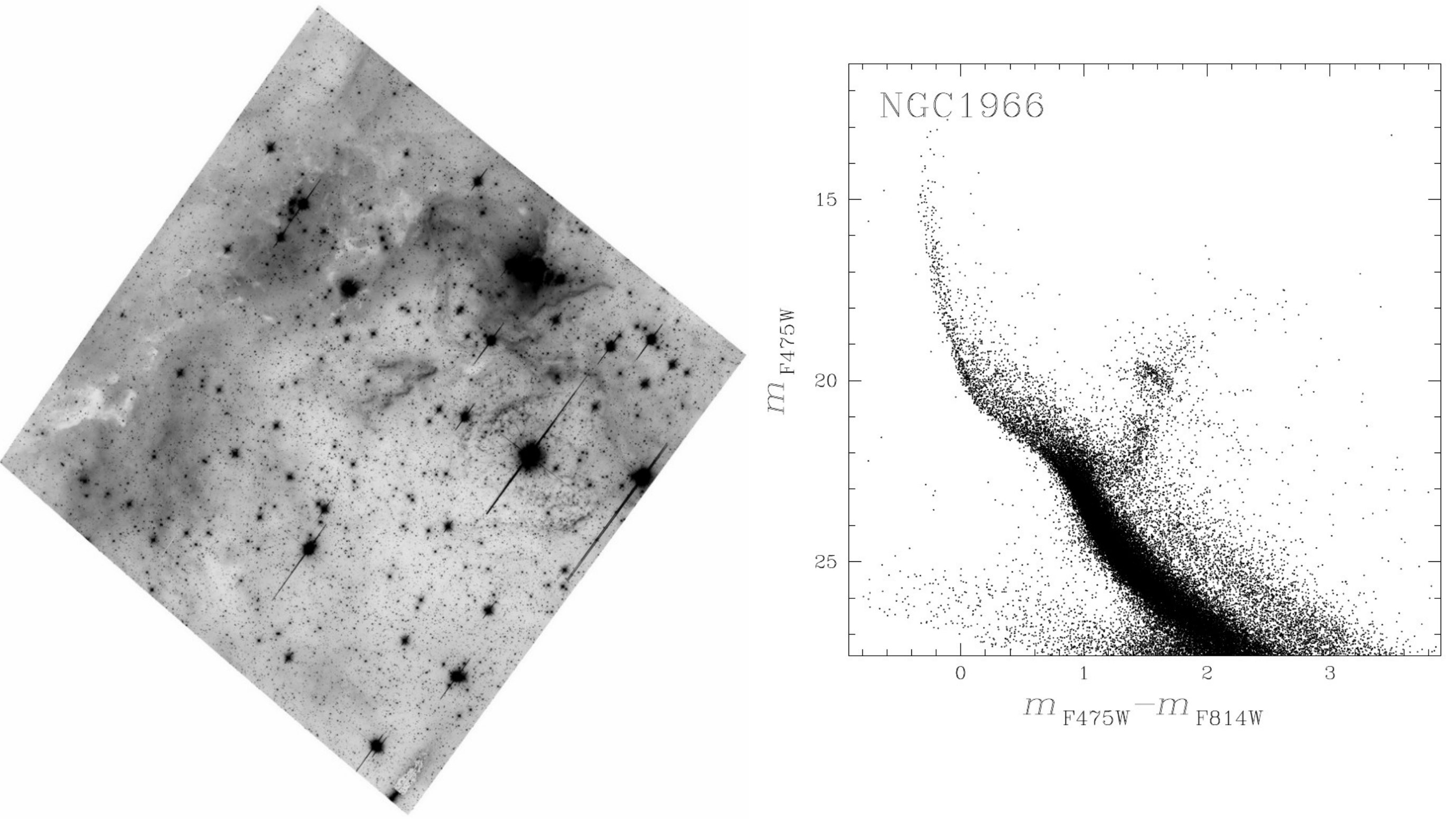} %\newline
\includegraphics[height=7.78cm,trim={0.0cm 0.0cm 0.0cm 0.0cm},clip]{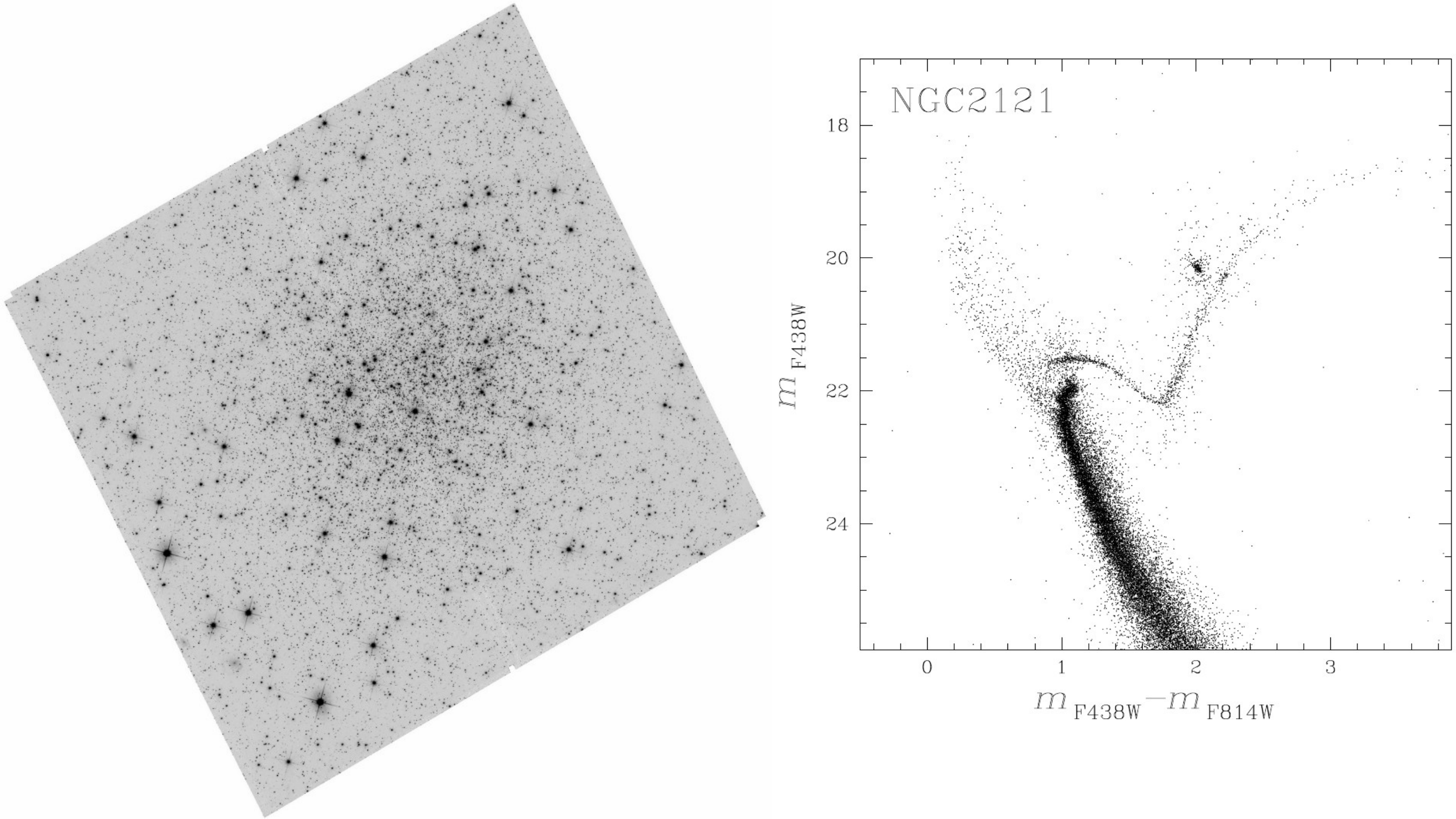} %\newline
\includegraphics[height=7.78cm,trim={0.0cm 0.0cm 0.0cm 0.0cm},clip]{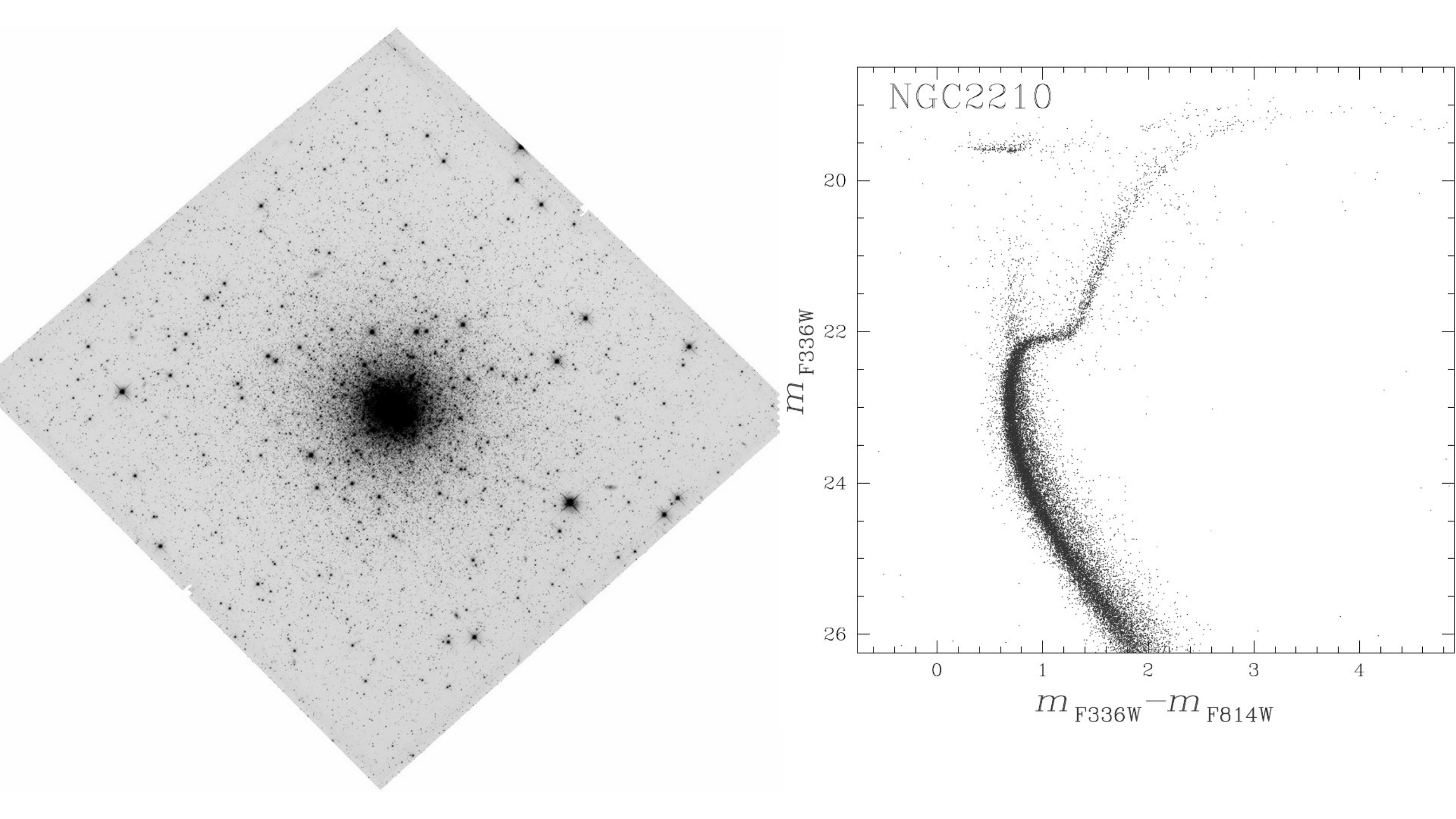}
 \caption{Stacked images and CMDs of stellar fields  with different ages and stellar densities.  North is up and East to the right.} The top panels show the F475W image and the $m_{\rm F475W}$ vs.\,$m_{\rm F475W}-m_{\rm F814W}$ CMD of stars in the star-forming region around the very young cluster NGC\,1966. The middle and bottom panels illustrate the F814W stacked images and the CMDs of the intermediate-age cluster NGC\,2121 (age $\sim 3$ Gyr) and the old GC NGC\,2210 (age $\sim 12$ Gyr), respectively. 
 \label{fig:CMDs} 
\end{figure*}

To further illustrate the variety of stellar populations and environments contained in the dataset of this paper, we show in Figure\,\ref{fig:CMDs} the stacked images and the CMDs of stars in three distinct fields that host stellar populations with different stellar densities, ages, and metallicities. The F475W image and the CMD of stars in the field around the open cluster NGC\,1966 are plotted in the top panels.  This region, which has never been studied with {\it HST}, hosts a conspicuous population of very young stars %associated with the star-forming region that hosts and surrounds NGC\,1966. 
 % Clearly, young stars 
  that populate the %sequences of 
   upper MS and the pre-MS. %stars in the $m_{\rm F475W}$ vs.\,$m_{\rm F475W}-m_{\rm F814W}$ CMD.
   The CMD also reveals old-RGB and red-clump stars that likely belong to foreground and background LMC old stellar populations. 
   Notably, the region hosts various nebula like NGC\,1965 and the gas nebula around the Wolf-Rayet star HD-269546 \citep[ the brightest star visible in the stacked image, ][]{westerlund1964a}, which are visible here in unprecedented detail.
%HD 269546 -- Wolf-Rayet Star   Bengt E. Westerlund, Lindsey F. Smith
The figures in the middle and bottom panels refer to the regions around the intermediate-age cluster NGC\,2121 (age $\sim 3$ Gyr) and the dense and old GC NGC\,2210 (age $\sim 12$ Gyr), respectively. 

%%%%%%%%%%%%%%%%%%%%%%%%%%%%%%%%
%%%%%%%%%%%%%%%%%%%%%%%%%%%%%%%%

The CMDs are used to estimate age, distance modulus, (m$-$M$)_{0}$, metallicity, [M/H], and reddening, E(B$-$V), by using isochrones from the Padova database \citep{marigo2017a}.  
 To minimize the contamination of field stars, we excluded from the analysis the stars at a large distance from the cluster center. Moreover, we statistically subtracted the field stars from the CMD of cluster members by using the method of \citet{gallart2003a}, in close analogy with what is done in previous papers from our group \citep[e.g.\,][]{marino2014a, milone2018a}. In a nutshell, we defined by eye a region that is centered on the cluster and includes the bulk of cluster stars (hereafter cluster field) and a reference field with the same area, and at a large distance from the cluster center, which is mostly composed of field stars. 
 We associated with each star in the reference field, the star in the cluster field at the smallest distance in the CMD, where the distance is defined as \\
 $
 {\rm distance}=\sqrt{(k \times {\Delta \rm color})^{2} + {(\Delta \rm magnitude)^{2}} }
 $\\
 where $\Delta$color and $\Delta$magnitude are the color and magnitude differences, respectively, and $k$ is a factor enhancing the difference in color with respect to the magnitude difference, which is derived as in \citet[][see their section 3.1]{marino2014a}. These stars are excluded from the comparison with the isochrones.
The cluster parameters and the best-fitting isochrones are used in Section\,\ref{sec:reddening} to estimate differential reddening maps and to investigate the eMSTO phenomenon (Section\,\ref{sec:secrets}). 
To find the best-fitting isochrone, we used the CMD from the  UVIS/WFC3 and/or WFC/ACS photometry providing the widest color baseline, thus maximizing the sensitivity to metallicity. 
However, when photometry in optical filters is available, we excluded the UV filters F225W, F275W, F336W, and F343N from the analysis to minimize the effect of multiple populations \footnote{These UV filters encompass various molecular bands that include carbon, nitrogen, and oxygen.
Their fluxes are sensitive to the abundances of these elements, which are not constant within clusters with multiple populations.  Hence, the CMDs made with UV filters  may lead to less  accurate determinations of the cluster parameters than optical CMDs.}. Indeed, these UV filters are sensitive to the potential effects due to stellar populations with different nitrogen and oxygen abundances, \citep[e.g.][]{marino2008a, milone2020a, dondoglio2021a}, which are typical features of GCs older than $\sim$2.3 Gyr, and to stellar populations with different rotation rates \citep[e.g.][]{dantona2015a, milone2016a, li2017a, marino2018a}, which are present in all clusters younger than $\sim$2 Gyr.

The observed CMDs are compared with grids of isochrones with different reddening values, distances, metallicities, and ages. The resulting best-fitting parameters are provided in  Table\,\ref{tab:info} and are estimated as follows.

 We first determined the isochrone and the values of reddening and distance modulus that, based on the visual comparison with the CMD,  provide the best match with the CMD. Then, we improved the determination of the best-fitting parameters using the following iterative approach. We fixed the values of age, distance, and reddening and better constrained the cluster metallicity by comparing the slopes of the fiducial lines of the observed RGB and the MS of the CMDs, and the slope of the corresponding magnitude intervals of the isochrones. 

  Then, we assumed the metallicity value corresponding to the minimum difference between the slopes of the observed CMDs and the isochrones to improve the estimates of reddening, age, and distance modulus. To do this, we 
   adopted the criteria of obtaining the best match between the isochrone and the observed CMD, {which may change for clusters with different ages.}

The best-fitting parameters of clusters older than $\sim 2.3$\,Gyr were estimated by determining the isochrone that best fits the CMD from the MSTO through the SGB \citep[e.g.][]{dotter2010a}.  Specifically, we calculated the $\chi^{2}$ values of the distances in the CMD between the fiducial lines of the MSTO and SGB stars and the isochrone. The best-fitting values of age, reddening and distance modulus are derived by means of $\chi^{2}$ minimization.
A visual inspection at the CMDs reveals that all clusters between $\sim 10$ Myr and $\sim$2.5 Gyr exhibit eMSTOs. Since the eMSTOs challenge their age determinations we provide two age values.
%Age determination of clusters between $\sim 10$ Myr and $\sim$2.5 Gyr is challenged by the eMSTOs, so we decided to provide two age values. 
We list in column 11 of Table\,\ref{tab:info} the age of the isochrone that best fits the lower part of the eMSTO. Clearly, this age value would represent the oldest cluster stars, if the eMSTO is entirely due to age variation. 
Alternatively, if the eMSTO is entirely due to rotation, our age estimate would provide an upper limit to cluster age, as the fast-rotating stars populate the lower part of the eMSTO \citep[e.g.][]{dupree2017a, marino2018a, marino2018b, kamann2020a}. Hence, we provide in column 12 of Table\,\ref{tab:info} the age of the isochrone that best fits the upper part of the eMSTO.
 In the clusters younger than $\sim 10$\,Myr, where it is challenging to identify the MSTO, our age determination is largely based on evolved stars.   In these young clusters and in the clusters with the eMSTO, the age, distance modulus, and reddening were derived by eye.

 To quantify the typical precision of the values of metallicity, age, reddening, and distance modulus inferred by the isochrones we applied the following procedure to four couples of clusters with different ages. The photometry of the clusters of each pair comes from datasets with large differences in the number of images and in the total exposure times. Hence, the range of uncertainties on the fitting parameter inferred from each couple of clusters would comprise the parameters' uncertainties of all studied clusters with similar ages.

 We first linearly added to the slopes of the fiducial lines of the RGB and MS stars that were used to constrain the metallicity, the corresponding errors. Hence, we derived the best value of [Fe/H] that corresponds to the isochrone that provides the best match with the used slope. We repeated the same  procedure but by using the slopes of the MS and RGB fiducials after subtracting the errors. 
  We consider the semi-difference between the maximum and the minimum [M/H] value, $\Delta$[M/H]  as a quantity indicative of the precision of our metallicity estimate.

Similarly, we shifted each point of the fiducial line of  the MS and the SGB to the bright and blue side of the CMD,  perpendicular to the isochrone. We indicate the resulting line as blue-shifted fiducial. The shift is applied in such a way that 68.27\% of the stars on the blue side of the original fiducial line are located on the red side of the blue-shifted fiducial. We applied a similar procedure to derive a red-shifted fiducial line.
Hence, we repeated four times the procedure described above to estimate the values of age, reddening, and distance modulus but by assuming the various combinations of the largest and minimum values of [M/H] and the blue-shifted and red-shifted fiducials.
We consider the semi-differences between the maximum values of age ($\Delta$age), distance modulus, ($\Delta$(m$-$M)$_{0}$), and reddening ($\Delta$E(B$-$V)), as a proxy of the precision of the estimates of the corresponding quantities.

The results are listed in Table\,\ref{tab:errori} for the pairs of clusters of old GCs  NGC\,2005 and NGC\,1939 (ages of $\sim 13$\,Gyr), intermediate age clusters Kron\,3 and Kron\,1 (ages of $\sim$6--7 Gyr). We also investigated the $\sim$2 Gyr-old clusters NGC\,1846 and Hodge\,7 and the young clusters NGC\,1866 and BSDL\,1650 (ages of $\sim$300 Myr).

\section{Differential reddening}\label{sec:reddening}
\begin{figure*} 
\centering
\includegraphics[width=21.0cm,trim={0.5cm 1.2cm 0.0cm 0.0cm},clip]{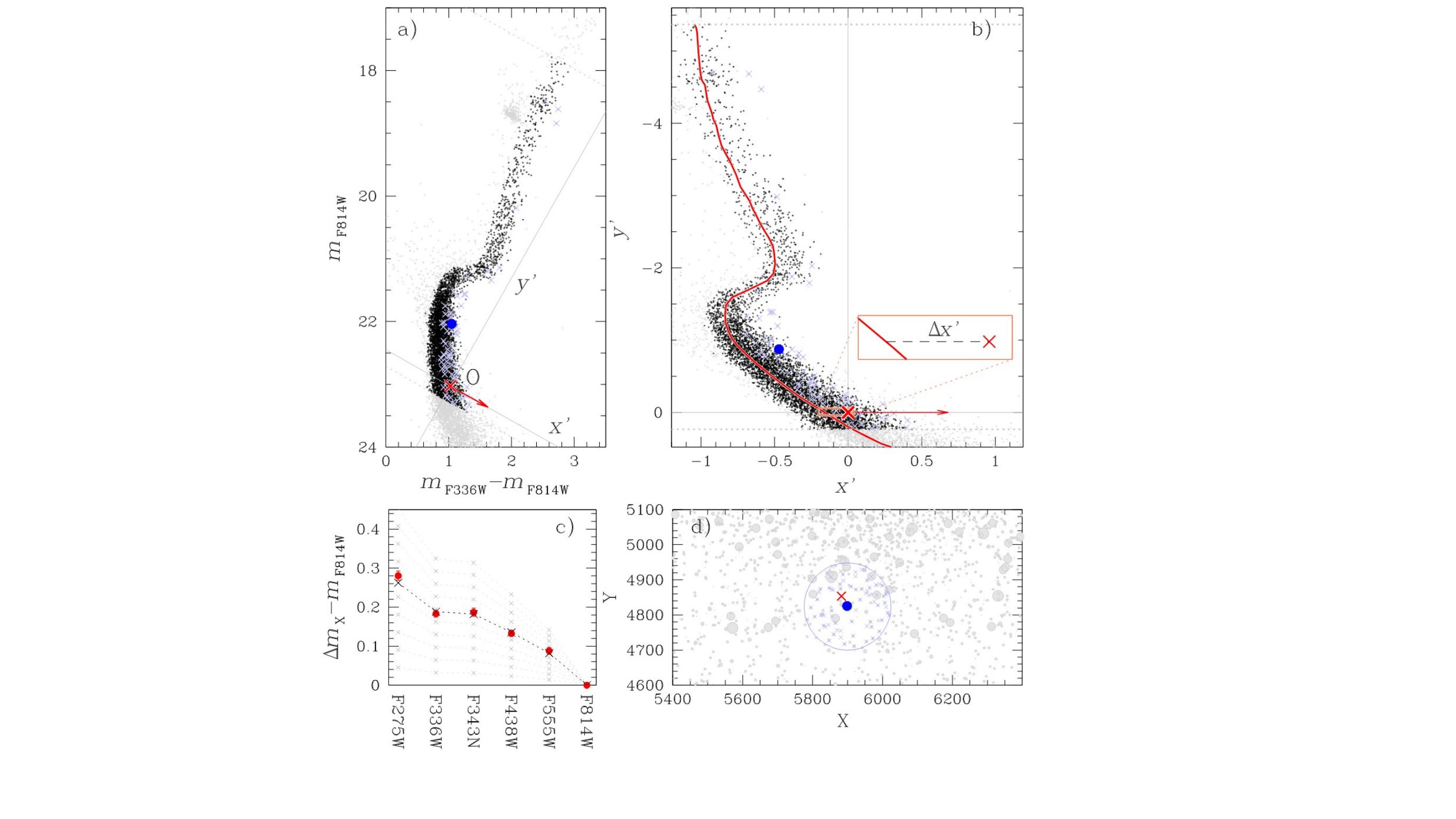}
%\newline
%\includegraphics[width=12.0cm,trim={0.5cm 5.5cm 0.0cm 17.0cm},clip]{red2.pdf}
 %/home/milone/NUBI/NGC0416/F814W/
 \caption{This figure illustrates the procedure to estimate the amount of differential reddening associated with the target star represented with the large blue dot. 
 Panel a shows the $m_{\rm F814W}$ vs.\,$m_{\rm F336W}-m_{\rm F814W}$ CMD of all the stars. Reference stars, are located between the two dotted gray lines and are colored black, whereas the neighboring reference stars are marked with light-blue crosses.  The gray continuous lines are the abscissa and the ordinate of the rotated reference frame centered on the reference star marked with the large red cross, while the red arrow indicates the reddening direction. Panel b shows the same stars as panel a but in the rotated reference frame. The red continuous line is the fiducial of reference stars and the inset highlights the relative position between one reference star and the fiducial.   
  Panel c represents the values of $\Delta x'$ inferred from different filters (red dots). Gray crosses are the corresponding values derived for $\Delta$ E(B$-$V) ranging from 0.01 to 0.10 mag in steps of 0.01 mag, while the black crosses provide the best fit to the observations and correspond to $\Delta$ E(B$-$V)=0.058 mag.  Finally, the finding chart zoomed in around the target is illustrated in panel d. See the text for details.
 }
 \label{fig:red} 
\end{figure*}

To derive high-resolution reddening maps, we applied to our dataset the method originally developed by \citet{milone2012a} to correct the ACS/WFC F606W and F814W magnitudes of Galactic GCs for differential reddening \citep[see also][]{bellini2017a, jang2022a}. 
The main difference of the adopted procedure is that the catalogs of several GCs comprise photometry in more than two bands. 
The main steps of our iterative method, which is illustrated in Figure \ref{fig:red} for NGC\,416, can be summarized as follows:
\begin{itemize}
    \item We built the $m_{\rm F814W}$ vs.\,$m_{\rm X}-m_{\rm F814W}$ diagrams, where X=F275W, F336W, F343N, F438W, F555W, and F814W. Each diagram has been used to gather information on differential reddening from a sample of reference stars. %The amount of differential reddening has been estimated for each reference stars independently.
    Reference stars are selected in the CMD region where the reddening direction defines a wide angle with the cluster fiducial line in such a way that we can easily disentangle the effect on stellar colors and magnitudes due to differential reddening from the shift due to photometric uncertainties.
    As an example, panel a of Figure \ref{fig:red} highlights in black the selected reference stars of NGC\,416 in the $m_{\rm F814W}$ vs.\,$m_{\rm F336W}-m_{\rm F814W}$ CMD.
    
    \item 
    We first derived the reddening direction corresponding to each star as $\theta={\rm arctan} \frac{A_{\rm X}}{A_{\rm X}-A_{\rm F814W}}$, where $A_{\rm X}$ and $A_{\rm F814W}$ are the absorption coefficients in the X and F814W bands, respectively. To derive them, we identified the point on the best-fitting isochrone with the same $m_{\rm X}$ magnitude as the reference star and calculate the $m_{\rm X}$ and $m_{\rm F814W}$ magnitude differences with the corresponding point of the isochrone with E(B$-$V)=0 mag. This procedure allows us to account for the dependence of reddening direction from the total amount of reddening and  from its spectral type.   
    As an example, panel a of Figure\,\ref{fig:red} shows the reddening direction associated with the reference star indicated by the red cross.
    
    \item We translated the CMD into a new reference frame where the origin corresponds to the reference stars as illustrated in panels a and b of Figure \ref{fig:red}. This CMD is rotated  counterclockwise by an angle $\theta$ so that the abscissa and the ordinate of the new reference frame are parallel and orthogonal, respectively, to the reddening direction. 
    
    \item We generated the fiducial line of MS, SGB, and RGB stars, which we plot as a continuous red line in panel b. To do this, we divided the sample of MS stars into 'ordinate' intervals. For each bin, we calculated the median abscissa associated with the median 'ordinate' of the stars in the bin. The fiducial line has been derived by linearly interpolating these median points.
     
    \item We calculated the distance of the reference star from the fiducial line along the reddening direction, $\Delta x'$ as shown in panel b for a reference star of NGC\,416 that we marked with a large red cross \footnote{ The differential reddening is responsible for shifting the stars along the reddening line. The amount of such shift, which is proportional to the amount of reddening along the line of sight, depends on the star's position in the field of view.  As a consequence, the stars in the different regions of the field are systematically shifted towards larger or lower values of x’ with respect to the cluster fiducial line depending on whether they are affected by a larger or smaller amount of reddening with respect to the median cluster reddening (see Figure 5 for an example). On the contrary, photometric errors are responsible for a random scatter along the fiducial line, but such a scatter is essentially not dependent on the reddening direction and the position of the star in the field of view.}.
    
    \item We calculated the projection of $\Delta x'$ along the $m_{\rm X}-m_{\rm F814W}$ color direction, $\Delta$ ($m_{\rm X}-m_{\rm F814W}$) and plotted this quantity for the available X filters as shown in panel c  of Figure \ref{fig:red}.  The observed values of $\Delta$ ($m_{\rm X}-m_{\rm F814W}$) are compared with corresponding quantities derived from the isochrones and corresponding to reddening variations ranging from $\Delta$ E(B$-$V)=$-$0.3 to 0.3 mag in steps of 0.001 mag. The value of $\Delta$ E(B$-$V) that provides the minimum $\chi^{2}$ is assumed as the best differential-reddening estimate associated with the reference star marked with the red cross.
    
    % fiducial line in the rotated RF
    %
    % Distance from the fiducial line along the reddening line. 
    %
\end{itemize}

To derive the amount of differential reddening associated with each star in the catalog, we selected   a sample of N spatially nearby reference stars, (light-blue crosses in Figure\,\ref{fig:red}) as shown in panel d. The best determination of differential reddening is provided by the median of the $\Delta$ E(B$-$V) values of these  $N$ neighbors. We excluded the target star from its own differential reddening determination.
  We derived various determinations of differential reddening  by assuming different values of $N$, from 35 to 95 in steps of 5 and from 100 to 150 in steps of 10. For each determination, we   calculated the pseudo-color distances between the value of $x'$ of the reference stars, corrected for differential reddening, and the fiducial line of Figure\,\ref{fig:red}. 
  We assumed that our best determination of differential reddening is given by the value of $N$ that provides the minimum value of the r.m.s of these distances. In particular, we used N=75 for NGC\,416.

\begin{figure*} 
\centering
%\includegraphics[width=8.0cm,trim={0.5cm 5.5cm 0.0cm 9.5cm},clip]{red.pdf}
%\newline
% \includegraphics[width=12.0cm,trim={2.2cm 7.0cm 0.0cm 5.0cm},clip]{redmap.pdf}
  \includegraphics[width=16.0cm,trim={3cm 0.0cm 3.0cm 0.0cm},clip]{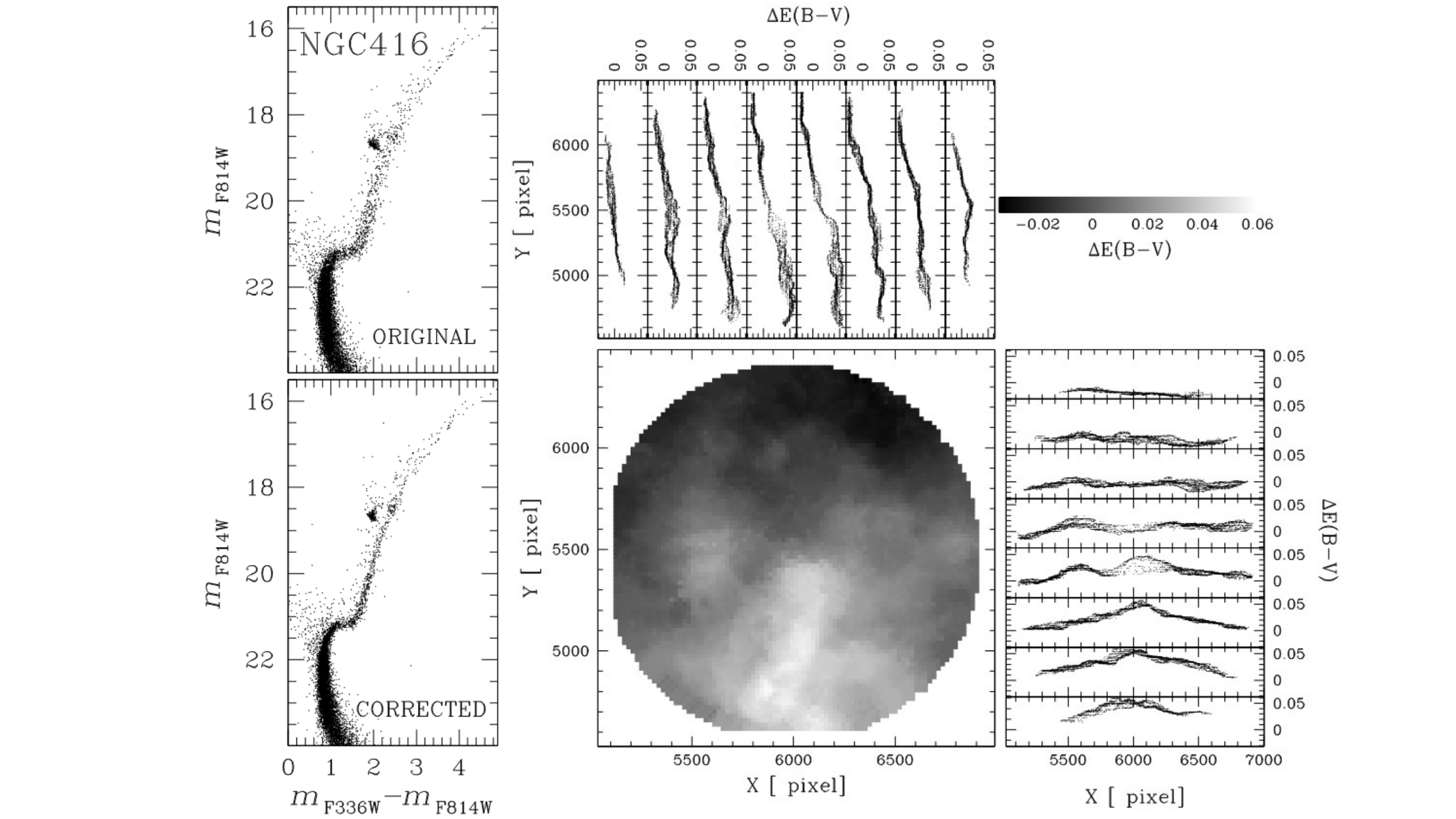}
 %/home/milone/
 \caption{\textit{Left.} Comparison between the original $m_{\rm F814W}$ vs.\,$m_{\rm F336W}-m_{\rm F814W}$ CMD of NGC\,416 (top) and the CMD corrected for differential reddening (bottom). \textit{Right.} Differential-reddening map in the direction of NGC\,416. The levels of gray are proportional to the reddening variation as indicated on the top-right. The panels on the right show $\Delta$ E(B$-$V) against the abscissa for stars in eight ordinate intervals. Similarly, the panels on the top represent the reddening variation as a function of the ordinate for stars in eight intervals of X. The field is centered around the center of NGC\,416 (X,Y=6019,5507) and the X and Y axis are parallel to the right ascension and declination direction, respectively. We adopted a scale of 0.04 arcsec per pixel.}
 \label{fig:redmap} 
\end{figure*}

\begin{figure*} 
\centering
  \includegraphics[width=16.0cm,trim={3cm 0.0cm 3.0cm 0.0cm},clip]{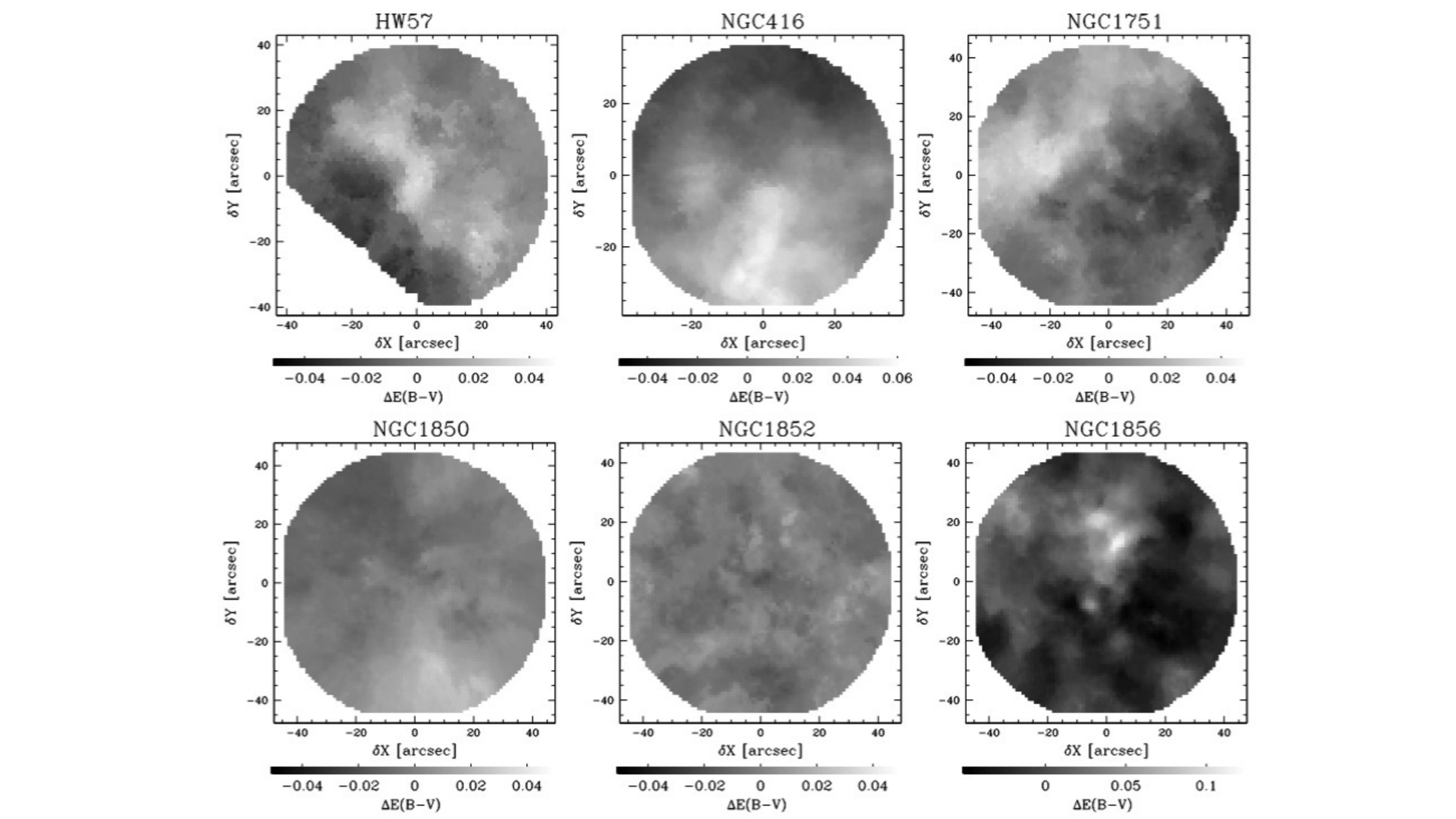}
%\includegraphics[width=5.0cm,trim={2.5cm 5.5cm 8.0cm 11.7cm},clip]{RedMapHW57.pdf}
%\includegraphics[width=5.0cm,trim={2.5cm 5.5cm 8.0cm 11.7cm},clip]{RedMapNGC0416.pdf}
%\includegraphics[width=5.0cm,trim={2.5cm 5.5cm 8.0cm 11.7cm},clip]{RedMapNGC1751.pdf}
%\includegraphics[width=5.0cm,trim={2.5cm 5.5cm 8.0cm 11.7cm},clip]{RedMapNGC1850.pdf}
%\includegraphics[width=5.0cm,trim={2.5cm 5.5cm 8.0cm 11.7cm},clip]{RedMapNGC1852.pdf}
%\includegraphics[width=5.0cm,trim={2.5cm 5.5cm 8.0cm 11.7cm},clip]{RedMapNGC1856.pdf}
 %/home/milone/
 \caption{Differential-reddening maps of the regions in front of HW\,57, NGC\,416, NGC\,1751, NGC\,1850, NGC\,1852 and NGC\,1856.}
 \label{fig:redmaps} 
\end{figure*}

As an example, Figure\,\ref{fig:redmap} shows the reddening map in the direction of NGC\,416 and compares the original CMD to the CMD corrected for differential reddening. A collection of reddening maps for six clusters is provided in Figure \ref{fig:redmaps}. 

%%%%%%%%%%%%%%%%%%%%%%%%%%%%%%%%

%%%%%%%%%%%%%%%%%%%%%%%%%%%%%%%%

\section{Cluster centers}\label{sec:centri}

To determine the coordinates of the center of each star cluster we followed the procedure described in \citet{cordoni2020b}. In a nutshell, we first selected by eye a sample of probable cluster members based on their location in the CMD and smoothed their stellar spatial distribution with a Gaussian kernel of fixed size. The kernel size has been chosen with the criteria of favoring the overall shape of the cluster, instead of the small-scale structures. 
We derived five contour lines within 50 arcsec from the cluster center and interpolated each of them with an ellipse by using the algorithm by \citet{halir1998a}.
Our best cluster-center determination corresponds to the median value of the centers of the ellipses, while the corresponding uncertainty has been estimated as the dispersion of the center determinations inferred from each ellipse. 
 Due to the low number of stars, it was not possible to apply the method above in 13 poorly-populated star clusters, namely BRHT\,5b, BSDL\,1650, KMK\,8827, KMHK\,1073, KMHK\,8849, OGLE-CL-LMC390, NGC\,1749, NGC\,290, NGC\,1850A, NGC\,1858, NGC\,1938 and NGC\,1966. For these clusters, we provide raw center determination based on the peak of the histogram distributions of the coordinates of the probable cluster members.
Results are provided in Table \ref{tab:info}.
%We typically used five ellipses per cluster.
%A visual inspection of Figure XX reveals that some HST fields contain multiple clusters.
%In such cases, we divided the field of view by hand, and we repeated the procedure described above on each cluster individually.
%The results of the procedure are shown in Figure XX.

%%%%%%%%%%%%%%%%%%%%%%%%%%%%%%%%

%%%%%%%%%%%%%%%%%%%%%%%%%%%%%%%%

%%%%%%%%%%%%%%%%%%%%%%%%%%%%%%%%

%%%%%%%%%%%%%%%%%%%%%%%%%%%%%%%%

%%%%%%%%%%%%%%%%%%%%%%%%%%%%%%%%
\begin{figure} 
\centering
\includegraphics[height=12cm,trim={1.0cm 5.8cm 5.2cm 4.5cm},clip]{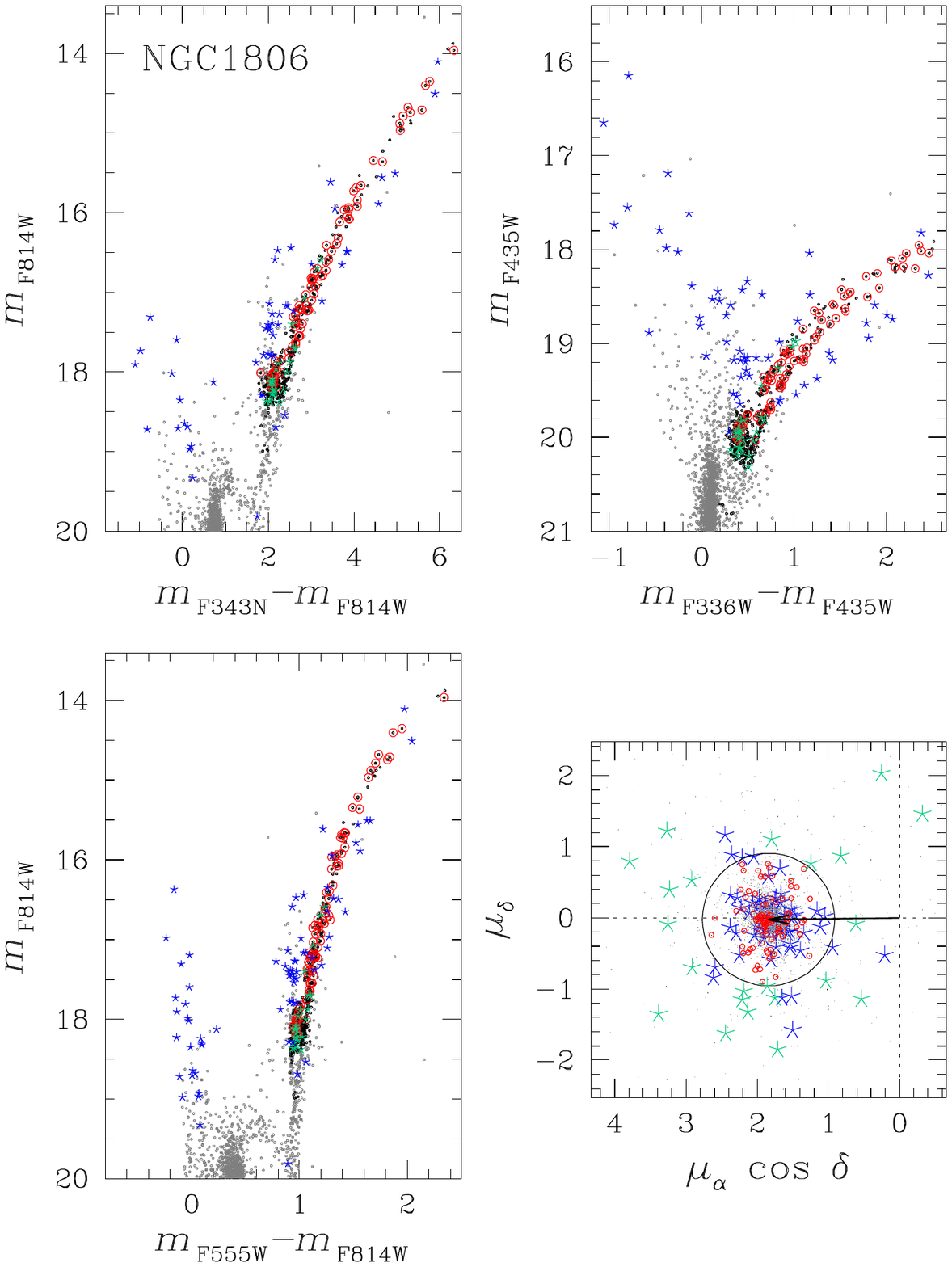} %\newline
\caption{This figure illustrates the procedure to identify stars that, based on the position in the CMDs from {\it HST} photometry and in the proper-motion diagram from GAIA eDR3, are probable members of NGC\,1806. The probable members are represented with red circles in the CMDs plotted in the top panels and    the bottom-left panel and in the proper motion diagram shown in the bottom-right panel. Stars that are not located on the main evolutionary sequences in at least one CMD are represented with blue-starred symbols. The arrow plotted in the proper-motion diagram indicates the mean cluster motion, while the circle is used to select the stars that are not included in the sample of probable cluster members, due to their large proper motions (aqua-starred symbols). See the text for details.  }
 \label{fig:PMGAIA} 
\end{figure} 
%%%%%%%%%%%%%%%%%%%%%%%%%%%%%%%%
%%%%%%%%%%%%%%%%%%%%%%%%%%%%%%%%%%%%%%%%%%%%%%%%%%%%%%%%%%%%%%%%%%%%%%%%%%%%%
\begin{figure*} 
\centering
%/home/milone/
%\includegraphics[height=8.5cm,trim={0.5cm 5.5cm 0.5cm 3.5cm},clip]{PMs.pdf}
%\includegraphics[height=8.5cm,trim={0.5cm 5.5cm 0.5cm 3.5cm},clip]{VPD.pdf}
\includegraphics[height=9cm,trim={0.0cm 1cm 0.0cm 1cm},clip]{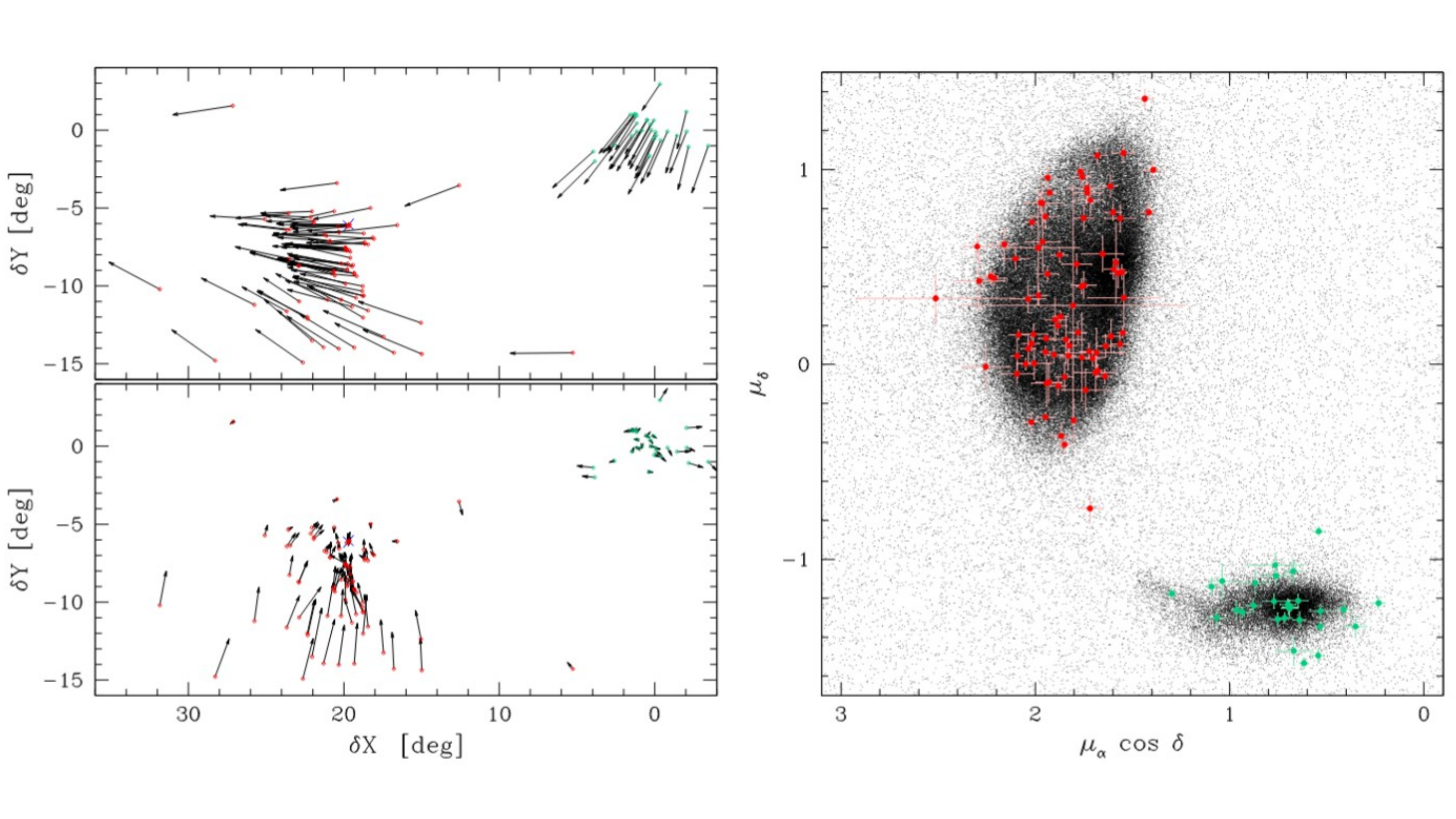}
 \caption{ \textit{Left.} Coordinates, in degrees, relative to the SMC center of the studied SMC and LMC clusters. The arrows in the top panel are indicative of the absolute proper motions of each cluster, while the  bottom panel represents  the proper motions of LMC and SMC clusters after subtracting the average motion of the corresponding galaxy.  \textit{Right.} Proper motions of stars brighter than g$_{\rm BP}=16.0$ mag in the region around the LMC and the SMC (black points). The studied LMC and SMC clusters are plotted in all panels with red and aqua dots, respectively.  }
 \label{fig:PMs} 
\end{figure*} 
\section{Proper motions}\label{sec:pms}

To estimate the absolute proper motions of the studied clusters, we combined information from {\it HST} photometry and Gaia eDR3 proper motions. Specifically, for each cluster, we selected by eye stars that,  based on their positions in all available CMDs, are probable cluster members.  Then, we used the Gaia eDR3 catalog to select stars with magnitude $g_{\rm BP}<19.0$ mag, which according to the criteria by \citet{cordoni2018a} have high-quality proper motions. The average proper motion of each cluster has been calculated as the 3-$\sigma$ clipped average of the proper motions of selected cluster members for which are available both {\it HST} photometry and Gaia eDR3 high-quality proper motions.  
We estimated the corresponding uncertainty by following the method of \citet{vasiliev2019a}, which accounts for systematic errors. %as the ratio between the proper-motion dispersion and the square root of the number of used stars minus one.  

 The main steps of the procedure used to derive the absolute proper motion are illustrated in Figure\,\ref{fig:PMGAIA} for NGC\,1806. For this cluster, we have photometry in five photometric bands of UVIS/WFC3 and WFC/ACS. We constructed ten CMDs of stars in the FoV of NGC\,1806 including four $m_{\rm F814W}$ vs.\,$m_{\rm X}-m_{\rm F814W}$ CMDs, where X=F336W, F343N, F435W, and F555W, three $m_{\rm F555W}$ vs.\,$m_{\rm X}-m_{\rm F555W}$ CMDs, where X=F336W, F343N, and F435W, two $m_{\rm F435W}$ vs.\,$m_{\rm X}-m_{\rm F435W}$ CMDs, where X=F336W and F343N, and the    $m_{\rm F343N}$ vs.\,$m_{\rm F336W}-m_{\rm F343N}$ CMD. For each CMD, we selected by eye the stars that, based on their colors and magnitudes, are located on the main cluster evolutionary sequences. As an example, the  stars that, based on their positions in all CMDs,  likely belong to the RGB, AGB, and red clump of NGC\,1806 are colored black in the three CMDs of Figure\,\ref{fig:PMGAIA}. The colored symbols mark  stars with available Gaia eDR3 proper motions in both the CMDs and in the proper-motion diagram. The stars that do not belong to the RGB, AGB, and red clump of NGC\,1806 in at least one CMD are represented with blue-starred symbols and are not included in the determination of the cluster proper motion. 
 We also excluded the selected stars with proper motions that differ from the average cluster motion by more than three times the proper motion dispersion (i.e.\, the stars outside the black circle shown in the bottom-right panel of Figure\,\ref{fig:PMGAIA} represented with aqua-starred symbols). The remaining stars are marked with red open dots.

Results are provided in Table\,\ref{tab:data}. The left panels of Figure\,\ref{fig:PMs} show the positions of the studied LMC and SMC clusters relative to the SMC center. In the top-left panel, we associate to each cluster the corresponding proper-motion vector, while in the bottom-left panel we show the proper-motion residuals after subtracting to LMC and SMC clusters the average motion of the corresponding Magellanic Cloud from  \citet{gaia2018a}\footnote{Although the investigation of the Magellanic Clouds' rotation is beyond our scope, we note that no clear rotation pattern is evident from the bottom-left panel of Figure\,\ref{fig:PMs}. This statement, which is based on a visual inspection of this figure, seems to contrast with the evidence of the LMC rotation pattern shown by \citet{vandermarel2014a, helmi2018a}. We also note that the right panel of Figure\,\ref{fig:PMs} highlights the relative motions within the SMC following the pattern of the SMC tidal expansion along the bridge and counter-bridge as detected in previous works \citep[e.g.][]{zivick2018a, piatti2021a, dias2021a, schmidt2022a}.}. 
The proper motion diagram is plotted in the right panel of Figure\,\ref{fig:PMs} and reveals that, based on proper motions, all clusters are consistent with being either LMC or SMC members. 
%\subsection{Proper motions from GAIA eDR3}

\subsection{Proper motions from Gaia eDR3 and {\it HST} }
For  thirteen GCs, we take advantage of having more than one epoch observations with appropriate signal-to-noise ratio and temporal baselines to disentangle the internal kinematics of Magellanic Cloud stars and separate cluster members and field stars by using {\it HST} data alone. Detailed information on the {\it HST} images available for these clusters are provided in Table\,\ref{tab:dataPM}. Relative {\it HST} motions are then transformed into absolute motions based on Gaia eDR3 proper motions.

To derive relative proper motions we applied to our dataset the procedure described by \citet[][]{piotto2012a} and described in the following for NGC\,1978. 
In a nutshell, we first identified the distinct groups of images collected at the same epoch through the same filter and camera. We reduced each group of images, separately, as described in Section\,\ref{sec:data}, and obtained the corresponding astrometric and photometric catalogs. 

 The reference frame defined by the first-epoch images collected through the reddest filter is adopted as a master frame. The coordinates of stars in each catalog are transformed into the master frame by means of six-parameter linear transformations \citep{anderson2006a}. 
 To minimize the effect of possible small residual distortions we applied local transformations based on the nearest 70 reference stars. Target stars are never  included in the calculation of their own transformations.   
 
\begin{figure*} 
\centering
 \includegraphics[height=9.5cm,trim={2cm 0cm 1.0cm 0cm},clip]{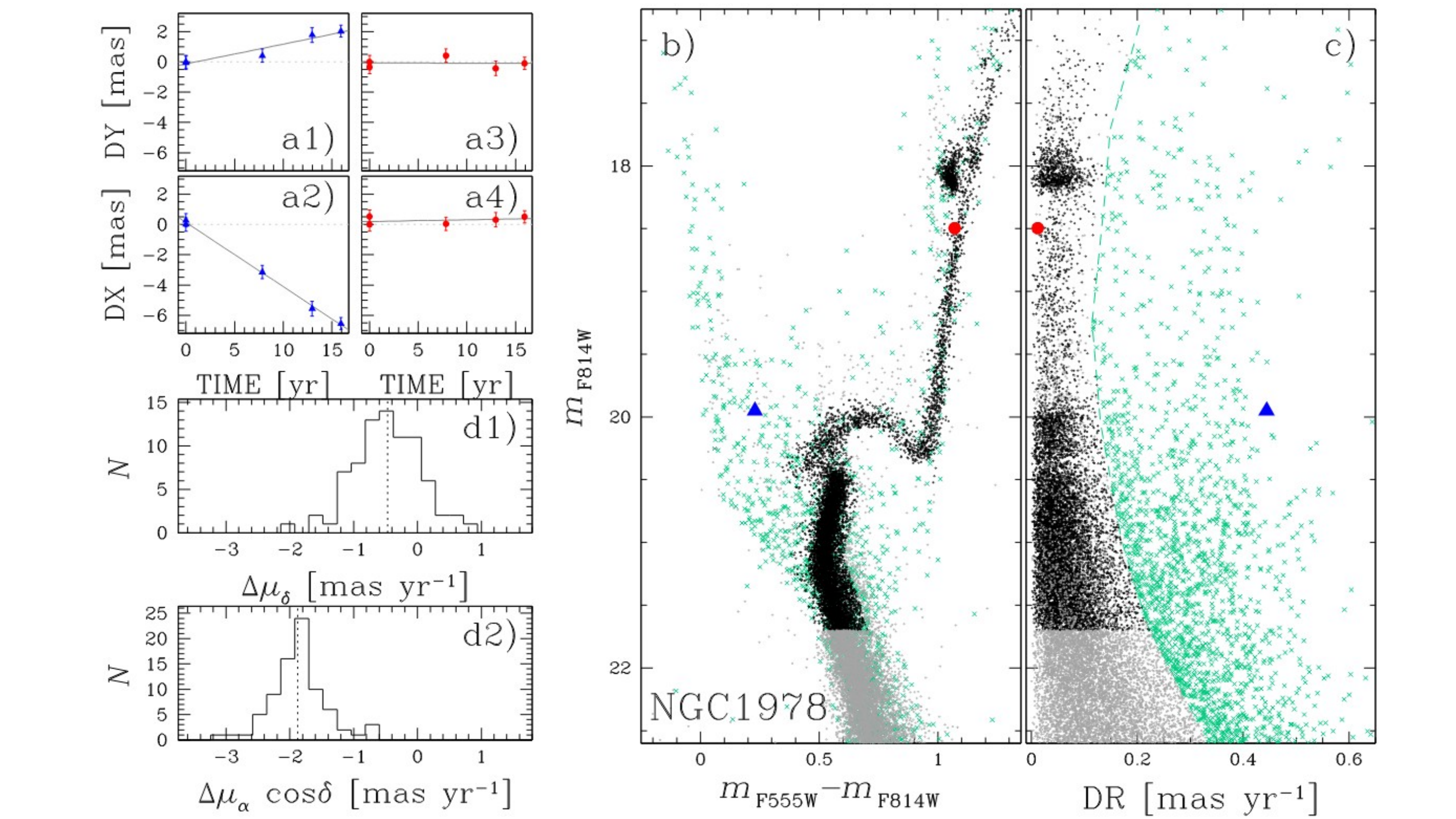}
%  \includegraphics[height=9.5cm,trim={0.5cm 5cm 10.0cm 5.2cm},clip]{MetPM1.pdf}
%degraphics[height=9.5cm,trim={0.5cm 5cm 0.2cm 4.5cm},clip]{MetPM2.pdf}
%/home/milone/NUBI/NGC1978/PMs
\caption{Procedure to estimate absolute proper motions. Panels a1 and a2 show the displacements along the X and Y directions in four epochs of a probable field star (blue triangles) relative to the mean motion of NGC\,1978. Similarly, panels a3 and a4 show the displacements of a candidate cluster member (red dots). The $m_{\rm F814W}$ vs.\,$m_{\rm F555W}-m_{\rm F814W}$ CMD of stars in the field of view of NGC\,1978 is plotted in panel b, while panel c shows relative stellar proper motions against $m_{\rm F814W}$. Aqua crosses are probable field stars selected on the basis of their proper motions. The stars used as references to calculate relative proper motions are colored black, while the remaining stars with cluster-like proper motions are gray. Panels d1 and d2 show the histogram of the difference between our relative proper motions and the absolute proper motions from Gaia eDR3. }
 \label{fig:MetPM} 
\end{figure*}

 The abscissa and the ordinate of each star, expressed in milliarcsec, are plotted against the epoch, expressed in years, as shown in panels a1--a4 of Figure\,\ref{fig:MetPM} for two stars in the field of view of NGC\,1978. For simplicity, in this figure, we show the displacements DX and DY, and the time relative to the stellar position and time at the first epoch. 
  These points are finally fitted with a weighted least-squares straight line, whose slope corresponds to the best proper motion estimate.   
 
The selection of the stars used to derive the transformation is a critical step for accurate proper-motion determination. Hence, we selected bright and unsaturated stars that pass the criteria of selection  discussed in Section\,\ref{subsec:quality}. We derived proper motions relative to a sample of cluster members that have been selected iteratively. As a consequence, the average relative motion of the cluster is set to zero.
We first identified probable cluster stars that, based on all available CMDs, lie on the main evolutionary sequences and used them to derive initial proper motion estimates.
Then, we iteratively excluded those stars that do not share the same motion as the bulk of cluster members (i.e.\,stars with proper motions greater than three times the proper-motion dispersion of cluster stars).
Panels b and c of Figure\,\ref{fig:MetPM} mark with black points the probable cluster members that we selected for deriving stellar proper motions in the $m_{F814W}$ vs.\,$m_{\rm F555W}-m_{\rm F814W}$ CMD and in the $m_{\rm F814W}$ vs.\,$DR=\sqrt{DX^{2}+DY^{2}}$ plane,  whereas aqua crosses are probable cluster members. Gray points mark the remaining stars with cluster-like proper motions, i.e.\, the saturated stars, the faint stars, and the stars that do not lie on the main evolutionary sequences in the CMDs.

To transform relative proper motions into absolute ones we derived the difference between the relative proper motions derived from {\it HST} images and the absolute proper motions from Gaia eDR3 for an appropriate sample of stars.
  Specifically, this selected sample includes stars with high-quality relative proper motions (i.e.\,bright, unsaturated stars that pass the criteria of selection of Section\,\ref{subsec:quality}).  In addition, the selected sample includes stars that, based on the proper motion uncertainties and on the values of the Renormalized Unit Weight Error (\texttt{RUWE}), the astrometric\_gof\_al
(\texttt{As\_gof\_al}) parameters of the Gaia eDR3 catalog   have accurate Gaia eDR3 absolute proper motions. We refer to papers by \citet{cordoni2018a, cordoni2020a}, for details on the procedure.
  The sample includes both cluster and field stars, with the exception of a few stars with parallaxes significantly larger than zero.

The $3\sigma$-clipped mean differences of the proper motion along each direction  ($\mu_{\alpha} {\rm cos}{\delta}$ and $\mu_{\delta}$)  are considered as the best estimate of the zero points of the motions and are used to convert relative proper motions into absolute ones.
As an example, panels d1 and d2 of Figure\,\ref{fig:MetPM} show the histogram distributions of the quantities $\Delta \mu_{\alpha} {\rm cos} {\delta}$=$\mu_{\alpha} {\rm cos} \delta$-DX and $\Delta \mu_{\delta}$=$\mu_{\delta}$-DY for stars in the field of view of NGC\,1978.

%%%%%%%%%%%%%%%%%%%%%%%%%%%%%%%%%%%%%%%%%%%%%%%%%%%%%%%%%%%%%%%%%%%%%%%%%%%%%

\begin{figure} 
\centering
 \includegraphics[height=9.0cm,trim={0.1cm 5.cm 0.0cm 3.5cm},clip]{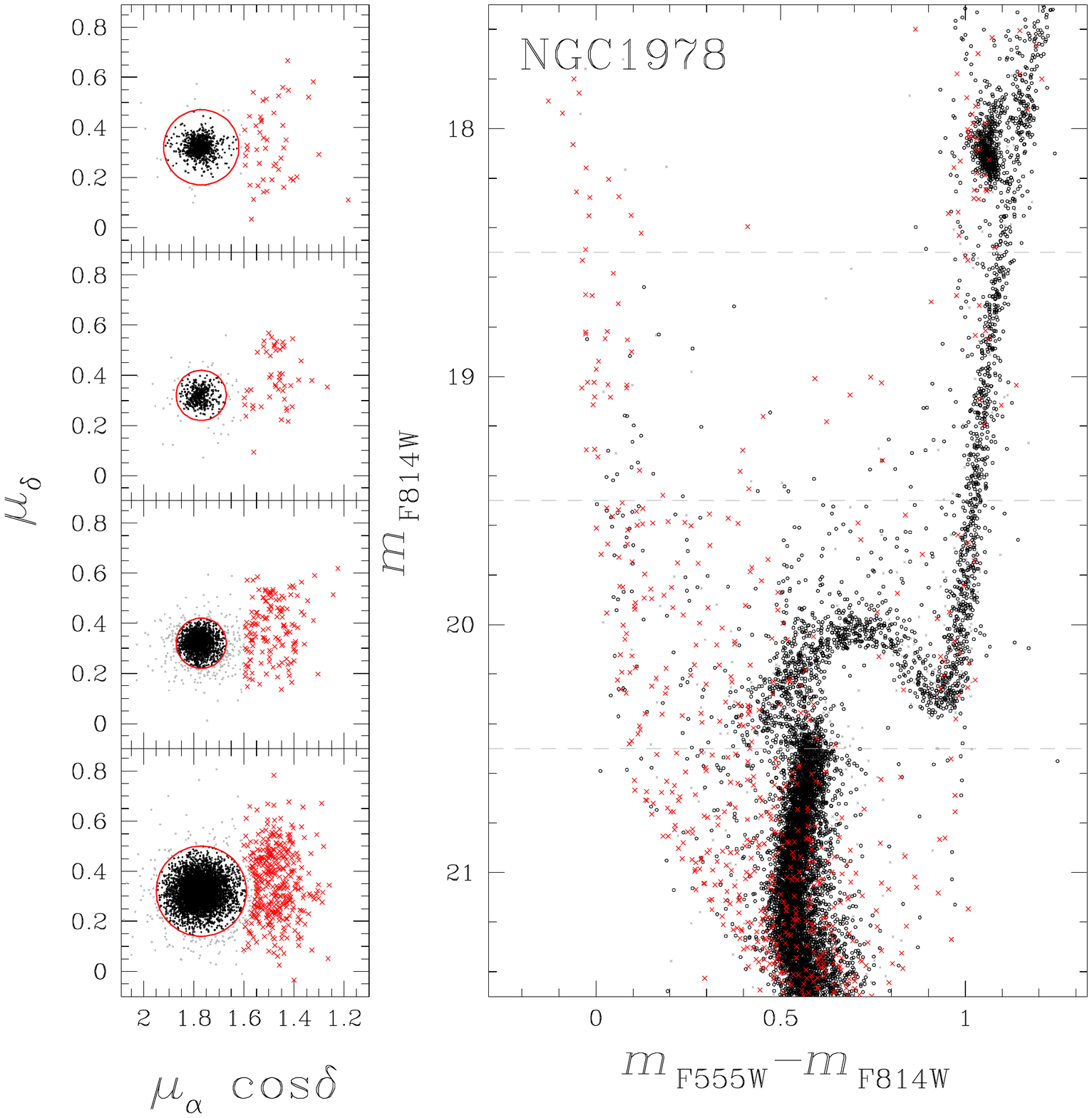}
 %/home/milone/NUBI/WFC3UVIS/NGC1978/DR2/cmd2.macro figpmV3 --- Canarie
 \caption{Proper motion diagrams of stars in the field of view of NGC\,1978 in four F814W magnitude intervals (left). The $m_{\rm F814W}$ vs.\,$m_{\rm F555W}-m_{\rm F814W}$ CMD of stars in the left panels is plotted on the right. Stars within the red circles plotted in the left panels are considered probable cluster members and are colored black, whereas  the most-evident field stars are represented with red crosses. The remaining stars are colored gray. See the text for details.}
 \label{fig:NGC1978vi} 
\end{figure}

The proper motion diagrams for NGC\,1978 stars are plotted in the left panels of Figure\,\ref{fig:NGC1978vi} in four distinct magnitude bins. These diagrams can be  used to separate the bulk of cluster members (black dots) from  probable field stars (red crosses). Here, the red circles that enclose the NGC\,1978 stars have radii equal to 2.5$\sigma$, where $\sigma$ is the average between the $\sigma-$clipped dispersion values of $\mu_{\alpha} {\rm cos} \delta$ and $\mu_{\delta}$.  For illustration purposes, we only mark in red the most-evident field stars with $\mu_{\rm \alpha} \cos{\delta} >1.6$ mas/yr and a distance of more than 0.2 mas/yr from the average motion of NGC\,1978, while the remaining stars are colored gray. 
 The  $m_{\rm F814W}$ vs.\,$m_{\rm F555W}-m_{\rm F814W}$ CMD of probable cluster members and field stars is shown in the right panel of  Figure\,\ref{fig:NGC1978vi}.
%% %%%%%%%%%%%%%%%%%%%%%%%%%%%%%%%%%%%%%%%%%%%%%%%%%%%%%%%%%%%%%%%%%%%%%%%%%%
%%%%%%%%%%%%%%%%%%%%%%%%%%%%%%%%
%\begin{centering} 
%%%%%%%%%%%%%%%%%%%%%%%%%%%%%%%%

\section{A saucerful of secrets}\label{sec:secrets}
The photometry and astrometry of this work are  exquisite tools  to investigate various astrophysical topics. In this section, we provide further examples of science outcomes that arise from visual inspections of the photometric diagrams and of the proper-motion diagrams.  
Specifically, in Section\,\ref{sec:NEWeMSTO} we report the discovery of eMSTOs in the clusters KMHK\,361 and NGC\,265. Section\,\ref{sec:AGEemsto} compares the CMDs of LMC clusters younger than $\sim$2.3 Gyr and investigates the color and magnitude distribution of eMSTO in clusters with different ages. 
Gaps and color discontinuities along the MS of NGC\,1783 are investigated in Section\,\ref{sub:zigzag} while Section\,\ref{sec:NGC1783} provides evidence of new features along the eMSTO and the upper MS of NGC\,1783. 
Finally, Section\,\ref{sec:pm} is focused on the proper motions of the star clusters and of Magellanic Cloud stellar populations in eleven fields.

\subsection{Clusters without previous evidence of eMSTO}\label{sec:NEWeMSTO}
\begin{figure*} 
\centering
  \includegraphics[height=6.5cm,trim={0.5cm 5cm 0.2cm 4.5cm},clip]{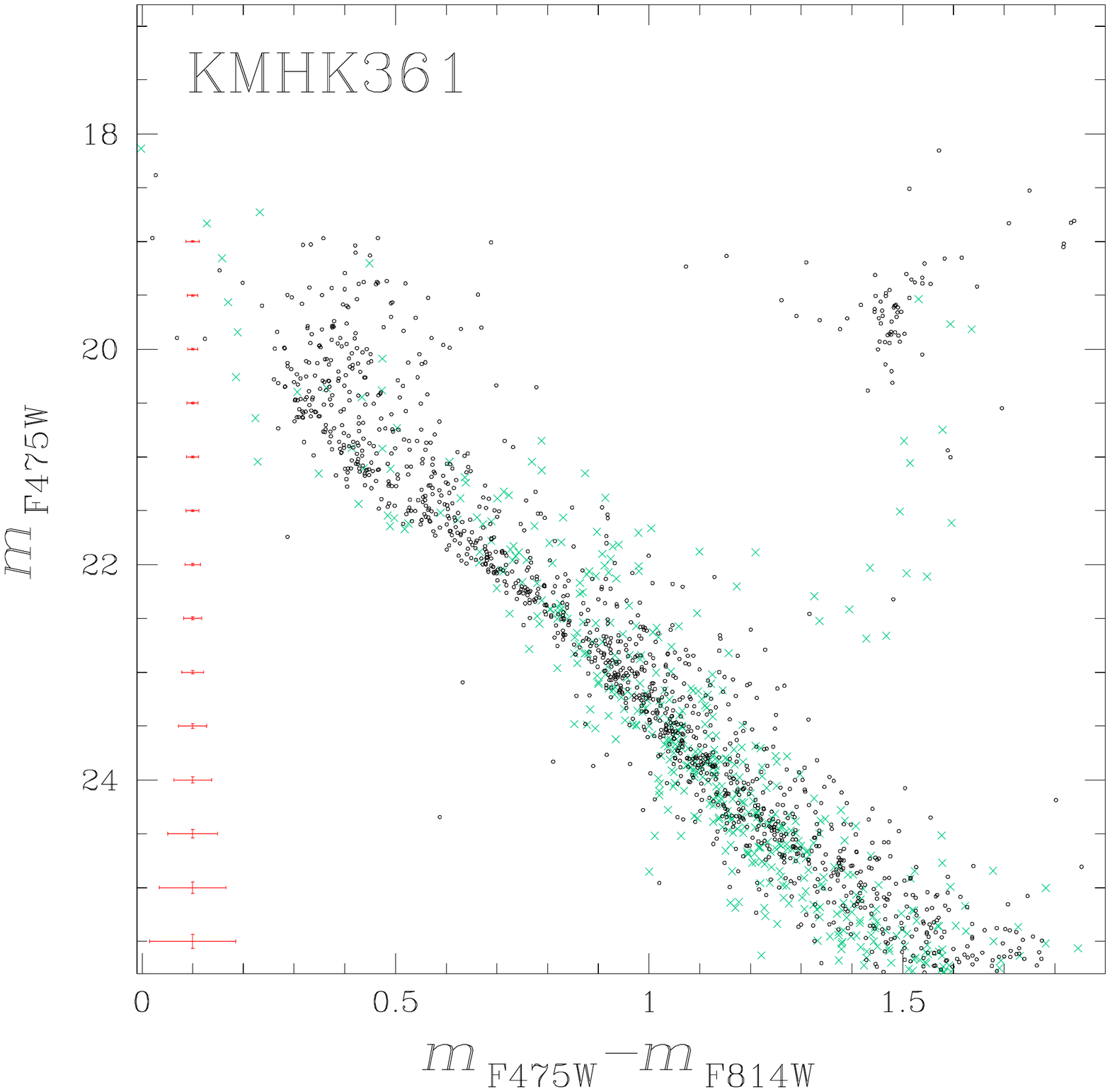}
    \includegraphics[height=6.5cm,trim={0.5cm 5cm 0.2cm 4.5cm},clip]{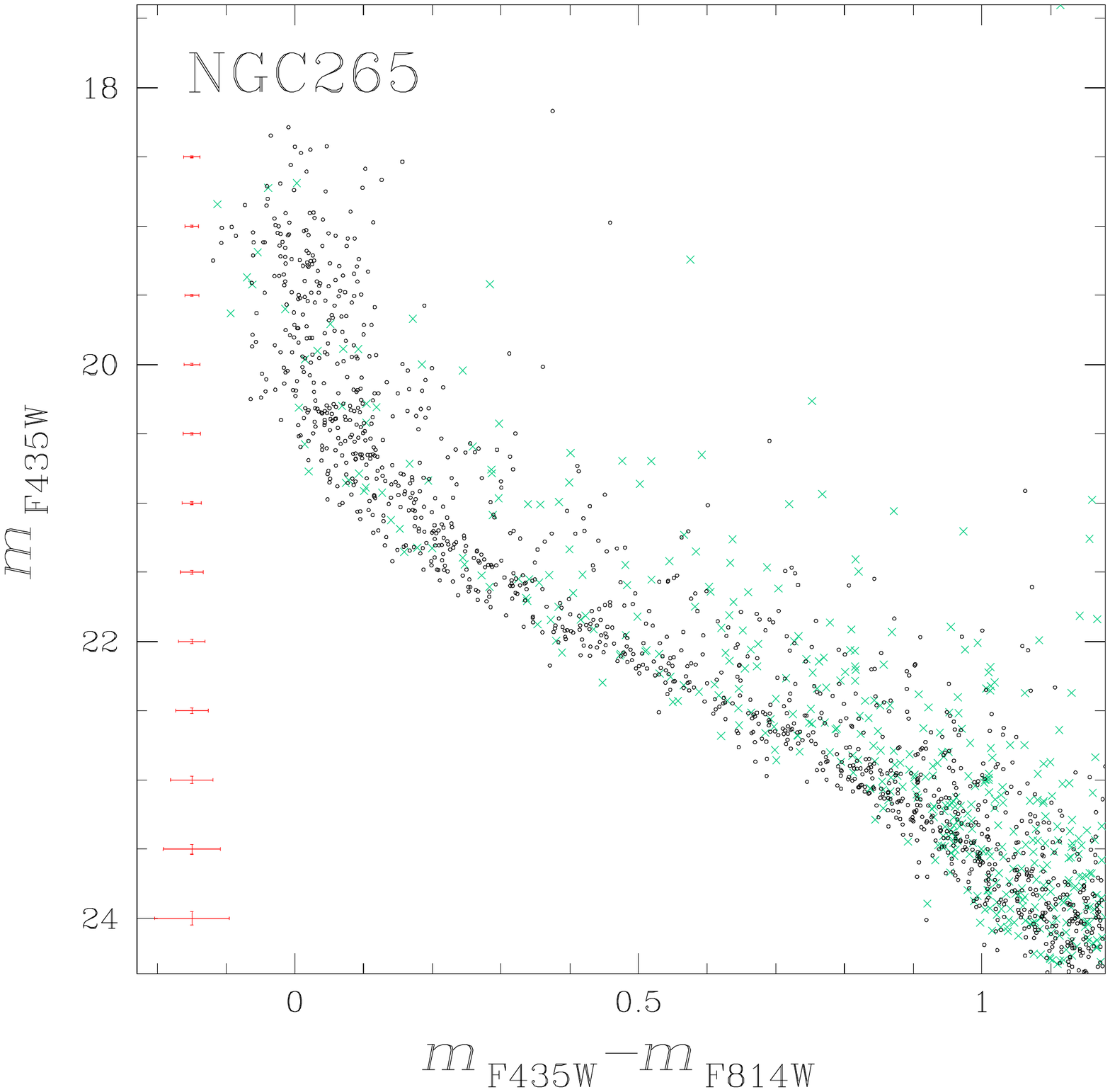}
 %/home/milone/
\caption{CMDs of the clusters KMHK\,361 and NGC\,265 without previous evidence of eMSTOs. Stars in the cluster field and reference field of each cluster are represented with black points and aqua crosses, respectively. See text for details.}
 \label{fig:NEWeMSTO} 
\end{figure*}

Figure\,\ref{fig:NEWeMSTO} provides evidence that the CMDs of the star clusters KMHK\,361 (age of 1.35 Gyr) and NGC\,265 (age of 450 Myr) are not consistent with a single isochrone.
 In this figure, we compare the CMDs of stars in circular fields centered on the cluster (hereafter cluster fields) and in reference fields of the same area. We adopted radii of 20 and 24 arcsec for KMHK\,361 and NGC\,265, respectively, enclosing the bulk of cluster stars.  To minimize the contamination from cluster stars, the reference fields are as
  far away from the cluster centers as possible, while still being within the FoV.
 %placed at the maximum distance from cluster centers to minimize the contamination from cluster stars. 
 By assuming a uniform distribution of field stars in the small {\it HST} field of view, the distribution of stars in the reference-field CMD is indicative of the  contamination due to field stars. 
 
 Clearly, KMHK\,361 exhibits an eMSTO, which cannot  be explained by field-star contamination alone. 
 Similarly, NGC\,265 shows an intrinsic eMSTO. The upper MS is split in the F435W magnitude interval between $\sim$21 and 22 mag, with the red MS hosting about two-thirds of MS stars. The two MSs merge around $m_{\rm F435W} \sim 22.5$ mag. The comparison between the CMDs of stars in the field and reference fields reveals that the split MS and the eMSTO are not due to field-star contamination.
 
 The visual inspection of the CMDs from our survey suggests that all clusters with ages between $\sim 0.1$ and $\sim$2.3 Gyr exhibit the eMSTO (Cordoni et al.\,in preparation).
 These findings corroborate the evidence that eMSTOs are common features of clusters younger than $\sim$2.3 Gyr, while split MSs are widespread phenomena among clusters younger than $\sim$0.8 Gyr \citep[e.g.][]{milone2009a, milone2022a, niederhofer2015a, goudfrooij2011a, li2017a, correnti2017a}. 
%% %%%%%%%%%%%%%%%%%%%%%%%%%%%%%%%%%%%%%%%%%%%%%%%%%%%%%%%%%%%%%%%%%%%%%%%%%%

%%%%%%%%%%%%%%%%%%%%%%%%%%%%%%%%%%%%%%%%%%%%%%%%%%%%%%%%%%%%%%%%%%%%%%%%%%%%%

 %%%%%%%%%%%%%%%%%%%%%%%%%%%%%%%%%%%%%%%%%%%%%%%%%%%%%%%%%%%%%%%%%%%%%%%%%%%%

\subsection{The eMSTO in clusters of different age}\label{sec:AGEemsto}
%%%%%%%%%%%%%%%%%%%%%%%%%%%%%%%%
\begin{centering} 
\begin{figure*} 
 \includegraphics[height=17cm,trim={0.1cm 5cm 0.0cm 4.5cm},clip]{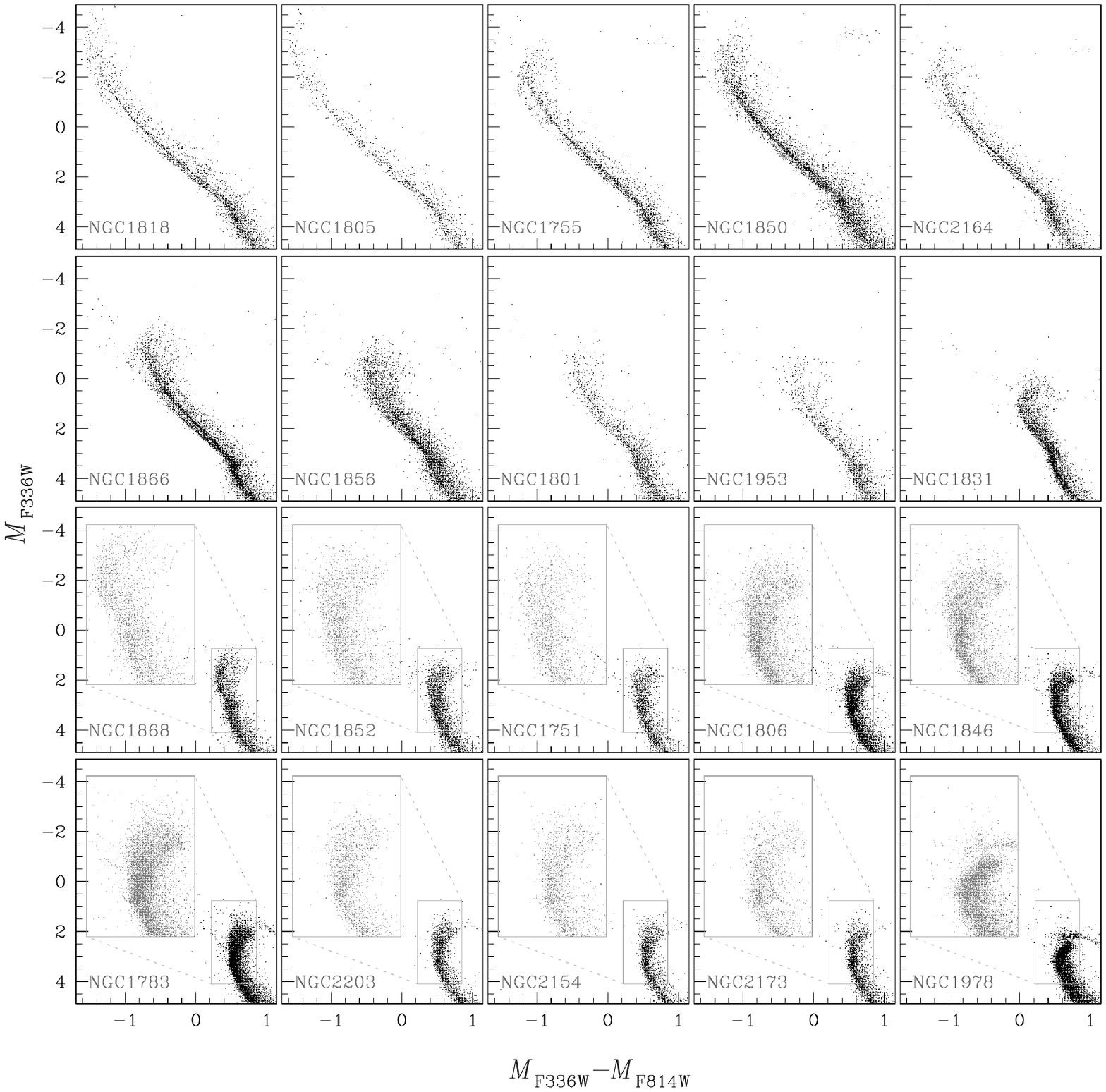}
 %/home/milone/NUBI/CATALOGHI/emsto.macrp go336l
 \caption{Collection of $M_{\rm F336W}$ vs.\,$M_{\rm F336W}-M_{\rm F814W}$ CMDs of LMC clusters younger than 2.5 Gyr. All panels have the same scale and are zoomed around the MS, while the insets highlight the MSTO. Clusters are sorted by age.}
 %APM correggere per DR
 %APM NGC1850 stelle brillanti
 %APM Hess diagram ?
 \label{fig:emstoLMC} 
\end{figure*} 
\end{centering} 

Our dataset provides a unique opportunity for comparing CMDs of clusters with different ages derived with homogeneous methods. As an example, we take advantage of the collection of $M_{\rm F336W} vs.\, M_{\rm F336W}-M_{\rm F814W}$ CMDs shown in Figure \ref{fig:emstoLMC} to investigate how the eMSTO phenomenon changes as a function of cluster age. In this figure, star clusters %of Figure \ref{fig:emstoLMC} 
 are sorted by age, from $\sim$10 Myr (NGC\,1818) to $\sim$2.5 Gyr (NGC\,1978). 
 The observed magnitudes have been converted into absolute ones by  adopting the values of distance modulus and reddening listed in Table\,\ref{tab:info}. 

A visual inspection of this figure corroborates the previous conclusion that the split MS is visible in all LMC clusters younger than $\sim$800 Myr (from NGC\,1818 to NGC\,1953) and seems to disappear at older ages \citep{milone2018a}.
The color separation between the blue and red MSs approaches its maximum value around the MSTO  and decreases towards faint luminosities \citep{milone2016a}.
As pointed out by \citet{wang2022a}, the gap between the blue and red MSs significantly narrows down around $M_{\rm F336W}=1.0$ mag,  %which corresponds to 
 which is the luminosity level where the fraction of blue-MS stars approaches its minimum value \citep{milone2018a}.
Noticeably, this magnitude value  corresponds to an MS mass of $\sim$2.5 $\mathcal{M}_{\odot}$, where the slowly rotating component of MS field stars disappears \citep{zorec2012a}.

We also confirm that the eMSTO is a ubiquitous feature of LMC clusters younger than $\sim$2.3 Gyr. It is visible in all clusters where the turn-off is brighter than the MS bending around $M_{\rm F336W}=3.0$ mag and disappears in NGC\,1978 \citep[e.g.][]{milone2009a, goudfrooij2014a}. 
Since the MS bending is due to a change in the stellar structure, the eMSTO is  associated with stars with radiative envelopes alone. In addition, the split MS is visible among stars brighter than the MS bend.

\begin{centering} 
\begin{figure} 
 \includegraphics[height=8.0cm,trim={1.0cm 5.5cm 0.0cm 4.5cm},clip]{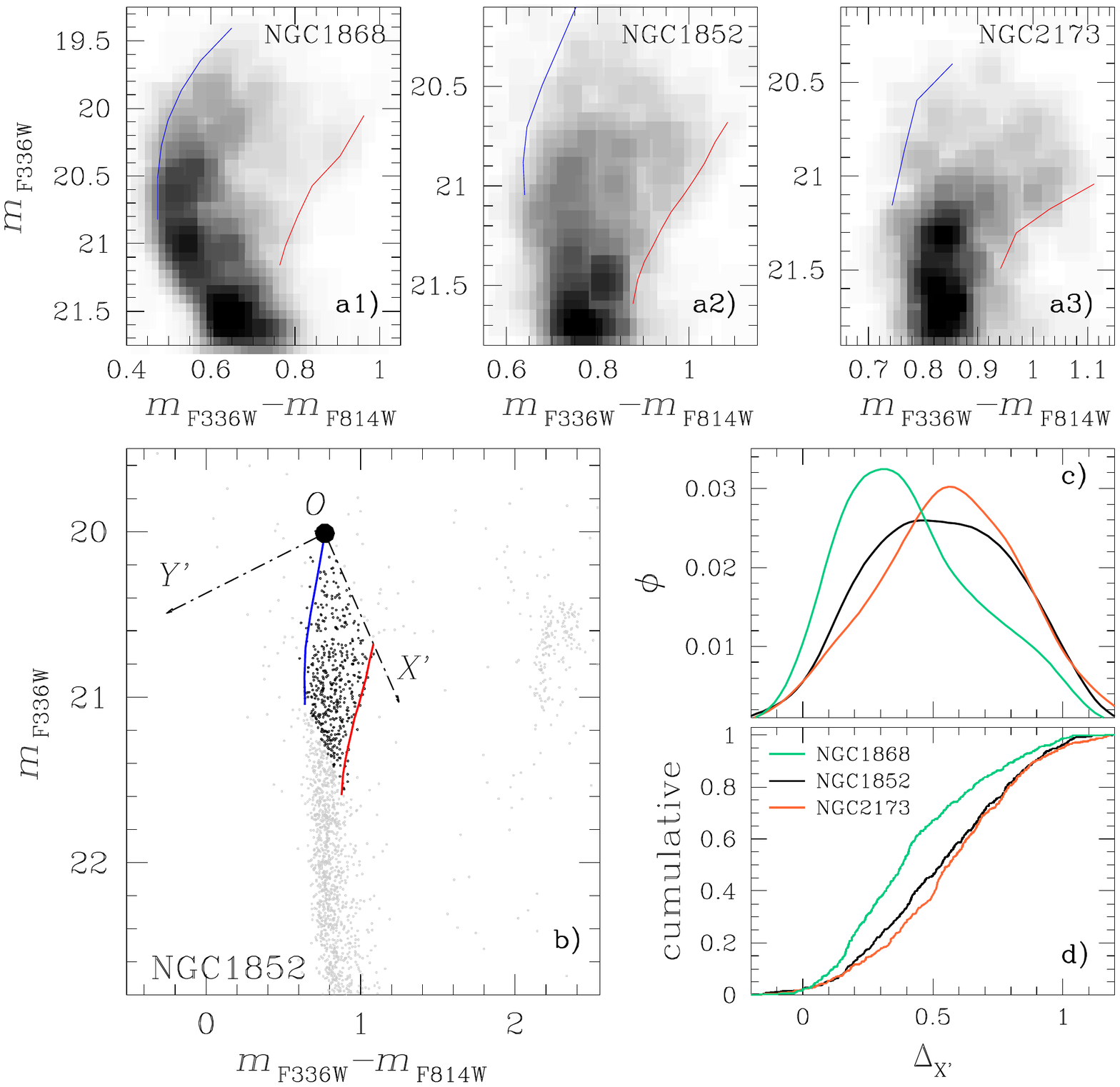}
 %/home/milone/NUBI/CATALOGHI/NGC1852/age.macro fig0
 \caption{$m_{\rm F336W}$ vs.\,$m_{\rm F336W}-m_{\rm F814W}$ Hess diagrams of NGC\,1868 (a1), NGC\,1852 (a2), and NGC\,2173 (a3) zoomed around the eMSTO. 
 The blue and red lines are the boundaries of the eMSTOs.  Panel b illustrates the scheme to derive the $\Delta_{X'}$ quantity for eMSTO stars, while the corresponding kernel-density distributions and cumulative distributions are plotted in panels c and d, respectively, for NGC\,1868 (aqua), NGC\,1852 (black) and NGC\,2173 (orange). See the text for details.}
 \label{fig:cumulative} 
\end{figure} 
\end{centering}

Figure\,\ref{fig:emstoLMC} reveals that the color and magnitude distributions of stars across the eMSTO significantly change from one cluster to another. As an example, the Hess diagrams plotted in the top panels (a1, a2, and a3) of Figure \ref{fig:cumulative} suggest that most TO stars of NGC\,1868 populate the bright and blue region of the eMSTO, whereas NGC\,2173 shows higher stellar density on the bottom-red side of its eMSTO. NGC\,1852 seems to show an intermediate distribution. 
 
  To parametrize the stellar distribution of eMSTO stars in the CMDs, %of clusters older than $\sim 20$Myr, 
   we adopted the procedure illustrated in Figure \ref{fig:cumulative}b for NGC\,1852.
  We defined a new reference frame where the origin, {\it O}, is set by hand on the bright and blue side of the eMSTO, and the abscissa, {\it X'}, envelopes the bright part of the eMSTO and points towards the red.
  We derived the red and blue fiducials of the eMSTO in the new  reference frame and represented them as red and blue lines in the CMD of Figure \ref{fig:cumulative}b. To derive the fiducials, we follow the recipe by \citet{milone2017a}, which is based on the naive estimator \citep{silverman1986a}. We first divided the eMSTO into a series of bins with fixed  pseudo-magnitude,  {\it $\delta$Y'}. The bins are defined over a grid of points separated by intervals of  fixed pseudo-magnitude (s=$\delta$Y/3). For each interval, we calculate the 4$^{\rm th}$ and the 96$^{\rm th}$ percentile of the {\it X'} distribution and associated these values with the mean pseudo-magnitude {\it Y'} of stars in the bin. These values are then linearly interpolated to derive the red and blue boundaries of the eMSTO. These lines are used to calculate the quantity
  \begin{equation}
\Delta_{\rm X'}=\frac {X'-X'_{\rm blue\,fiducial}} {X'_{\rm red\,fiducial}-X'_{\rm blue\,fiducial}}  
  \end{equation}
   that is defined in such a way that the stars on the blue and red fiducials have $\Delta_{X'}$=0 and 1, respectively.
  
   Figure\,\ref{fig:cumulative} compares the kernel density  (panel c) and the cumulative distributions of $\Delta_{X'}$ (panel d) for NGC\,1852 (black), NGC\,1868 (aqua), and NGC\,2173 (orange).
   We confirm the visual impression of a predominance of blue eMSTO stars in NGC\,1868, whereas the $\Delta_{ X'}$ distribution of NGC\,2173 is peaked towards the red. NGC\,1852 has an intermediate distribution.

%%%%%%%%%%%%%%%%%%%%%%%%%%%%%%%%
%%%%%%%%%%%%%%%%%%%%%%%%%%%%%%%%%

\begin{figure} 
\centering 
 \includegraphics[height=5.0cm,trim={0.0cm 5.5cm 0.0cm 12.0cm},clip]{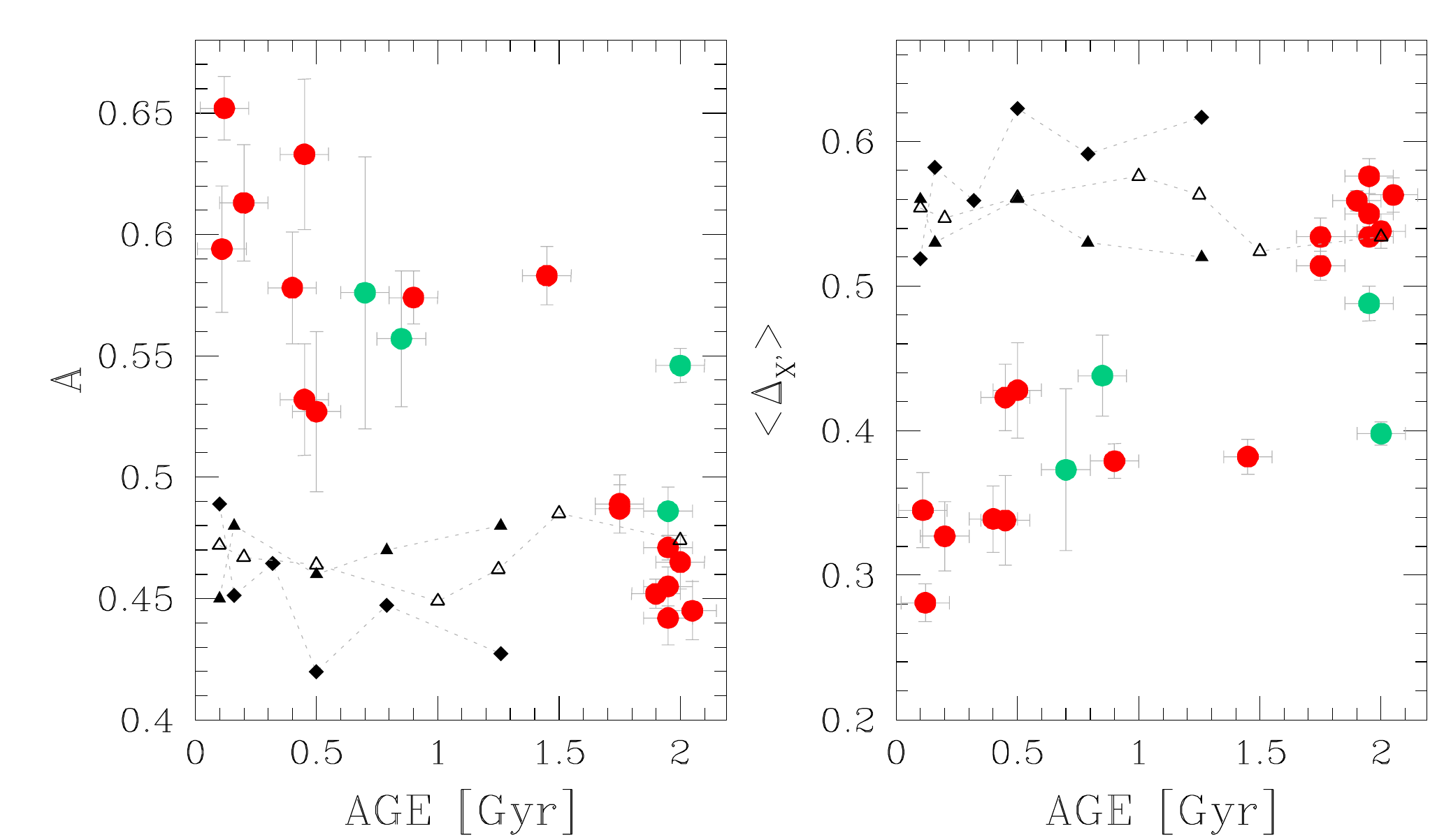}
 %/home/milone/NUBI/CATALOGHI/NGC1852
 %
 % /home/milone/PAPERS/draft/APMnubi/VARIE
 \caption{Area below the $\Delta_{\rm X'}$ cumulative curve, A, (left) and median $\Delta_{\rm X'}$ value as a function of cluster age for LMC (red dots) and SMC (aqua dots) clusters with the eMSTO.  Open and filled triangles are inferred from simulated CMDs of non-rotating stellar populations with different ages  derived from the Padova and Geneva database, respectively. The diamonds correspond to coeval stellar populations with different rotation rates from the Geneva database. 
See text for details.}
 \label{fig:RelEMSTO} 
\end{figure}

To quantify the $\Delta_{X'}$ differences among the various clusters, we define two quantities: i) the area, A, below the cumulative curve shown in Figure\,\ref{fig:cumulative} and ii) the median value of $\Delta_{\rm X'}$, $<\Delta_{\rm X'}>$. If the distribution is dominated by blue and bright MSTO stars we would expect large values of A and small values of $<\Delta_{\rm X'}>$, while a predominance of faint and red eMSTO stars corresponds to small A and large $<\Delta_{\rm X'}>$. Results are shown in Figure\,\ref{fig:RelEMSTO}, where we plot both quantities against cluster age. 
%XXXXXXXXXXXXXX
  LMC clusters % older than 1 Gyr 
  (red dots in Figure\,\ref{fig:RelEMSTO}) exhibit a strong anti-correlation between A and age and a correlation between  $<\Delta_{\rm X'}>$ and age, as also indicated by the values of the Spearman's rank correlation coefficients of $-$0.91 and 0.92, respectively.  
Intriguingly, the $\sim$2 Gyr old SMC clusters
 %(red dots in Figure \ref{fig:RelEMSTO})
  NGC\,411 and NGC\,416 exhibit larger values of A and smaller values of  $<\Delta_{\rm X'}>$ than LMC clusters with similar ages, with NGC\,411 having the largest differences. The small statistical sample of clusters prevents us from reaching a firm conclusion on whether NGC\,411 is an outlier or SMC and LMC clusters exhibit different trends. %, possibly due to a dependence on the environment or the different metallicity.

The multiple populations of young and  intermediate-age star clusters share common features that have been instrumental to shed light on the origin of split MSs and eMSTOs \citep[see][for a recent review]{milone2022a}. 
As an example, the eMSTO width depends on cluster age. Specifically, if the eMSTO is interpreted as an age spread, the resulting age range is proportional to cluster age \citep[e.g.][]{niederhofer2015a, cordoni2018a}.
Moreover, the fractions of stars along the blue and the red MS correlate with stellar mass. The fraction of blue-MS stars varies from $\sim 40\%$ among stars with masses of $\sim 1.5 \mathcal{M_{\odot}}$ to $\sim 15\%$  among $\sim 2.5-3.0 \mathcal{M_{\odot}}$ stars. It  arises again in more massive stars, up to $\sim 40\%$ in $\sim 5.0 \mathcal{M_{\odot}}$-stars.
The fractions of blue- and red-MS stars do not depend on other properties of the host cluster like the global cluster's mass \citep[][]{milone2018a}. These results have been instrumental to demonstrate that rotation plays a major role in shaping the eMSTOs and the split MSs of Magellanic-Cloud clusters.

The evidence that the  $\Delta_{X'}$ distribution of stars along the eMSTO depends on cluster age provides a potential further constraint to the eMSTO phenomenon.
To start investigating the physical reasons responsible for the relations shown in Figure \ref{fig:RelEMSTO}, we used stellar models from the Padova database  \citep{marigo2017a} to simulate a group of CMDs of non-rotating stellar populations with ages of 100, 200, 500, 1,000, 1,250, 1,500 and 2,000 Myrs and internal age spreads. We assumed a flat distribution and maximum width corresponding to the average age variations inferred by \citet{cordoni2018a} for Magellanic Cloud clusters with the same age.  We derived the A and  $<\Delta_{\rm X'}>$ quantities for each simulated CMD by using the same procedure adopted for real stars and plotted the resulting values against the oldest age of the simulated stellar population (open triangles of Figure\,\ref{fig:RelEMSTO}). 

 Similarly, we simulated another group of CMDs for coeval stellar populations where 33\% of stars have no rotation, whereas the remaining  67\% of stars have rotation equal to 0.9 times the breakout value. The simulated diagrams have ages of  100, 150, 500, 800, and 1,250 Myr and are derived by means of Geneva models \citep{ekstrom2012a, ekstrom2013a, mowlavi2012a, wu2016a}. 
 We assumed random viewing-angle distributions and adopted the gravity-darkening model by \citet{espinosa2011a} and the limb-darkening effect \citep{claret2000a}. Stellar magnitudes for the available {\it HST} filters have been derived using the model atmospheres by Castelli \& Kurucz (2003). 
 The resulting A and  $<\Delta_{\rm X'}>$ quantities are represented with  filled diamonds in Figure\,\ref{fig:RelEMSTO}. 
  For completeness, we used the Geneva models to simulate non-rotating stellar populations with internal age spreads, in close analogy with what we did with the Padova models. Results are represented with filled triangles.

 Clearly, the  A and  $<\Delta_{\rm X'}>$ quantities inferred from both groups of simulated diagrams provide poor fits to the observations. This fact indicates that internal age variation alone is not responsible for the eMSTO when we assume a flat age distribution for all clusters.
 Similarly, rotation alone is not responsible for the eMSTO when we assume two populations for all clusters: one of non-rotating stars and one of fast rotators with $\omega=0.9\omega_{c}$.
 
  It is now widely accepted that the luminosity of eMSTO stars depends on gravity darkening and that its  effect is strong for large values of the ratio between the rotational velocity and the critical velocity. Our results could indicate that this ratio increases when stars age on the MS as suggested by \citet{hastings2020a}.
 
 To properly constrain the contribution of rotation and age variation on the eMSTO, it is mandatory to extend the analysis to simulated diagrams that account for different internal age distributions, different rotation-rate distributions \citep[e.g.][]{huang2006a, huang2010a, goudfrooij2018a}, and that account for both age variations and stellar populations with different rotation rates. 
 
  The interpretation of the eMSTO phenomenon should also account for binary evolution effects \citep[e.g.][]{wang2022a}. As an example, the stellar models by \citet{wang2020a} show that the fraction of evolutionary-driven mergers rises for smaller stellar masses, at the expense of the binaries that survive the mass transfer and produce spun-up accretors. An appropriate comparison between the observations illustrated in  Figures\,\ref{fig:cumulative} and \ref{fig:RelEMSTO} and the predictions of stellar models that account for binary evolution is mandatory to shed light on the effect of binary evolution on the eMSTO.  
%APM sarebbe utile confronto in Fig. 9 con CMD simulati di diversa eta' e stesso spread di eta' e poi rotazione.

%%%%%%%%%%%%%%%%%%%%%%%%%%%%%%%%

%%%%%%%%%%%%%%%%%%%%%%%%%%%%%%%%%

\subsection{A population of UV-dim stars along the eMSTO of NGC\,1783}\label{sub:1783emsto}
  %%%%%%%%%%%%%%%%%%%%%%%%%%%%%%%%
\begin{centering} 
\begin{figure*} 
 \includegraphics[height=9.0cm,trim={0.1cm 4.cm 0.0cm 3.5cm},clip]{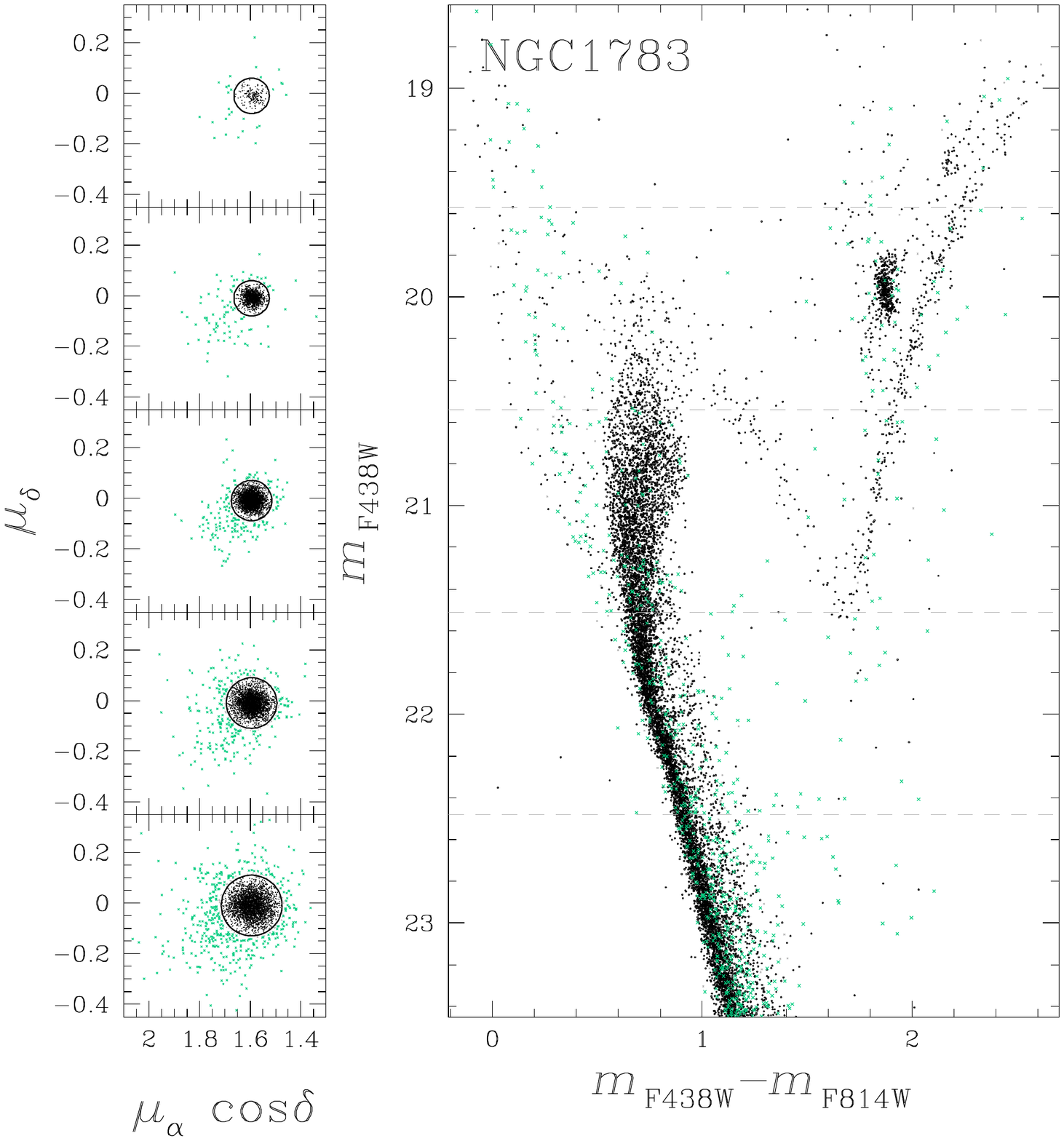}
  \includegraphics[height=9.0cm,trim={0.1cm 4.cm 0.0cm 3.5cm},clip]{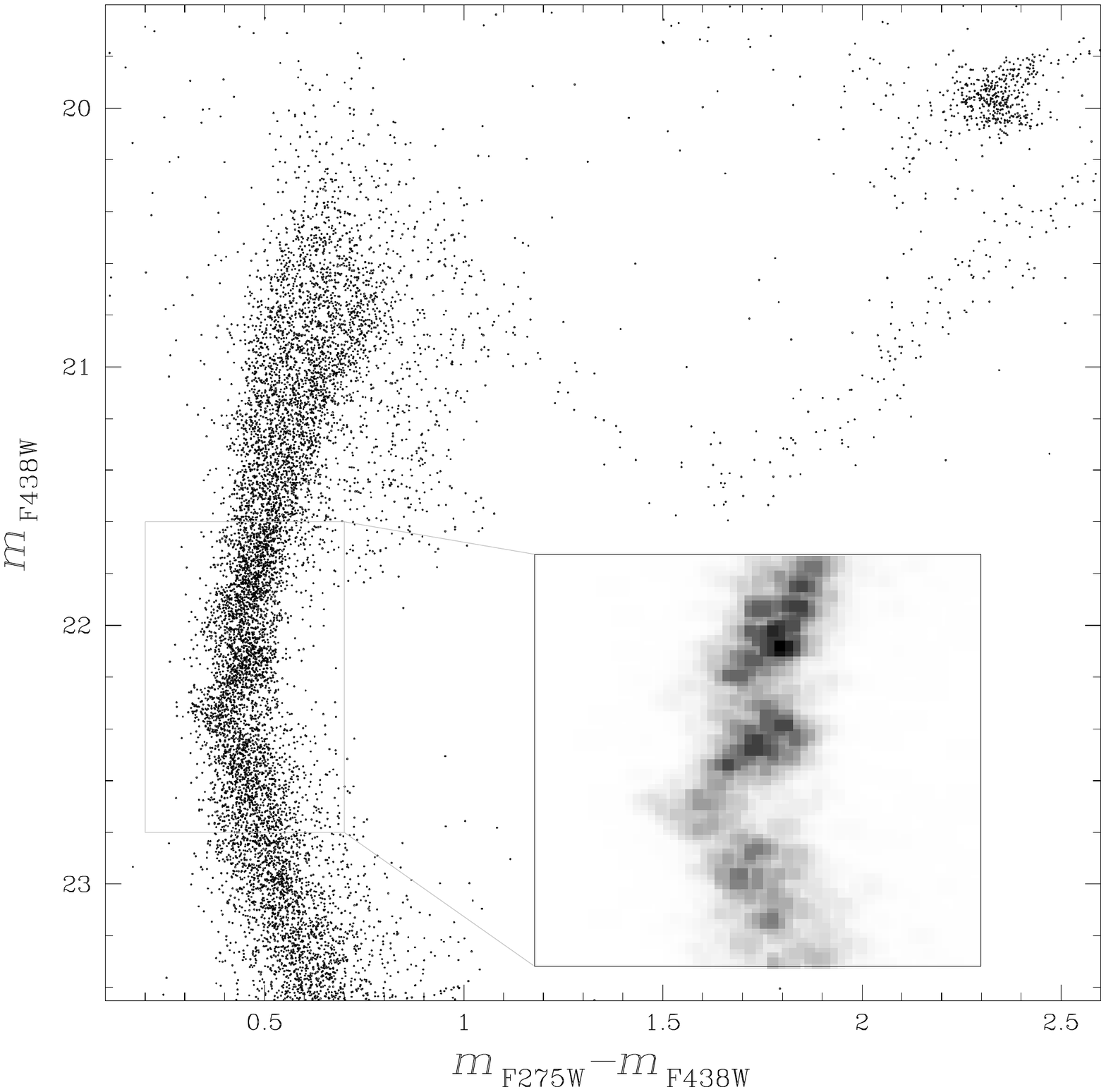}
 %/home/milone
 \caption{Proper motion diagrams of stars in the field of view of NGC\,1783 in five F438W magnitude intervals (left). The $m_{\rm F438W}$ vs.\,$m_{\rm F438W}-m_{\rm F814W}$ CMD for stars in the left panels is plotted on the middle. Stars within the black circles plotted in the proper motion diagrams are considered probable cluster members and are colored black, whereas field stars are represented with aqua crosses. The right panel shows the $m_{\rm F438W}$ vs.\,$m_{\rm F275W}-m_{\rm F438W}$ CMD for probable cluster members, while the inset represents the Hess diagram of the CMD region around the upper MS.} 
 \label{fig:NGC1783pm} 
\end{figure*} 
\end{centering} 
%%%%%%%%%%%%%%%%%%%%%%%%%%%%%%%%
The stellar proper motions derived from our dataset allow the partial separation of bright field stars from NGC\,1783 cluster members, thus providing new insights on its stellar populations. 
The left panels of Figure\,\ref{fig:NGC1783pm} show the proper-motion diagram for stars in the field of view of NGC\,1783 in five magnitude bins. The black circles are centered on the absolute proper motion of NGC\,1783, and are used to separate probable cluster members (black points) from field stars (aqua crosses). 

The corresponding $m_{\rm F438W}$ vs.\,$m_{\rm F438W}-m_{\rm F814W}$ CMD (middle panel) highlights several characteristics of NGC\,1783 in unprecedented detail. These include the eMSTO \citep{mackey2008a, milone2009a, goudfrooij2014a} together with a well-populated sequence of MS-MS binaries with large mass ratio \citep{milone2009a}. The SGB also exhibits intrinsic broadening in color and magnitude, with the majority of stars populating the upper SGB.  
Moreover, the CMD reveals a broad, possibly dual, sequence of stars  brighter and bluer than the turn-off.
 This blue sequence, which will be investigated in detail in Section\,\ref{sec:NGC1783} was first identified by \citet{li2016a} who associated it with the young stellar populations within NGC\,1783. Their result has been challenged by \citet{cabreraziri2016a} who suggested that the blue sequence is composed of field stars. 

\begin{figure} 
\centering
 \includegraphics[height=5.0cm,trim={1.2cm 6.0cm 0.0cm 12.0cm},clip]{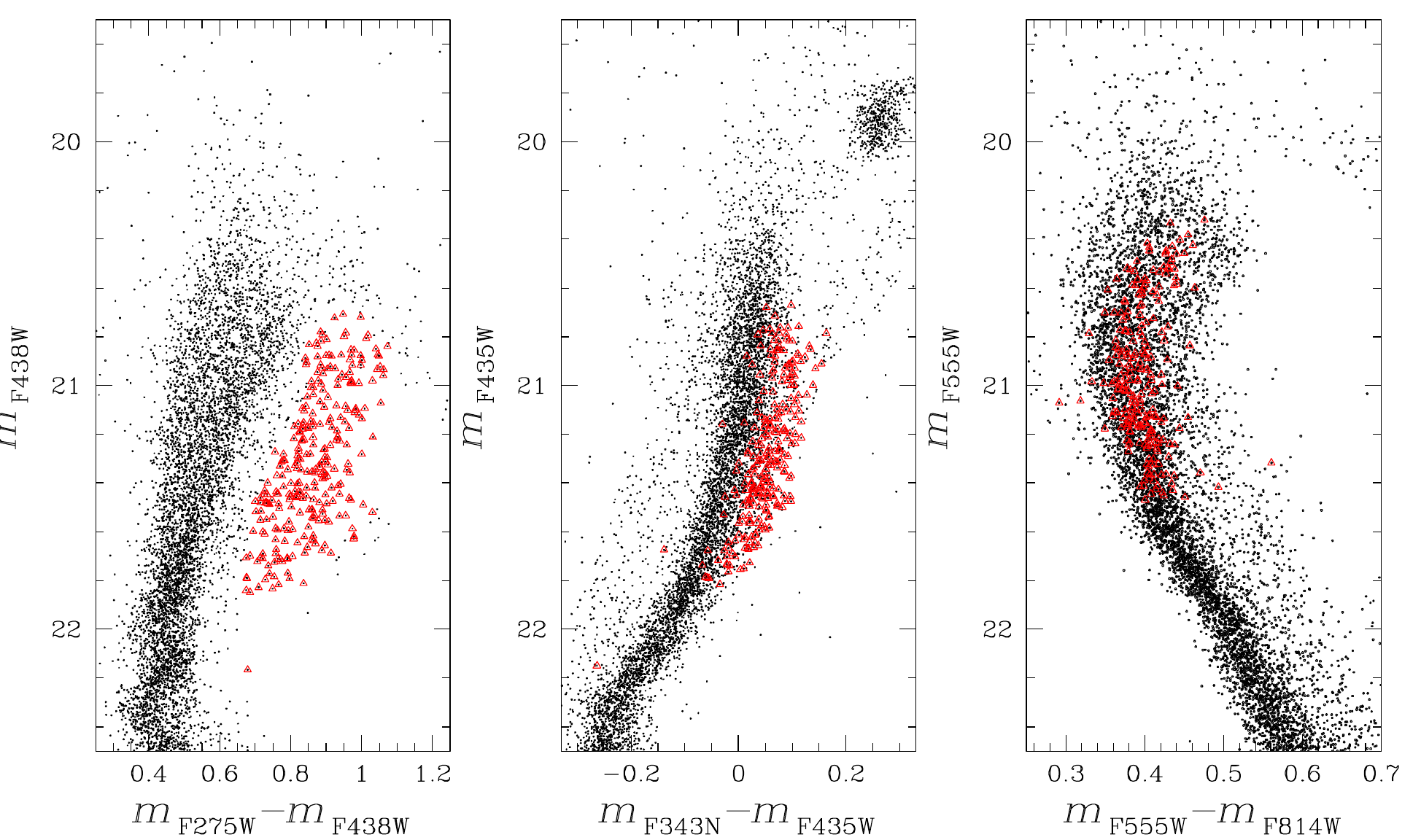}
 %/home/milone
 \caption{$m_{\rm F438W}$ vs.\,$m_{\rm F275W}-m_{\rm F438W}$ (left), $m_{\rm F435W}$ vs.\,$m_{\rm F343N}-m_{\rm F435W}$ (middle), and $m_{\rm F555W}$ vs.\,$m_{\rm F555W}-m_{\rm F814W}$ (right) CMDs of proper-motion selected NGC\,1783 stars. Stars in the red tail of the eMSTO, selected from the left-panel CMD, are colored red.}
 \label{fig:NGC1783emsto} 
\end{figure}

Here, we focus on the $m_{\rm F438W}$ vs.\,$m_{\rm F275W}-m_{\rm F438W}$ CMD of NGC\,1783, which is illustrated in the right panel of Figure\,\ref{fig:NGC1783pm}. 
An unexpected feature of this CMD is the sparse cloud of stars on the red side of the eMSTO. These stars, which we dub UV-dim, are marked with red triangles in the left panel of Figure\,\ref{fig:NGC1783emsto} where we reproduce the $m_{\rm F438W}$ vs.\,$m_{\rm F275W}-m_{\rm F438W}$ CMD  zoomed around the eMSTO.
UV-dim stars comprise a small fraction of $\sim$7\% of the total number of eMSTO stars with $20.4<m_{\rm F555W}<21.5$ mag.
We used the same colors to represent these stars in the other panels of Figure\,\ref{fig:NGC1783emsto} and showed that they define distinct sequences in both $m_{\rm F435W}$ vs.\,$m_{\rm F343N}-m_{\rm F435W}$ and $m_{\rm F555W}$ vs.\,$m_{\rm F555W}-m_{\rm F814W}$ CMDs. If the extreme position in the left-panel CMD is due to observational errors alone, the selected stars would have the same probability of having redder or bluer $m_{\rm F343N}-m_{\rm F435W}$ and $m_{\rm F555W}-m_{\rm F814W}$ colors than the bulk of MSTO stars. On the contrary, the presence of distinct sequences demonstrates that the extreme red $m_{\rm F275W}-m_{\rm F438W}$ colors of the selected stars are intrinsic.
%Qualitatively, UV-dim stars share some similar behaviors as 
 We note that the Be stars, which are commonly observed in Magellanic Cloud clusters younger than $\sim 300$ Myr \citep{keller2000a, bastian2017a, correnti2017a, milone2018a},  also exhibit redder colors  than the bulk of eMSTO stars in CMDs composed of F275W and F336W filters.

%In an attempt to investigate the physical reasons responsible for this new feature of the eMSTO, we speculate on the origin of eMSTO with extreme F275W$-$F438W colors.  
In the following, we explore the possibility that the extreme F275W$-$F438W colors of UV-dim stars are an effect connected to the stellar rotation.
%  
% APM Comparison with simulated CMDs (marcella)
% rotation
% effect 
  It is well known that stellar rotation diminishes the effective temperature and the luminosity of a star, with fast-rotating MSTO stars being redder and dimmer than slow rotators. The position of a star along the eMSTO depends on the effects of limb and gravity darkening and on the viewing angle of the stellar rotation axes with respect to the line of sight. In this context, UV-dim stars 
   %the cloud of eMSTO stars with large $m_{\rm F275W}-m_{\rm F438W}$ 
   would comprise of fast rotators that are seen equator-on, as these stars appear colder and fainter than pole-on fast rotators. 
  
\begin{figure} 
\centering
 \includegraphics[height=8.0cm,trim={0.5cm 6.0cm 0.0cm 5.0cm},clip]{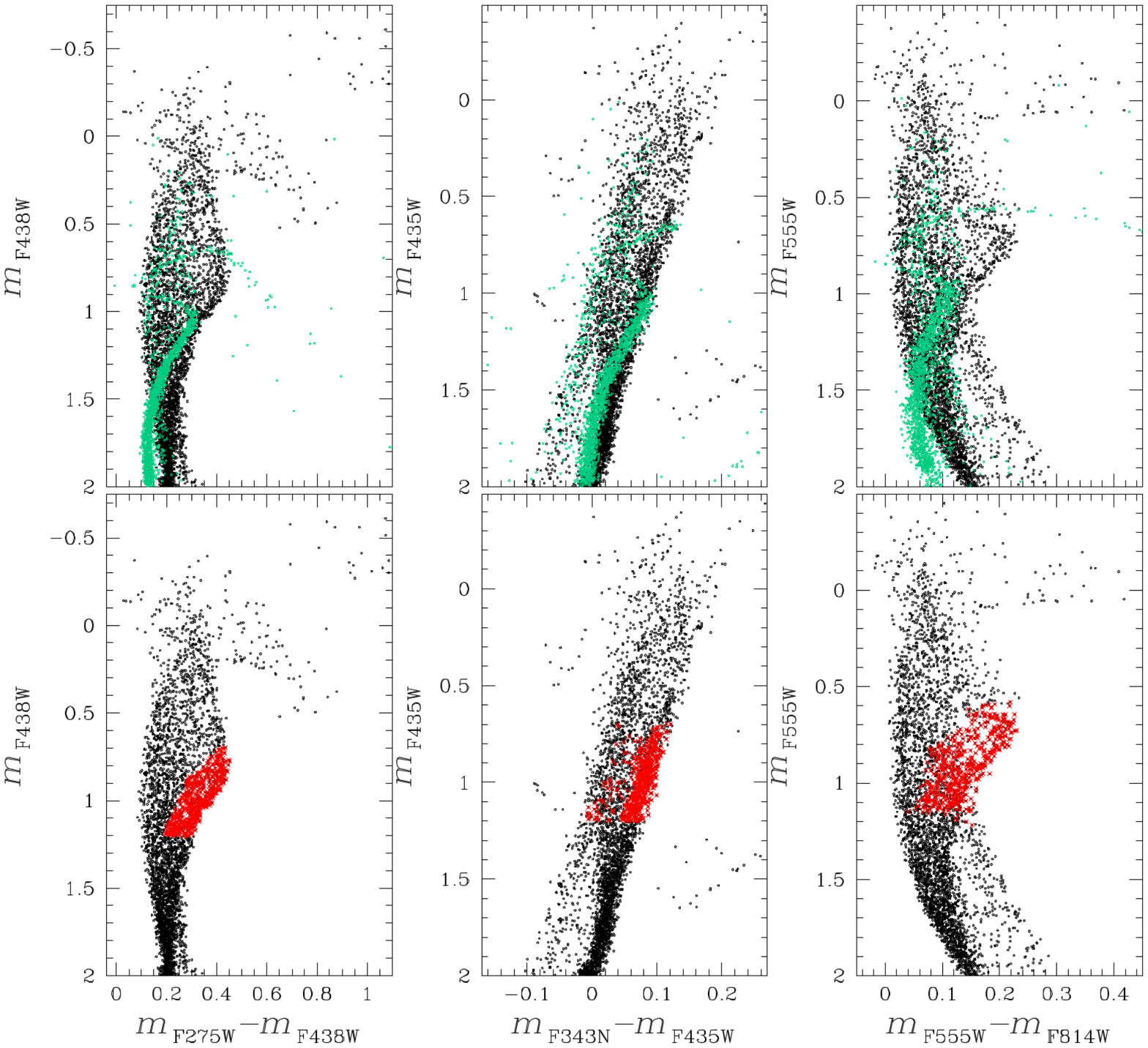}
 %/home/milone
 \caption{Simulated CMDs of two stellar populations of fast-rotating stars ($\omega/\omega_{\rm c}=0.9$, black points) and non-rotating stars (aqua points). Red points in the bottom-panel diagrams mark the sample of fast rotating stars selected by hand and located on the red side of the eMSTO in the $m_{\rm F438W}$ vs.\,$m_{\rm F275W}-m_{\rm F438W}$ CMD. Simulations are derived from Geneva isochrones.}
 \label{fig:CLOUDsim} 
\end{figure}

  %To qualitatively investigate the effect of rotation on the colors of eMSTO stars we used isochrones from the Geneva data base 
  To qualitatively explore this suggestion, we produced simulations based on the isochrones  from the Geneva database \citep[][]{mowlavi2012a, ekstrom2012a, ekstrom2013a,  georgy2014a} in close analogy with what we did in Section\,\ref{sec:AGEemsto}. 
 % generate the CMDs plotted in Figure\,\ref{fig:CLOUDsim} 
   In the top panels of Figure \ref{fig:CLOUDsim} we simulated a population of non-rotating stars (aqua points), which includes the 33\% of the total number of stars, and a population of fast rotators, where the stellar rotation corresponds to 0.9 times the breakout value ($\omega/\omega_{\rm c}=0.9$, black points). 
%%% 
%   We assumed random viewing-angle distributions, adopted the gravity-darkening model by \citet{espinosa2011a} and the limb-darkening effect \citep{claret2000a}. Stellar magnitudes into the ACS/WFC and UVIS/WFC3 filters have been derived by using the model atmospheres by Castelli \& Kurucz (2003). %\citet{castelli2003a}. 
   The simulated fraction of binaries is 0.3 and is similar to the observed binary fractions of intermediate-age LMC star clusters \citep{milone2009a}.

   We note that non-rotating stars are located on the red and faint side of the eMSTO in the $m_{\rm F438W}$ vs.\,$m_{\rm F275W}-m_{\rm F438W}$ and $m_{\rm F435W}$ vs.\,$m_{\rm F343N}-m_{\rm F435W}$ CMDs, similarly to the clouds of UV-dim stars observed in the corresponding diagrams of NGC\,1783. However, these stars define a narrow sequence that overlaps with the red portion of the eMSTO, in contrast with what is observed in the cloud of stars of NGC\,1783 where we observe a broad color distribution that extends towards the red of the bulk of eMSTO stars.  In addition, simulated stars exhibit fainter $m_{\rm F555W}$ magnitudes and redder $m_{\rm F555W}-m_{\rm F814W}$ colors than the bulk of eMSTO, in disagreement with what is observed in the $m_{\rm F555W}$ vs.\,$m_{\rm F555W}-m_{\rm F814W}$ CMD. Hence, we conclude that the  cloud of stars in NGC\,1783 is not consistent with a population of non-rotating stars. 
   
   In the bottom panels of Figure\,\ref{fig:CLOUDsim} we limit the analysis to fast-rotating stars only, where the position of a star in the eMSTO strongly depends on the limb and gravity darkening and on the viewing angle.
    We select by hand a sample of eMSTO stars with red $m_{\rm F275W}-m_{\rm F438W}$ colors and faint $m_{\rm F438W}$ magnitudes (red points in Figure\,\ref{fig:CLOUDsim}).
  Clearly, the hypothesis that stars with a certain range of viewing angle correspond to the cloud of NGC\,1783 stars is challenged by the position of the selected stars in the optical CMD, where they define a narrow sequence in the middle of the eMSTO.
  Further, it seems unlikely that the spread of the turnoff can be entirely attributed to the viewing angle of stars rotating close to breakout value $\sim$2 Gyrs after their formation.

   %As an alternative, we speculate that circumstellar disks are the responsible for the absorption of the UV radiation and for the consequent cloud of stars on the red side of the eMSTO.      In this context we notice that debris disks, possibly associated with planet formation, are frequently observed around A-type stars 
   As an alternative, circumstellar disks could be responsible for absorbing the UV radiation and the consequent cloud of stars on the red side of the eMSTO. Debris disks, possibly associated with planet formation, are frequently observed around A-type stars \citep[e.g.][and references therein]{eiroa2013a}.
   A challenge is the location of UV-dim stars in the $m_{\rm F555W}$ vs.\,$m_{\rm F555W}-m_{\rm F814W}$ CMD, which would imply that their disks poorly affect the emergent optical radiation. On the contrary, dust absorption is strongly dependent on the wavelength and is much more significant in the UV than in the optical. Hence, circumstellar dust in a disk could explain the location of these stars in the CMD (D’Antona et al., in preparation).
   
   In this scenario, the disks are associated with stars that are currently non-rotating so that they can distribute along a narrow sequence in the optical CMD. 
   %If the disk formation is associated with rotation, these stars should have experienced fast rotation in their lifetime. 
   % APM XXXXX questa frase sotto era rimossa nei commenti di Franca.
   If the disk formation is associated with rotation, these stars should have experienced fast rotation in their lifetime.
 % disk -- Be
% debree planets  

 %%%%%%%%%%%%%%%%%%%%%%%%%%%%%%%%
%%%%%%%%%%%%%%%%%%%%%%%%%%%%%%%%

 %%%%%%%%%%%%%%%%%%%%%%%%%%%%%%%%
%%%%%%%%%%%%%%%%%%%%%%%%%%%%%%%%

%\subsection{Gaps and discontinuities along the MS of NGC\,1783}
\subsection{A zigzag across the MS of NGC\,1783}\label{sub:zigzag}
A visual inspection of the CMD of Figure\,\ref{fig:NGC1783pm} shows another intriguing detail of the CMD of NGC\,1783. As highlighted by the Hess diagram in the inset, the upper MS runs in a zigzag, with two main discontinuities ($m_{\rm F438W} \sim 22.0$ and 22.4) and various sudden changes of slope around $m_{\rm F438W}=21.9, 22.1, 22.3$ and 22.5.
 
\begin{centering} 
\begin{figure} 
 \includegraphics[height=4.5cm,trim={0cm 0.0cm 0.0cm 0.0cm},clip]{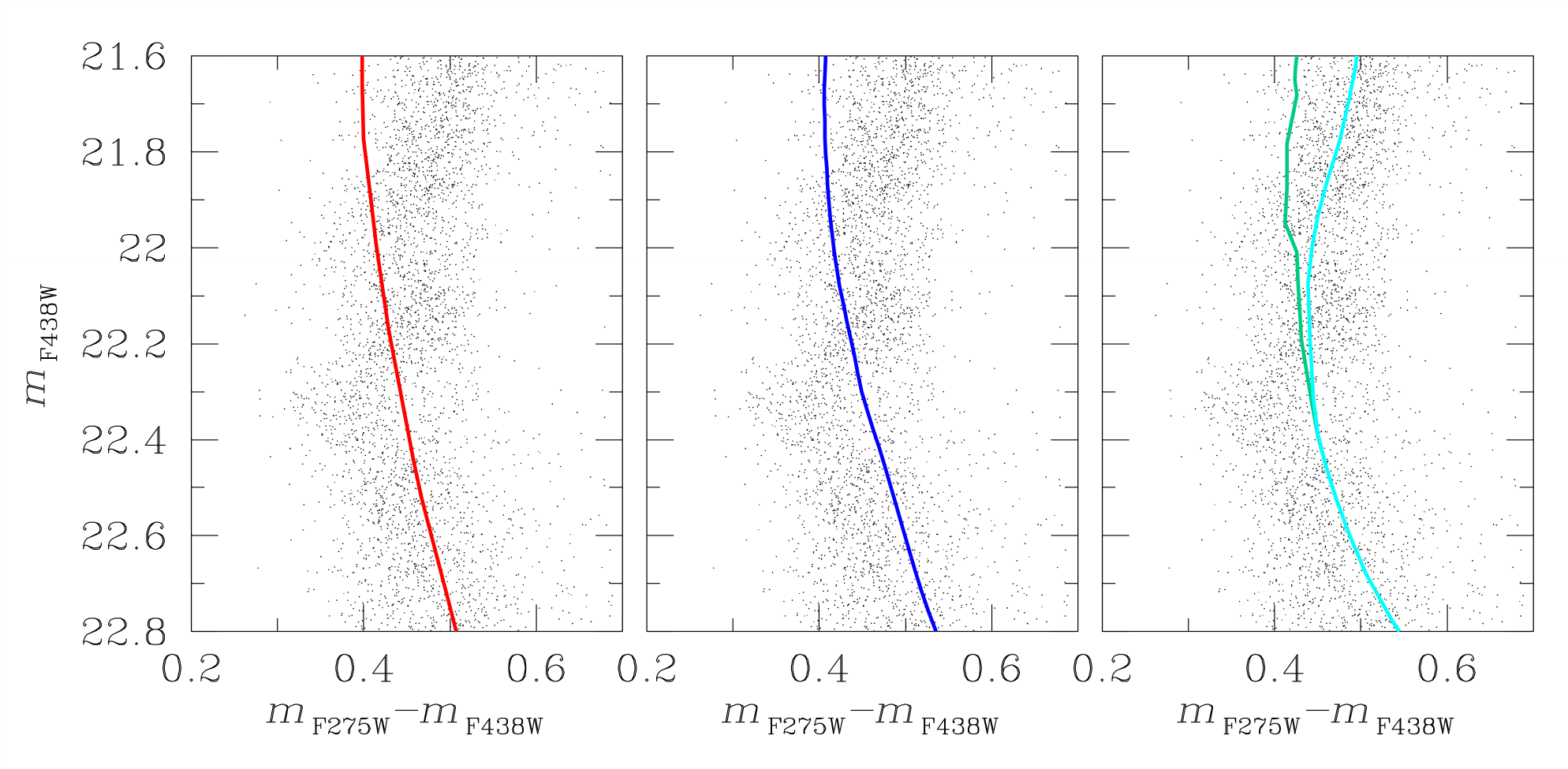}
 %/home/milone
 \caption{Reproductions of the $m_{\rm F438W}$ vs.\,$m_{\rm F275W}-m_{\rm F438W}$ CMD of Figure\,\ref{fig:NGC1783pm} zoomed around the upper MS. The red, blue, and green lines superimposed on each CMD are the best-fitting non-rotating isochrones from the Padova (left), BaSTI (middle), and MESA (right) databases. A rotating MESA isochrone with $\omega=0.4 \omega_{\rm c}$, with $\omega_{\rm c}$ being the breakout velocity, is plotted in the right panel in cyan.}
 \label{fig:NGC1783iso} 
\end{figure} 
\end{centering}

We compare in Figure\,\ref{fig:NGC1783iso} the observed upper MS of NGC\,1783 with the isochrones from Padova \citep[left,][]{marigo2017a}. The faint MS discontinuity corresponds to effective temperature $T_{\rm eff}=6,900$K and mass  $\mathcal{M}$=1.26 $M_{\odot}$, whereas stars on the bright MS discontinuity have $T_{\rm eff}=7,250$K and $\mathcal{M}$=1.19 $M_{\odot}$. 

We tentatively associate the hotter gap of the NGC\,1783 MS with the original B{\"o}hm-Vitense gap. %whereas the.
A gap along the MS at $T_{\rm eff} \sim 7,500$K has been first predicted by \citet[][]{bohmvitense1970a} and observed in the nearest Galactic open clusters \citep[e.g.][]{bohmvitense1974a, debruijne2000a, debruijne2001a}. The B{\"o}hm-Vitense gap is associated with sudden changes in the structure of convective atmospheres. It has been interpreted as a color effect, due to the fact that the temperature gradient in deep atmospheric layers becomes smaller than the radiative gradient. As an alternative, it is the effect of temperature inhomogeneities produced by photospheric granulation \citep[e.g.][]{bohmvitense1982a}.

The colder discontinuity of the NGC\,1783 MSs could correspond to a distinct MS gap, which was earlier investigated by \citet{dantona2002a}.
 Indeed, fainter MS gaps have been observed around $T_{\rm eff} \sim 7,000$K \citep[e.g.][]{rachford2000a}.
At this temperature, convection begins in stellar envelopes, and an increasing amount of the stellar exterior 
%external fractions of the star 
 becomes convective as the mass and effective temperature decrease. Also, the eMSTO disappears below $\sim 7,000$K, confirming that fast-rotating MS stars are only  present at hotter temperatures. Indeed, the external turbulence brakes the envelope rotation. 
Clearly, the change of stellar structure results in a variation of the MS slope. 
%This is the temperature location where convection begins in the stellar envelope, and increasing external fractions of the star become convective as the mass and Teff decreases. Also the extended turnoff feature disappears below this Teff, confirming a well known occurrence that fast rotating main sequence stars are present only until the early spectral type F (CONTROLLARE!) because external turbulence brakes the envelope rotation. The change of stellar structure results in a variation of slope of the MS. 
Recent works provide evidence of an MS kink at similar temperatures in several Galactic and Magellanic Cloud clusters, where the split MS, which is associated with stellar populations with different rotation rates, merges into a single MS \citep[e.g.][]{dantona2017a, milone2018a, marino2018a, goudfrooij2018a}.  

The fainter MS gap has been investigated in the Hyades by \citet{dantona2002a} based on the full spectrum of turbulence model by \citet{canuto1996a}, which  predicts that the depth of the convective envelope suddenly changes within a narrow range of stellar mass and around 
% in a tiny interval of mass of stars with 
 $T_{\rm eff} \sim 6,800$K. They concluded that the gap is an effective-temperature effect associated with the sharp effective-temperature difference between stars that are only convective in the surface layers  and stars with well-developed convective interiors. 

However, these results are based on poorly-populated CMDs of open clusters, which often make it challenging to assess the statistical significance of the gaps.
The high-precision {\it HST} photometry of populous star clusters may overcome this limitation and provides new insights on the MS region in the temperature range between $\sim$6,500K and 7,500K.

%%%%%%%%%%%%%%%%%%%%%%%%%%%%%%
As shown in Figure\,\ref{fig:NGC1783iso}, neither the isochrones from the Padova group nor those from the BaSTI \citep[middle,][]{pietrinferni2004a}, and MESA \citep[][]{choi2016a, dotter2016a, paxton2011a} databases reproduce the observed MS discontinuities.
This fact corroborates the conclusion that  these stellar models, poorly reproduce the CMD region where the stellar atmosphere changes from radiative to convective.
%This fact corroborates the conclusion that mixing-length theory, which is commonly used to model energy transport in the convective zone \citep[][]{bohmvitense1958a}, poorly reproduces the CMD region where the stellar atmosphere changes from radiative to convective.  

%Their conclusion can not be extended to the second MS gap that has been observed in the Hyades \citep[][]{debruijne2000a} and has been possibly related to a dip in the chromospheric activity \citep[][]{bohmvitense1995a}.
  %%%%%%%%%%%%%%%%%%%%%%%%%%%%%%%%
%%%%%%%%%%%%%%%%%%%%%%%%%%%%%%%%

\subsection{Search of multiple generations in NGC\,1783} \label{sec:NGC1783}

In the past years, astronomers have dedicated huge efforts to searching for young star clusters that are analogous to old GCs. Indeed, they may provide a snapshot of multiple populations shortly after their formation.
The two sequences of stars in the field of view of the $\sim$1.5 Gyr-old cluster NGC\,1783 are a hotly debated case because they are consistent with younger stellar populations of $\sim$440 and 520 Myr.
%stellar populations with different ages in young and intermediate-age clusters the  
%thus providing the opportunity of investigating the multiple-population phenomenon shortly after their formation. 
% The origin of the sequences of bright and blue MS stars in the field of view of NGC\,1783 has been recently debated in the literature.  
After statistically subtracting the contribution of field stars from the CMD of stars around NGC\,1783, \citet{li2016a}  concluded that the blue MSs are cluster members. They  suggested that these young stars are the signature of burst-like star formation and that NGC\,1783 has experienced multiple bursts of star formation with an age difference of a few hundred million years. In this scenario, NGC\,1783 is eventually the young counterpart of GCs with multiple populations. 
 
 This result has been challenged by \citet{cabreraziri2016a}, who suggested that the background subtraction method adopted by Li and collaborators may not remove contaminating field stars. Hence, they concluded that there is no evidence for multiple generations within NGC\,1783 and that the young populations are field LMC stars along the same line of sight of NGC\,1783.

\begin{figure*} 
\centering
 \includegraphics[height=11.0cm,trim={3.0cm 0cm 3.0cm 0cm},clip]{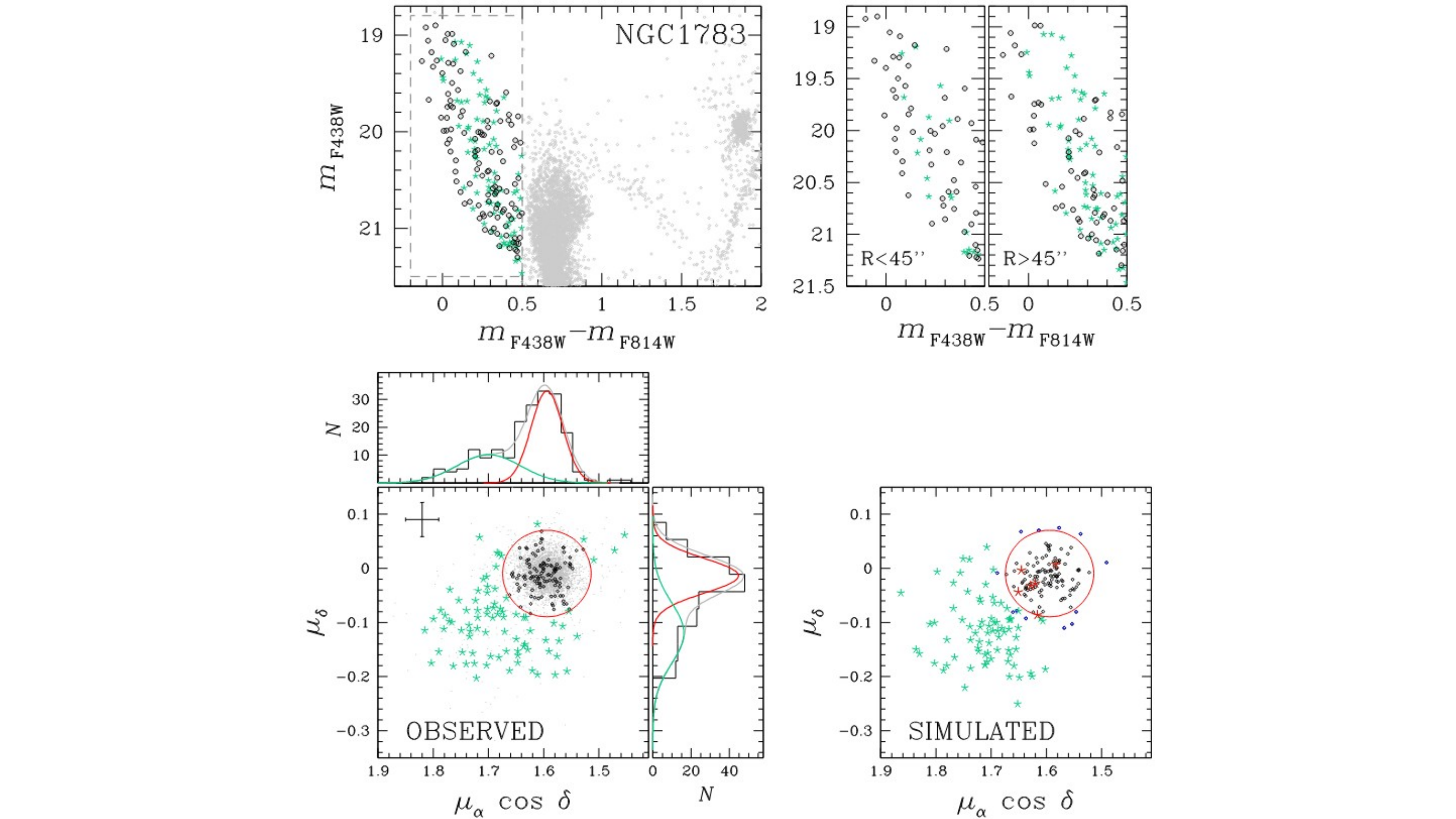}
 %/home/milone/
 \caption{ \textit{Top.} Reproductions of the CMDs of NGC\,1783 of Figure\,\ref{fig:NGC1783pm}. Right panels are zoomed-in views of the CMD region on the bright-blue side of the MSTO (dashed rectangle in the top-left panel) for stars with radial distances from the cluster center smaller and larger than 45 arcsec. 
 \textit{Bottom.} Proper motion diagram of stars plotted in the top-left panel (left). The red circle separates stars with a cluster-like motion from the bulk of field stars. Black points and aqua-starred symbols mark the selected bright-blue stars. The corresponding histogram distributions for $\mu_{\rm \alpha}$ cos $\delta$ and $\mu_{\delta}$ are also represented on the top and right side of the panel. The best fit bi-Gaussian functions are represented with gray lines, and the two Gaussian components are colored aqua and red. The bottom-right panel shows the simulated proper motions for cluster members and field stars.  Black and blue dots indicate NGC\,1783 stars with cluster-like and field-like proper motions, respectively, while red color is used to distinguish field stars with cluster-like proper motions from the remaining field stars (aqua starred symbols). 
 }
 \label{fig:NGC1783y} 
\end{figure*}

 As anticipated in Section\,\ref{sub:1783emsto}, the $m_{\rm F438W}$ vs.\,$m_{\rm F438W}-m_{\rm F814W}$ CMD plotted in Figure\,\ref{fig:NGC1783pm} clearly reveals the stellar sequence first investigated by \citet{li2016a}, whereas stellar proper motions allow disentangling the bulk of cluster members and field stars.
 To investigate whether NGC\,1783 hosts a population of bright and hot MS stars, we combined information from photometry and stellar proper motions as illustrated in Figure\,\ref{fig:NGC1783y}.
  We first used the dashed rectangle plotted in the top-left panel of Figure \ref{fig:NGC1783y} to select a sample of stars on the blue side of the cluster MSTO in the $m_{\rm F438W}$ vs.\,$m_{\rm F438W}-m_{\rm F814W}$ CMD. 
  %Top-right panels compare the CMD for selected stars in the internal region (within 45 arcsec from the cluster center) with the same CMD for stars in the $\sim$four-time wider external region. 
 In the top-right panels, we compare the CMDs for stars in the dashed-line rectangle located within and outside a radius equal to 45 arcsec from the cluster's center. The external region has a $\sim$four-time wider area than the internal one.
 
 The proper motion diagram plotted in the bottom-left panel of Figure\,\ref{fig:NGC1783y} shows that NGC\,1783 stars are partially separated from LMC stars.
  We draw the red circle to separate the bulk of cluster members from field stars and represent these stars with black circles and aqua-starred symbols, respectively, in the bottom-left and in the top panels of Figure\,\ref{fig:NGC1783y}.  
  
  We find that the sample of selected blue stars in the CMD comprises 18 field stars in the internal region, while 64 field stars belong to the external region. Their ratio of about four is comparable with the ratio of the corresponding field-of-view areas as expected if field stars have uniform spatial distribution. On the contrary, the number of stars with cluster-like proper motion in the internal and external field are 53 and 64 and are comparable with each other thus indicating that they unlikely belong to the LMC field population.
  
 Noticeably, the distribution of stars in the proper motion diagram (i.e.\,the stellar abscissas and ordinates and their density) is well reproduced by a function composed of the sum of two 3D Gaussian  functions, that we derived by means of least squares minimization. For illustration purposes, we show in Figure\,\ref{fig:NGC1783y} the histogram distributions of $\mu_{\rm \alpha}$ cos $\delta$ and $\mu_{\delta}$ together with the corresponding two-dimensional Gaussian functions.
 
    Clearly, the stars with cluster-like proper motions   selected in the bottom-left panel of Figure\,\ref{fig:NGC1783y}  may also include field stars. To estimate the fraction of field stars that contaminate the sample of probable cluster members we used 
 the best-fitting 3D Gaussian functions to simulate the proper motions plotted on the bottom-right panel of  Figure\,\ref{fig:NGC1783y}. Here, we show a subsample of 198 simulated stars, which is the same number of observed stars. Simulated field stars are represented with starred symbols and cluster stars with small dots. Clearly, a fraction of field stars (red starred symbols) have cluster-like proper motions while some cluster members (blue dots) lie outside the red circle.  In particular, the fraction of field stars within the red circle, with respect to the number of cluster members is $\sim$5\%. These facts demonstrate that the majority  ($\sim$ 95\%) of stars with cluster-like proper motions selected in Figure\,\ref{fig:NGC1783y}  are cluster members.
 
 In summary, our proper-motion-based results confirm the conclusion of Li and collaborators that the blue sequences are composed of genuine members of NGC\,1783.
More sophisticated analysis is mandatory to understand whether the blue sequence is associated with young stellar populations as suggested by Li and collaborators or whether it is composed of blue stragglers. 
  The cluster members on the bright and blue side of the red clump are consistent both with young red clump stars and with binary systems composed of red-clump stars.
 
%%%%%%%%%%%%%%%%%%%%%%%%%%%%%%%%

\subsection{Relative proper motions of stellar populations in the LMC and the SMC} \label{sec:pm}
\begin{figure*} 
\centering
\includegraphics[width=13.0cm,trim={7cm 1cm 9cm 0cm},clip]{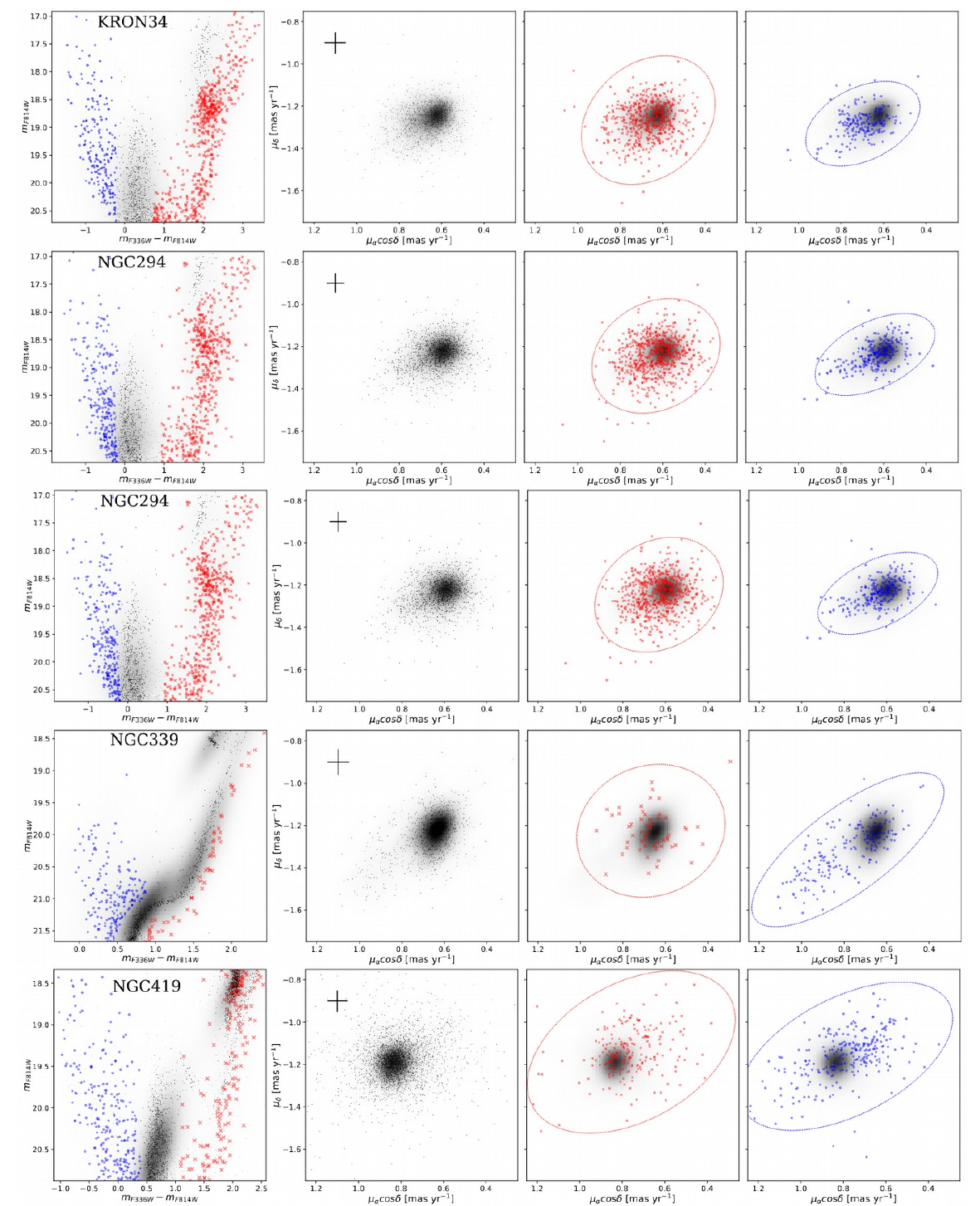}
 \caption{First-column panels show the CMDs of stars in the field of view of five SMC clusters with available proper motions.  The  corresponding proper motion diagrams  are plotted in the second column, while the third and fourth columns represent the proper motions of candidate old and young field stars selected in the CMDs and colored red and blue, respectively. 
  Red and blue ellipses provide the best-fitting of the distributions of candidate old and young field stars in the proper motion diagram. 
  The Hess diagrams of the proper-motion distributions are shown in all proper-motion diagrams.  }
 \label{fig:PMsSMC} 
\end{figure*}

\begin{figure*} 
\centering
\includegraphics[width=15.0cm,trim={3cm 0cm 3cm 0cm},clip]{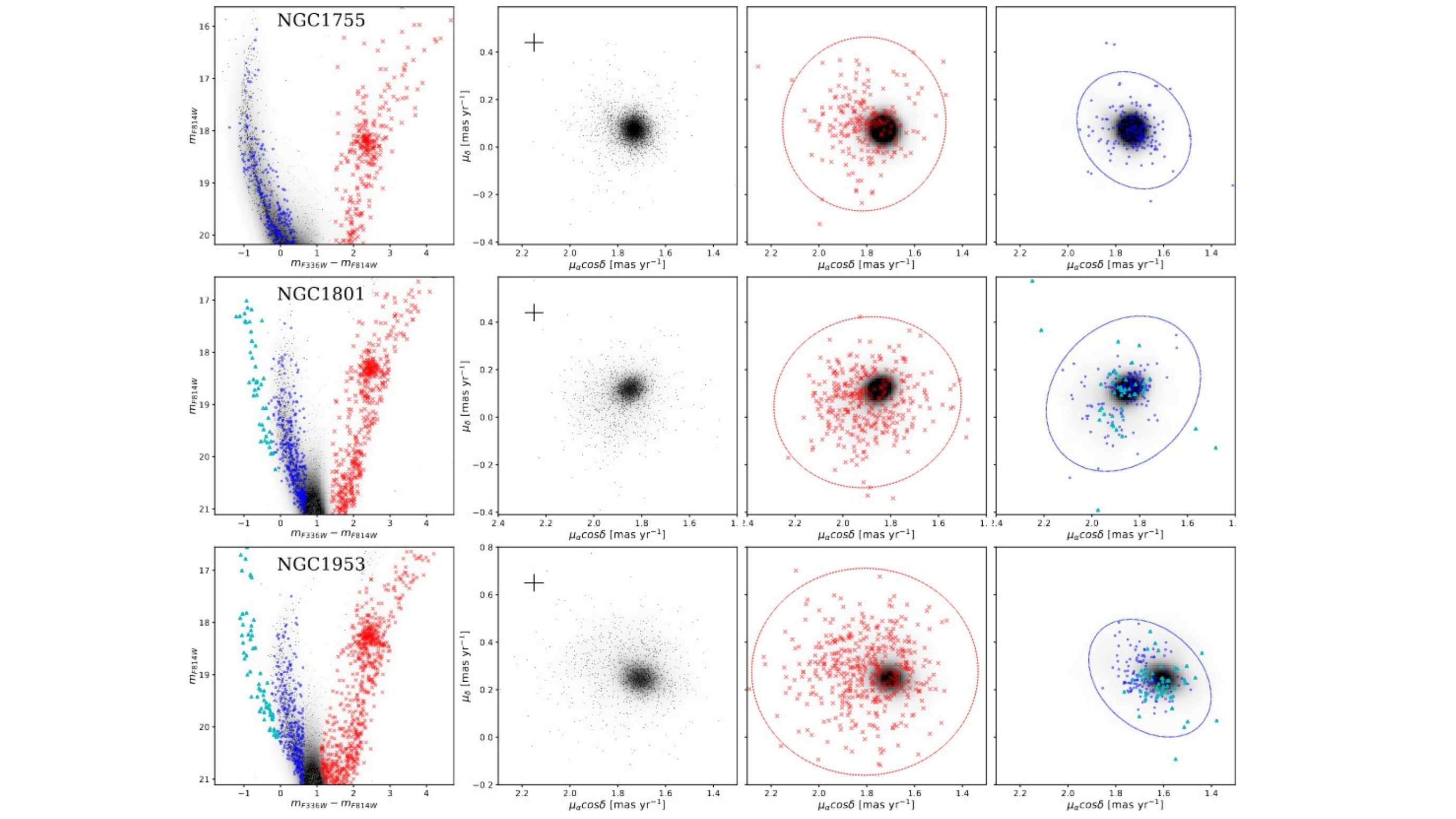}
 \caption{Similar to Figure\,\ref{fig:PMsSMC} but for the LMC clusters NGC\,1755, NGC\,1801, and NGC\,1953. Very young stars in the fields of view of NGC\,1801 and NGC\,1953 are represented with aqua triangles.}
 \label{fig:PMsLMC} 
\end{figure*}

Figures\,\ref{fig:PMsSMC} and \ref{fig:PMsLMC} show the CMDs and the proper motion diagrams  of stars in the fields of view of five SMC clusters, namely Kron\,34, NGC\,294, NGC\,339, NGC\,416, and NGC\,419, and three LMC clusters, namely NGC\,1755, NGC\,1801 and NGC\,1953.
Since we focus on the internal kinematics of LMC and SMC stars, we restrict the analysis to the magnitude interval that provides the most precise proper-motion determinations. 
The stellar concentrations around the center of each diagram are composed of cluster members and their broadening is mostly due to observational uncertainties. Indeed, the star-to-star scatter associated with the internal motions of cluster members is negligible with respect to proper motion errors at the distance of the Magellanic Clouds. 
On the contrary, field stars exhibit broad proper-motion distributions, which are significantly wider than what is expected from observational uncertainties alone. 

%To further investigate the kinematics of Magellanic-Cloud field stars, 
We identified in each CMD a group of stars with blue $m_{\rm F336W}-m_{\rm F814W}$ colors, which mostly comprise the young stellar populations of the host galaxy, and a group of old stars with red colors.   Furthermore, we selected a sample of very-young LMC stars in NGC\,1801 and NGC\,1953 that define the bluest and brightest MS in their CMD (aqua triangles). The selected groups of old and young stars, identified in the CMDs, are highlighted with red and blue symbols, respectively, in the proper motion diagrams plotted in the third and fourth columns of panels. 
%Clearly, the proper motions of young and old stars show different distributions.
  A visual inspection of these figures suggests that the proper motions of young and old SMC stars typically show different ellipticities, whereas the differences are less pronounced for LMC stellar populations. 

\begin{figure}
\centering
\includegraphics[height=8.5cm,trim={6.0cm 0cm 4cm 0cm},clip]{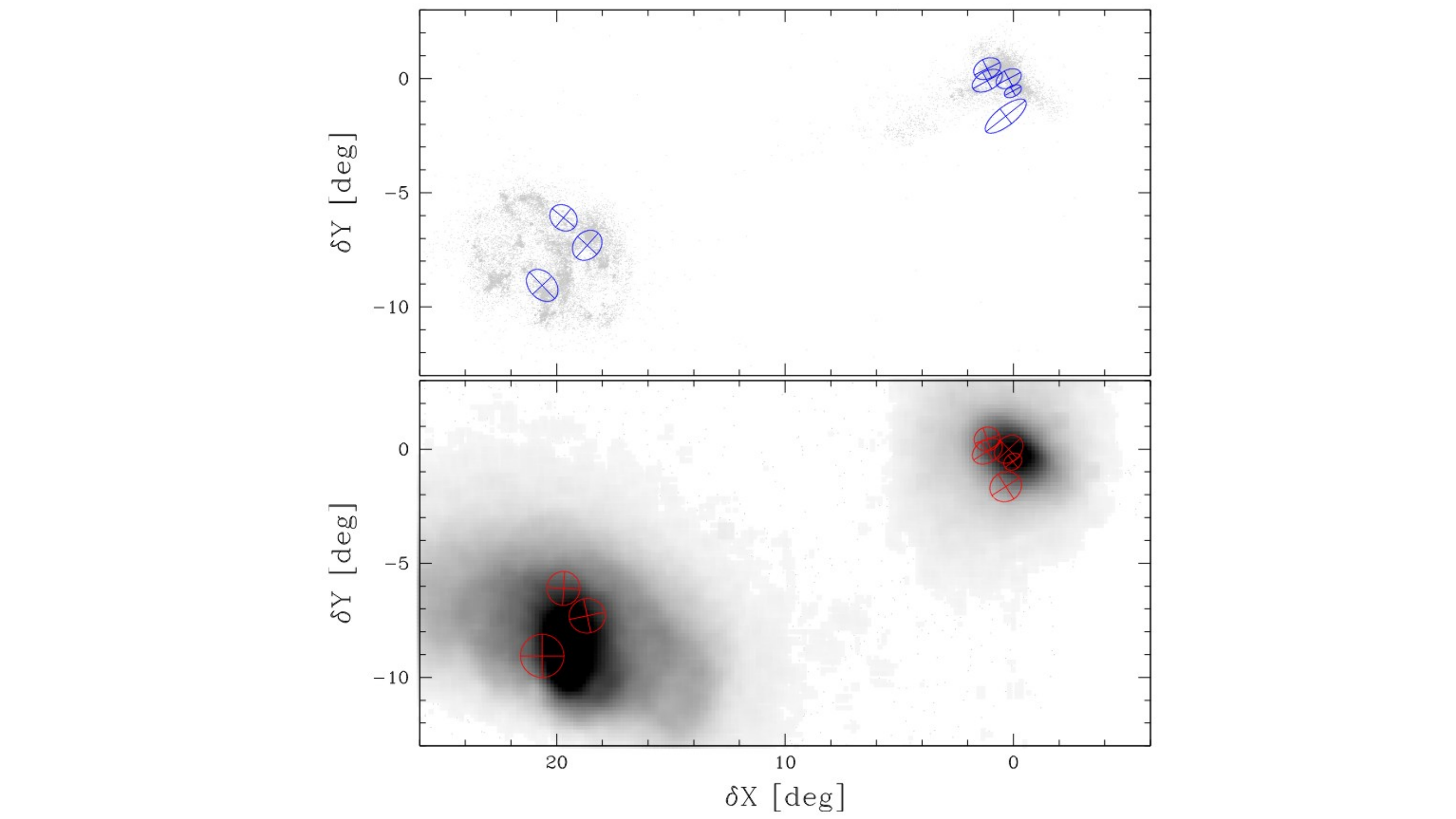}
%/home/milone/NUBI/CATALOGHI/coo.macro (gogaia) pm pm2
 \caption{Positions of candidate young LMC and SMC stars relative to the center of the SMC (top). The bottom panel shows the density distributions of candidate old LMC and SMC stars. 
 The ellipses that provide the best fit of proper-motion distributions in young and old stars in the fields of eight clusters are colored blue and red, respectively.   }
 \label{fig:ellissi} 
\end{figure}

Results are illustrated in Figure\,\ref{fig:ellissi} and  summarized in Table\,\ref{tab:ellissi}, where we provide for each population the median proper motions relative to the main cluster in the field, the ellipticity of the best-fitting ellipse that encloses 90\% of stars ($\epsilon=1-b/a$, where $a$ and $b$ are the minor and major axes of the ellipse), and the position angle $\theta$.
As shown in Figure\,\ref{fig:PMsSMC}, young SMC stars exhibit flat proper motion distributions, where the eccentricity of the best-fitting ellipses ranges from $\epsilon \sim 0.3$ in NGC\,416 to $\epsilon \sim 0.6$ in NGC\,339. The proper motion distributions of the old stellar populations have smaller eccentricity values, between $\epsilon \sim 0.1$ in NGC\,416 and NGC\,339 and $\epsilon \sim 0.3$ in NGC\,419.  In all cases, the major axis of the best-fitting ellipses roughly follows the direction North West - South East, thus pointing towards the LMC.   
Similar conclusions are derived by \citet{massari2021a} based on high-precision proper motions and stellar photometry from {\it HST} of NGC\,419. These authors demonstrated that it is possible to separate cluster members from SMC field stars  by using stellar kinematics. Moreover, they identified a kinematic stellar component that they associated with the Magellanic Bridge. Although our results do not provide evidence for populations of field stars with distinct kinematics, the flattened proper motion distributions would reflect the flow motion of stars from the SMC to the LMC.
 Further evidence of SMC star clusters showing a relative motion pointing towards the LMC is provided by \citet{zivick2018a, piatti2021a, dias2021a, schmidt2022a}.
%also the major-axis extension significantly changes from one cluster to another.  

%%%%%%%%%%%%%%%%%%%%%%%%%%%%%%%%%%%%%%%%%%%%%%%%%%%%%%%%%%%%%%%%%%%%%%%%%%%%

 Young LMC stars exhibit flatter proper-motion distributions than old LMC stars, in close analogy with what is observed for the SMC. However, the ellipses that best fit the proper motions of LMC stars in the fields of NGC\,1755, and NGC\,1953 have different orientations than the corresponding ellipses inferred for stars in the direction of NGC\,1801.  Very young stars in the field of view of NGC\,1953 have more clustered proper motions with respect to the remaining young stars. Interestingly,   the selected young stars show some hints of split MS in the CMD. A spectroscopic investigation is mandatory to understand whether the split is due to stellar populations with different rotation rates, similar to what is observed in the star clusters with similar ages, or to differences in distance, age, and/or chemical composition.
 The proper motion distributions of stars in the direction of the three analyzed LMC clusters are nearly circular, in contrast with what is observed for young stars in both Magellanic Cloud clusters and old SMC stars.

\subsubsection{The massive star-forming region NGC\,346}\label{sub:NGC346}

\begin{centering} 
\begin{figure*} 
 \includegraphics[height=7.5cm,trim={0.1cm 4cm 0.0cm 1.0cm},clip]{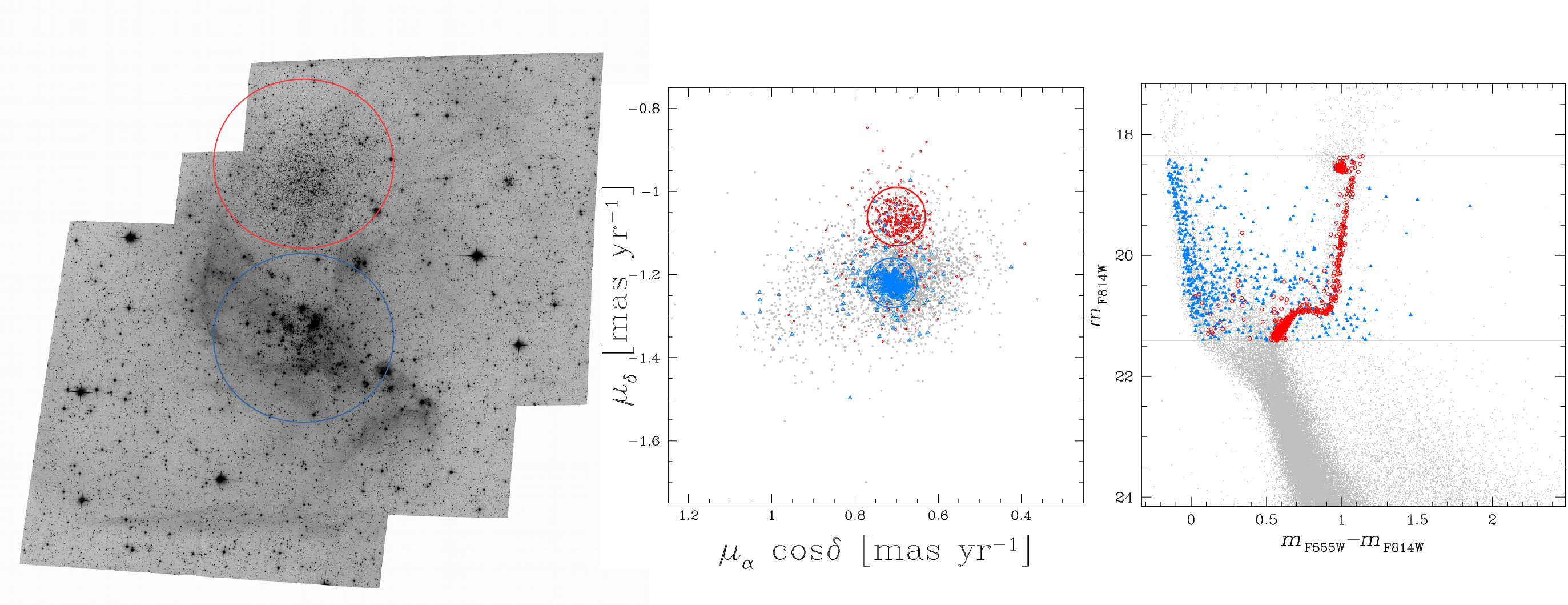}
 %/home/milone/NUBI/NGC0346/F814W/EI
 \caption{Stacked F814W WFC/ACS image of the SMC field that includes the clusters NGC\,346 and BS\,90, which are highlighted by the azure and red circle, respectively (left). The middle panel shows the proper motion diagram, while the $m_{\rm F814W}$ vs.\,$m_{\rm F555W}-m_{\rm F814W}$ CMD is represented in the right panel. 
  Only stars in the F814W magnitude interval between the two horizontal lines in the CMD are plotted in the proper-motion diagram, where the stars in the NGC\,346 and BS\,90 regions defined in the left panel are colored blue and red, respectively. Blue and red symbols in the right-panel CMD mark proper-motion selected cluster members in the NGC\,346 and BS\,90 regions. See text for details.}
 \label{fig:NGC346} 
\end{figure*} 
\end{centering}

NGC\,346 is a very young SMC star cluster \citep[age $\sim$ 3 Myr,][]{bauret2003a, sabbi2007a} that is responsible for the excitation of the surrounding H${\rm II}$ region N\,66. 
A stacked F814W image of NGC\,346 and its neighborhoods is shown in the left panel of Figure\,\ref{fig:NGC346} where we mark with an azure circle the central part of the NGC\,346 star-forming region.    

The intermediate-age star cluster BS\,90 is also visible to the north of NGC\,346 and is highlighted by the red circle in the left panel of Figure\,\ref{fig:NGC346}. The proper motion diagram of all stars with $m_{\rm F814W}$ between 18.4 and 21.4 mag is plotted in the middle panel of Figure\,\ref{fig:NGC346} and comprises stars with the best proper-motion quality \footnote{ Specifically, the selected stars pass the criteria of selection based on the $RADXS$ and $qfit$ parameters discussed in Section\,\ref{subsec:quality} and that are not saturated in the long-exposure F555W and F814W images (see Table\,\ref{tab:dataPM} for details on the dataset). Moreover, we only included stars with small proper-motion errors, when compared to the bulk of stars with similar magnitudes. To select them, we first plotted the proper-motion uncertainty against the F814W magnitude. Then, we divided the magnitude interval into various 0.25 mag bins and calculated the median proper-motion uncertainty for each bin. We computed the absolute values of the difference between the uncertainty of each star and the median value and estimated the 68.27$^{\rm th}$ percentile of the corresponding distribution ($\sigma$). We added three times $\sigma$ to the median uncertainty of each bin  and associated this value with the median magnitude of the stars in the bin. Finally, these points are linearly interpolated and the stars that are located below this line in the proper-motion uncertainty vs.\,magnitude plane are considered as well measured. }. Stars within the regions centered on NGC\,346 and BS\,90 (azure and red points, respectively) define two distinct clumps in the proper motion diagram, which are mostly composed of cluster members.    
We selected probable cluster members with proper motions smaller than four times the r.m.s of the proper motion distributions (stars within the circles) and calculated the median values of $\mu_{\rm \alpha} cos{\delta}$ and $\mu_{\rm \delta}$. Results are listed in Table\,\ref{tab:n346}. The fact that NGC\,346 and BS\,90 exhibit different proper motions demonstrates that they are distinct clusters projected onto the same field of view.

The probable members of NGC\,346 and BS\,90, selected from both stellar proper motions and positions, are marked with azure and red points in the CMD in the right panel of Figure\,\ref{fig:NGC346}. The sample of selected NGC\,346 stars defines a well-populated MS and upper pre-MS, whose large broadening is indicative of a significant amount of differential reddening.  On the contrary, BS\,90 exhibits narrow SGB and RGB sequences, and a well-defined red clump, thus confirming that this cluster is poorly affected by differential reddening \citep[][]{sabbi2007a}. Specifically, the average reddening variation in the field of view within 36 arcsec from the center of BS\,90 never exceeds $\Delta$E(B$-$V)$=0.013$ mag, with  $\sim 68 \%$ of the stars  having $\Delta$E(B$-$V) values within 0.004 mag from the average reddening.  This fact demonstrates that this cluster is in the foreground with respect to the region of NGC\,346 and N\,66.
%%%%%%%%%%%%%%%%%%%%%%%%%%%%%%%%
%%%%%%%%%%%%%%%%%%%%%%%%%%%%%%%%

\begin{figure*} 
\centering
 \includegraphics[height=14.0cm,trim={4.5cm 0cm 3.0cm 0cm},clip]{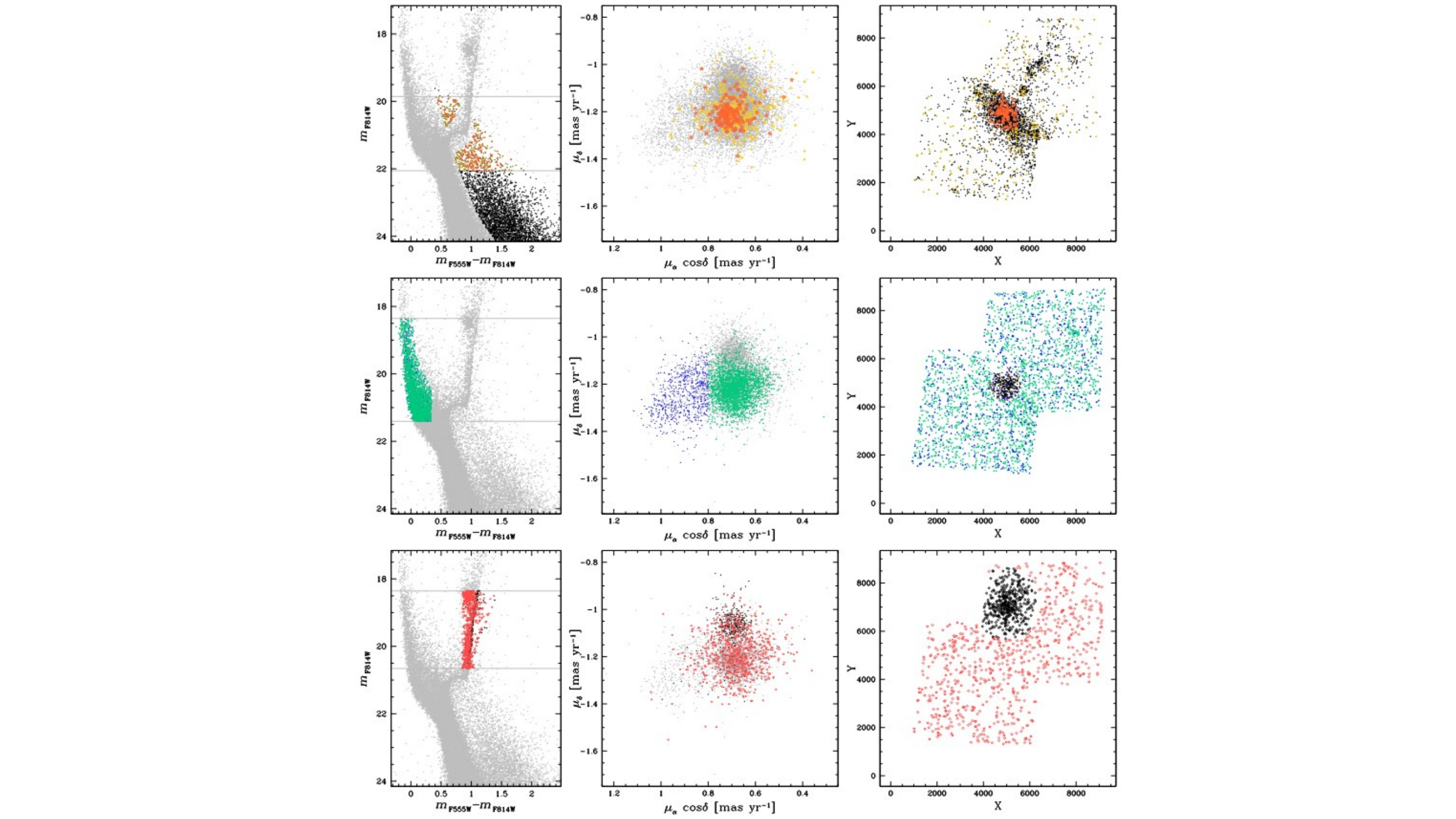}
 %/home/milone/NUBI/NGC0346/F814W/EI
 \caption{The three left panels are reproductions of the $m_{\rm F814W}$ vs.\,$m_{\rm F555W}-m_{\rm F814W}$ CMDs for stars in the field of view of NGC\,346 and are used to select various groups of stars that we represented with colored symbols. 
 The middle panels show the proper motions for stars in the magnitude interval delimited by the horizontal lines in the corresponding CMDs, whereas the right panels show the coordinates of the selected stars.
 The top panels are focused on candidate pre-MS stars. Specifically, bright pre-MS within 28 arcsec from the center of NGC\,346 (pre-MS\,I sample) are represented with orange symbols, the remaining bright pre-MS stars (pre-MS\,II sample) are colored yellow, while the black symbols in the top panels mark faint pre-MS stars.  Bright MS stars are investigated in the middle panels. Blue and aqua colors mark MS\,I and MS\,II stars, which have radial distances larger than 28 arcsec from the center of NGC\,346, but different proper motion distributions. The remaining bright-MS stars are colored black. The bottom panels highlight the selected RGB stars with radial distances from the center of BS\,90 larger (red) and smaller than 60 arcsec (black). }
 \label{fig:NGC346pop} 
\end{figure*}

%APM descrizione incompleta della figura 16.
 The CMD of stars in the field of view of NGC\,346 comprises stars in different evolutionary stages for which high-precision proper motions are available, including pre-MS stars, MS stars, and evolved stars in the SGB, RGB, and red-clump phases. 
  As widely discussed in literature works \citep[e.g.][]{sabbi2007a, cignoni2010a, cignoni2011a}, this field hosts a conspicuous population of pre-MS stars that are highlighted in the top-left panel of Figure \ref{fig:NGC346pop}. In particular, we used orange and yellow colors to represent two samples of pre-MS\,I and pre-MS\,II stars with high-precision proper motions, which lie respectively inside and outside the central region of NGC\,346. The remaining pre-MS stars are colored black.
   As indicated in Table\,\ref{tab:n346}, the proper motion distributions of the two groups of pre-MS stars share the same mean motion as NGC\,346 indicating that stars in both the central region and in the outskirts share the same mean motions (top-middle panel of Figure\,\ref{fig:NGC346pop}), although the latter  exhibits a wider proper motion dispersion. 
   The distribution of pre-MS stars across the field of view highlights the distinctive structure of the NGC\,346 region described by \citet{contursi2000a}, including the  low-density filament oriented to the north-east direction (spur), and the fan-shaped structure (bar), that hosts the majority of pre-MS stars.
   % comparison motion spur and bar
  
  MS stars are investigated in the middle panels of Figure\,\ref{fig:NGC346pop}, and comprise stars of the young population of the SMC. The proper motion diagram reveals that the bulk of MS stars (hereafter MS\,I stars, aqua points) exhibit a proper-motion distribution similar to the NGC\,346 pre-MS stars. In addition, we note a tail of stars in the proper motion diagram that points towards the LMC and that we colored blue and name MS\,II. 
  Both groups of selected MS stars seem diffused over the whole field, but the MS\,I stars define some stellar overdensities that trace the bar and possibly, some clumps of the Spur. 
We suggest that the MS\,II is mostly composed on SMC field stars that follow an elliptical proper motion distribution, in close analogy with what is observed for the other analyzed SMC young stars. On the contrary, MS\,I stars comprise both field SMC stars, and stars of the NGC\,346 region.
 
 Finally, the old SMC field stars are investigated in the bottom panels of Figure\,\ref{fig:NGC346pop}. Although this stellar population hosts some stars up to ages of more than 10 Gyr, it is mostly associated to a major star-formation episode that occurred between $\sim 3$ and 5 Gyr ago. The RGB stars of the old populations with radial distance smaller than 36 arcsec from the center of BS\,90 are marked with red points. These stars exhibit similar proper motion distribution as young stars and pre-MS stars in the star-forming region of BS\,90 (bottom-middle panel) and are uniformly distributed across the entire field of view (bottom-right panel of Figure \ref{fig:NGC346pop}). Note, the motion of the bulk of old SMC field stars is significantly different from that of BS\,90, despite this cluster having an age  similar to most old field stars. 

%%%%%%%%%%%%%%%%%%%%%%%%%%%%%%%%

\section{Summary and conclusions}\label{sec:summary}
We have used the {\it HST} archive to retrieve ACS/WFC, UVIS/WFC3, and NIR/WFC3 images of 101 fields in the direction of the LMC and the SMC. These images include  29 SMC clusters and 84 LMC clusters. We derived high-precision photometry and astrometry by using the methods and the computer programs developed by Jay Anderson and his collaborators and obtained high-resolution reddening maps in the direction of each  cluster. % that are significantly affected by differential reddening. 
We provide accurate determinations of cluster centers and estimate distance modulus, reddening, metallicity, and age by comparing the CMDs with Padova isochrones \citep{marigo2017a}.
Moreover, we calculated proper motions for cluster and field stars in twelve stellar fields that have been multi-epoch observations. 

The exquisite photometry, astrometry, and proper motions presented in this paper have the potential to shed light on a variety of astrophysical phenomena. As an example, we present here some results that are evident from visual inspection of the CMDs and the proper motion diagrams.

\begin{itemize}

    \item {\bf New insights into the eMSTO phenomenon.}
    
    The photometric catalogs, derived from homogeneous  data reduction, allow accurate comparison of clusters with different ages.  
    The analysis of the $m_{\rm F336W}$ vs.\,$m_{\rm F336W}-m_{\rm F814W}$ CMDs of 19 LMC clusters in a wide range of ages between $\sim$20 Myr and 2 Gyr, reveals that the distribution of stars along the eMSTO significantly changes from one cluster to another and depends on GC age. While the eMSTOs of young clusters are dominated by blue and bright eMSTO stars, the fraction of stars in the red and faint eMSTO increases in older clusters. 
    This property of eMSTO stars provides a new observational constraint to understand the physical mechanism that is responsible for the eMSTO. 
%APM. Add conclusive remarks
    
    We also provide the first evidence of eMSTO in the LMC intermediate-age cluster KMHK\,361 and in the SMC young star cluster NGC\,265, where we also detect a split MS, with the blue MS hosting about one-third of MS stars.  This finding corroborates the conclusion that the eMSTO is a universal feature of the CMD of clusters younger than $\sim$2 Gyr and the split MS is a common phenomenon that occurs in clusters younger than $\sim 1$ Gyr. 
    
\item {\bf A new feature along the eMSTO.} 
We find that about 7\% of eMSTO stars in NGC\,1783 exhibit redder $m_{\rm F275W}-m_{\rm F438W}$ and  $m_{\rm F343N}-m_{\rm F438}W$  colors than the remaining eMSTO stars. They show a wide color broadening up to $\sim$0.2 mag in the $m_{\rm F438W}$ vs.\,$m_{\rm F275W}-m_{\rm F438W}$ CMD, but the color spread decreases to less than 0.1 mag in the  $m_{\rm F435W}$ vs.\,$m_{\rm F343N}-m_{\rm F435W}$ diagram. On the contrary, when observed in optical CMDs, these stars define a narrow sequence and have intermediate colors relative to the remaining eMSTO stars.
    
    \item     {\bf Hunting for multiple star-formation bursts in intermediate-age star clusters.}
It has been suggested that the bright and blue stars of the CMD of NGC\,1783 are cluster members and correspond to young stellar generations \citep{li2016a}. This result, which would be a major step towards the understanding of multiple populations in GCs, has been challenged by \citet{cabreraziri2016a}, who suggested that the blue sequences are composed of field stars. 

The catalogs of the present work include proper motions of stars in the field of view of NGC\,1783 thus providing additional information on the origin of the blue sequences.  
%Above the MSTO, cluster members define a narrow clump of stars in the vector-point diagram of proper motions, while field stars show a broad proper-motion distribution. 
 In particular, we support the conclusion that most bright blue MS stars are consistent with being cluster members, thus excluding that the blue MS are artifacts produced by poor statistical subtraction of the field.
  Our results, based on proper motion analysis, do not allow us to infer whether the blue sequence is composed of MS stars of a young stellar generation or are the blue straggler sequences of NGC\,1783. 

\end{itemize}

In addition to disentangling cluster members and field stars, proper motions allow the investigation of the internal kinematics of stellar populations in the Magellanic Clouds.
 We have identified two groups of young and old Magellanic-Cloud field stars in the FoVs of five SMC clusters, namely KRON\,34, NGC\,294, NGC\,339, NGC\,416, and NGC\,419, and of three LMC clusters NGC\,1755, NGC\,1801 and NGC\,1953.

The proper motions of young SMC stars exhibit elliptical  distributions with high ellipticity values and major axes that point toward the LMC. 
The flattened proper-motion distributions would be associated with the Magellanic bridge and represent the dynamic signature of the flow motion of stars from the SMC to the LMC. Our results corroborate the evidence that SMC stars are affected by the LMC \citep[e.g.][]{piatti2015a}.
Old and young SMC stars exhibit different kinematics.
The proper motions of the old SMC stars are also oriented towards the LMC and have elliptical distributions  but with lower values of ellipticity and, in most cases, different centers.  The different proper-motion distributions of old and young stars could reflect, in part, the presence of different young and old bridges.
 
 The young and the old LMC stars also exhibit different motions on the plane of the sky. While the proper motions of the old populations show nearly circular distributions, young LMC stars have more flattened proper-motion distributions, with different orientations of the best-fitting ellipses. 

\section*{Acknowledgments} 
\small
We thank the anonymous referee for various suggestions that improved the quality of the manuscript.
This work has received funding from the European Research Council (ERC) under the European Union's Horizon 2020 research innovation programme (Grant Agreement ERC-StG 2016, No 716082 'GALFOR', PI: Milone, http://progetti.dfa.unipd.it/GALFOR).
%and the European Union's Horizon 2020 research and innovation programme under the Marie Sklodowska-Curie (Grant Agreement No 797100, beneficiary Marino). 
APM, MT, and ED acknowledge support from MIUR through the FARE project R164RM93XW SEMPLICE (PI: Milone). APM and MT  have been supported by MIUR under PRIN program 2017Z2HSMF (PI: Bedin). This research was supported in part by the Australian Research Council Centre of Excellence for All Sky Astrophysics in 3 Dimensions (ASTRO 3D) through project number CE170100013.
%C.\,L.\. acknowledges support from the one-hundred-talent project of Sun Yat-set University. C.\,L.\, was supported by the National Natural Science Foundation of China under grants 11803048.
This work is based on observations made with the NASA/ESA {\it Hubble Space Telescope}, obtained from data archive at the Space Telescope Science Institute (STScI). STScI is operated by the Association of Universities for Research in Astronomy, Inc. under NASA contract NAS 5-26555.
% LMC +1.872      +0.224    
% SMC +0.819      -1.177  
\section*{Data availability}
The data underlying this article will be shared on reasonable request to the corresponding author. \\

\section*{Appendix. Serendipitous discoveries}
\begin{figure*} 
\centering
 \includegraphics[height=11cm,trim={2cm 0cm 2cm 0.0cm},clip]{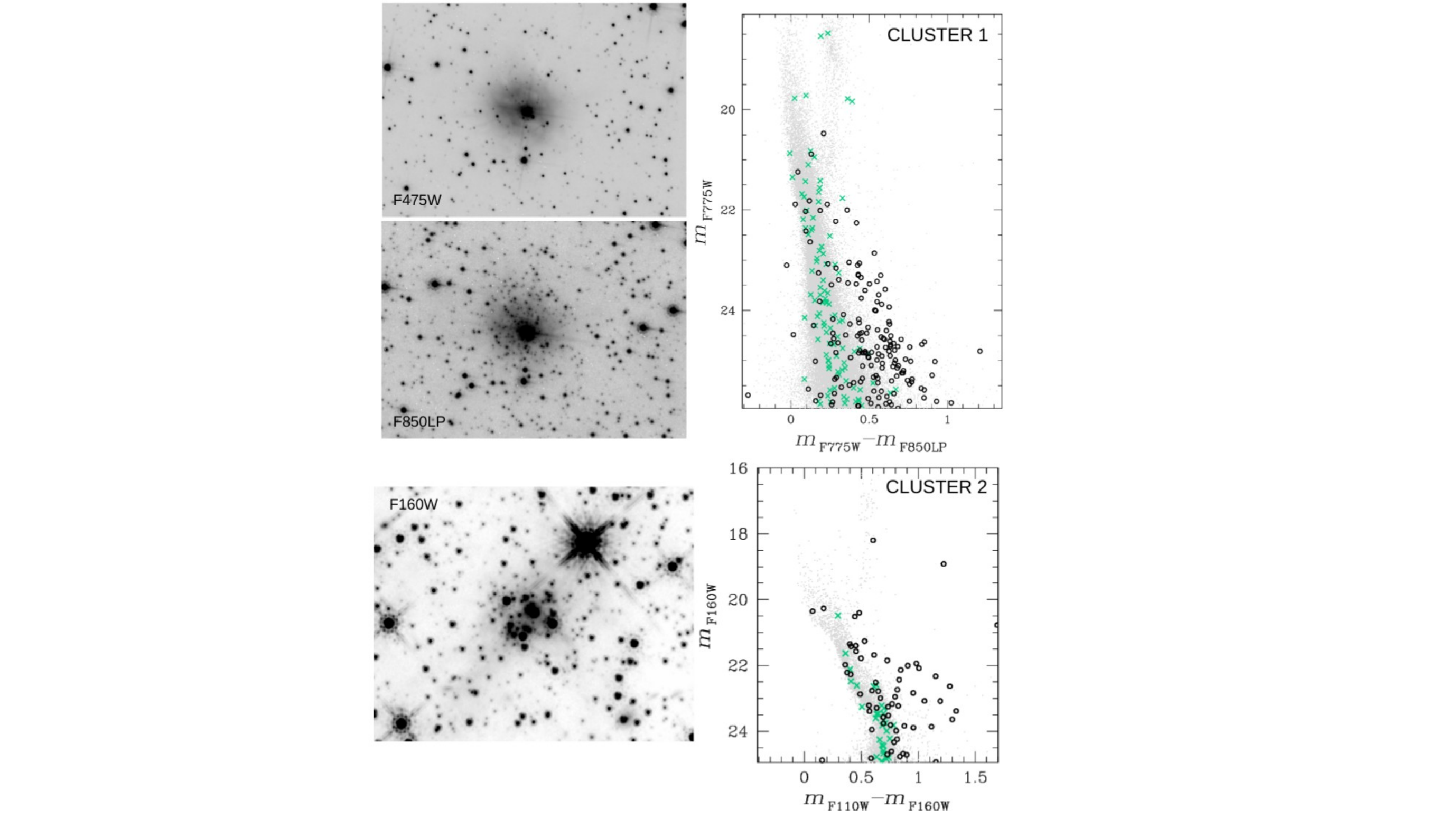}
% \includegraphics[height=8cm,trim={0cm 1.5cm 0cm 0.0cm},clip]{Galfor2.pdf}
% \newline
 %/home/milone/NUBI/GALFOR02/F850LP/CLEAN/cmd.macro
%  \includegraphics[height=6cm,trim={0cm 0cm 0cm 6.0cm},clip]{Galfor3.pdf}
 \caption{{\textit{Top panels}}. Stacked images in F475W and F850LP centred on the cluster\,1 discovered in this paper (left). The $m_{\rm F775W}$ vs.\,$m_{\rm F775W}-m_{\rm F850LP}$ diagram is plotted on the right for all stars in the ACS/WFC field (gray points), stars in the cluster\,1 field (black circles) and in the reference field (aqua crosses). {\textit{Bottom panels}}. Stacked image in F160W for the discovered cluster\,2 (left) and $m_{\rm F160W}$ vs.\,$m_{\rm F110W}-m_{\rm F160W}$ CMD (right). Gray, black, and aqua symbols indicate stars in the WFC3/NIR field, stars in the cluster field, and in the reference field, respectively. }
 \label{fig:galfor2} 
\end{figure*}

We report the serendipitous findings of two new star clusters, hereafter clusters 1 and 2. Two zoom-ins of the region around cluster 1 are provided in the top-left panels of Figure\,\ref{fig:galfor2}, where we provide the monochromatic images in the F475W and F850LP bands. The cluster, which is centered around RA=00:45:21.28, DEC=$-$73:12:18.1, J2000 is clearly visible in the F850LP image, while the most prominent feature of the F475W image is a nebula that envelopes it. The top-right panel shows the $m_{\rm F775W}$ vs.\,$m_{\rm F775W}-m_{\rm F850LP}$ CMD for all stars in the WFC/ACS field that includes cluster 1. 
The bright star close to the cluster center is classified as the H$\alpha$ emission-line star MA93-99 by \citet{meyssonnier1993a}.
 We defined a circle centered on the cluster with a 4 arcsec radius (hereafter cluster region) and represented with black circles all stars within this region. We also plot the stars in a randomly selected reference region with the same area as the cluster region with aqua crosses.
 Clearly, we note an overdensity of stars in the cluster region with red $m_{\rm F775W}-m_{\rm F850LP}$ colors and luminosity fainter than $m_{\rm F775W} \sim 23$. These stars are qualitatively consistent with a population of pre-MS stars.

Cluster\,2 is located around RA=05:03:55.87 DEC=$-$66:24:36.9, J2000 and is shown in the F160W image plotted on the bottom-left of Figure\,\ref{fig:galfor2}. The $m_{\rm F160W}$ vs.\,$m_{\rm F110W}-m_{\rm F160W}$ CMD plotted in the right panel reveals an overdensity of red stars in the cluster field, which suggests that cluster\,2 hosts a conspicuous population of pre-MS stars. 

%%%%%%%%%%%%%%%%%%%%%%%%%%%%%%%%

\begin{figure} 
\centering
 \includegraphics[height=7.0cm,trim={8cm 4cm 6cm 4.0cm},clip]{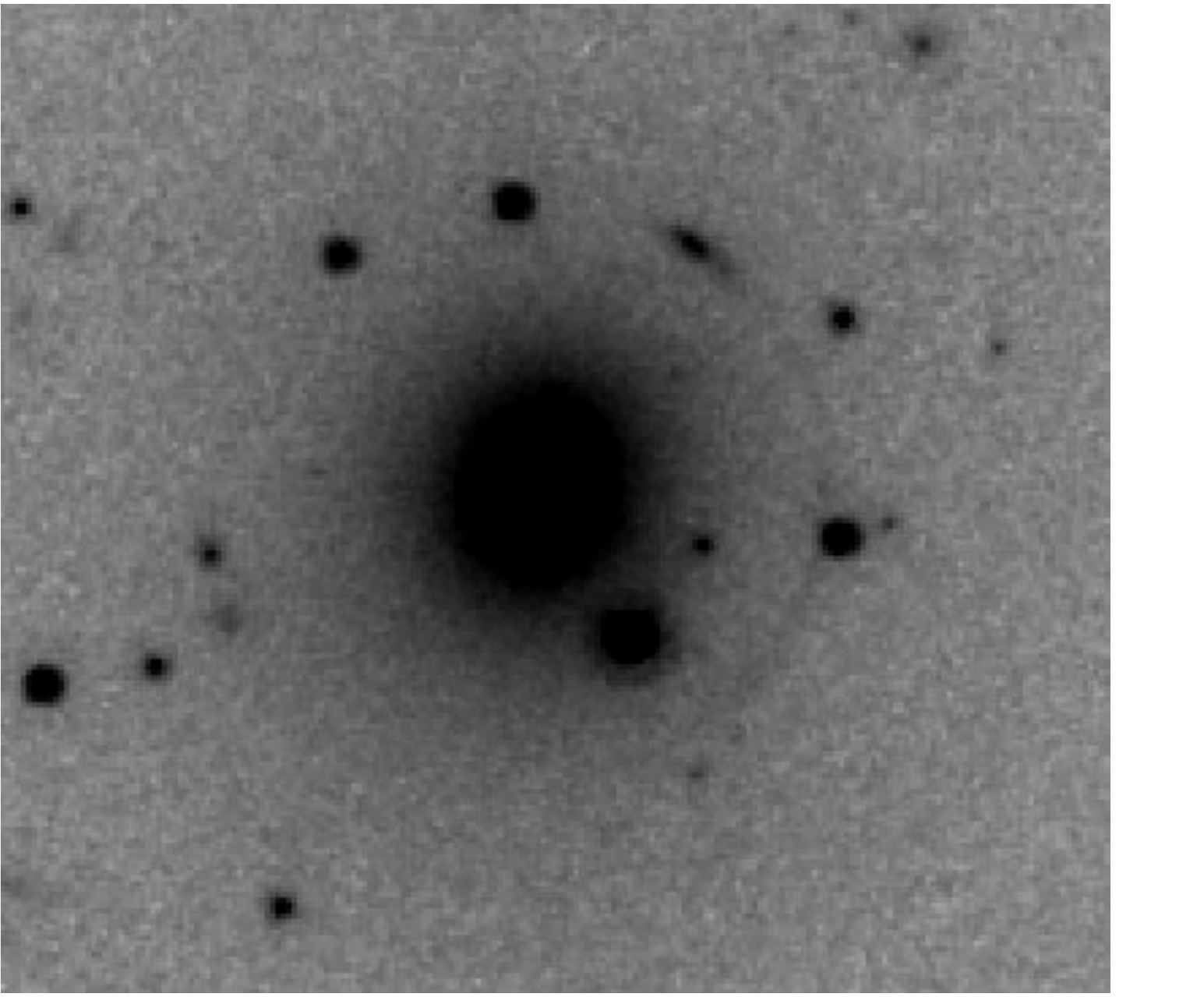}
 %/home/milone/
 \caption{Stacked F814W WFC/ACS image of the field around the   early-type galaxy centered at RA=00 48 57.01, DEC=$-$69 51 30.4, J2000. Note the elongated arc-like structure due to gravitational lensing. }
 \label{fig:lente} 
\end{figure}

%%%%%%%%%%%%%%%%%%%%%%%%%%%%%%%%
As a further outcome of the survey, we report the serendipitous discovery of a gravitational lens in the field of view of the SMC  star cluster Lindsay\,38.
 The F814W stacked image shown in Figure \ref{fig:lente} reveals an elongated arc-like structure around what appears to be a single early-type galaxy centered on the coordinates RA=00 48 57.01, DEC=$-$69 51 30.4, J2000. Its spatial extension is such that fine structures of the source are detectable in the arc. This object may correspond to a strong lensing configuration that implies remarkable alignment between lens and source along the line of sight.
 Due to their intrinsic symmetry, such a peculiar configuration would permit to constrain with great accuracy the total enclosed mass within the projected Einstein radius \citep[e.g.][and references therein]{bettinelli2016a}. 
%%%%%%%%%%%%%%%%%%%%%%%%%%%%%%%%
%%%%%%%%%%%%%%%%%%%%%%%%%%%%%%%%

\onecolumn

\begin{center}
\begin{longtable}{ccccccccccccccc}

\caption{Description of the {\it HST} images used in the paper. For each cluster, we provide the camera(s), the filters, and the total exposure times in seconds.} \label{tab:data} \\
\multicolumn{12}{c}%

%{{\bfseries \tablename\ \thetable{} -- continued from previous page}} \\
\endhead

\hline \multicolumn{3}{|r|}{{Continued on next page}} \\ \hline
\endfoot

\hline \hline
\endlastfoot
{\bf ESO121-03} &    &    &       ACS/WFC   & F814W  & 3832 &    {\bf NGC\,1868}   &  &    \\ 
ACS/WFC   & F435W &   1170&       {\bf NGC\,419}  &   &     &    WFC3/UVIS  & F336W & 2492 \\ 
ACS/WFC   & F555W &    330&       WFC3/UVIS & F336W  & 9056 &    WFC3/UVIS & F656N & 1429  \\ 
ACS/WFC   & F606W &    710&       WFC3/UVIS & F343N  &18563 &    WFC3/UVIS & F814W &  756  \\ 
ACS/WFC   & F814W &    908&       WFC3/UVIS & F438W  & 4028 &    {\bf NGC\,1872} &  &      \\ 
{\bf Hodge\,2}    &    &  &       ACS/WFC   & F555W  & 2024 &    ACS/WFC   & F555W &  115  \\ 
ACS/WFC   & F475W &   1440&       ACS/WFC   & F814W  & 4012 &    ACS/WFC   & F814W &  90   \\ 
ACS/WFC   & F814W &   1430&       {\bf NGC\,422}  &   &     &    {\bf NGC\,1898}   &  &    \\ 
{\bf Hodge\,6}    &   &   &       ACS/WFC   & F555W  & 73   &    WFC3/UVIS & F336W & 2070  \\  
WFC3/UVIS   & F475W & 1440&       ACS/WFC   & F814W  & 58   &    WFC3/UVIS & F438W &  400  \\ 
WFC3/UVIS   & F814W & 1430&       {\bf NGC\,602}  &   &     &    WFC3/UVIS & F814W &  100  \\ 
{\bf Hodge\,7}    &   &   &       ACS/WFC   & F555W  & 4338 &    ACS/WFC   & F475W & 1000  \\ 
ACS/WFC   & F555W &   330 &       ACS/WFC   & F814W  & 4450 &    ACS/WFC   & F814W & 1000  \\ 
ACS/WFC   & F814W &   200 &       {\bf NGC\,1466}  &   &    &    {\bf NGC\,1903}   &  &    \\ 
{\bf Hodge\,11}    &    & &       WFC3/UVIS & F336W  &11668 &    ACS/WFC   & F555W &   40  \\ 
WFC3/UVIS & F336W & 10848 &       ACS/WFC   & F606W  & 4336 &    ACS/WFC   & F814W &   50  \\ 
ACS/WFC   & F606W &  4390 &       ACS/WFC   & F814W  & 7082 &    {\bf NGC\,1917}   &  &    \\  
ACS/WFC   & F814W &  6932 &       {\bf NGC\,1644}  &   &    &    ACS/WFC   & F555W &  300  \\ 
{\bf HW\,57}    &    &    &       ACS/WFC   & F555W  &  250 &    ACS/WFC   & F814W &  200  \\ 
ACS/WFC   & F475W &   2490&       ACS/WFC   & F814W  &  170 &    {\bf NGC\,1928}   &  &    \\ 
ACS/WFC   & F814W &   2700&       {\bf NGC\,1651}  &   &    &    ACS/WFC   & F555W &  330  \\ 
{\bf IC\,1660}    &    &  &       WFC3/UVIS & F475W  & 1440 &    ACS/WFC   & F814W &  200  \\ 
ACS/WFC   & F555W &    73 &       WFC3/UVIS & F814W  & 1430 &    {\bf NGC\,1939}   &  &    \\ 
ACS/WFC   & F814W &    58 &       {\bf NGC\,1652}  &   &    &    ACS/WFC   & F555W &  330  \\ 
{\bf IC\,2146}    &    &  &       ACS/WFC   & F555W  & 300  &    ACS/WFC   & F814W &  200  \\ 
ACS/WFC   & F555W &   250 &       ACS/WFC   & F814W  & 200  &    {\bf NGC\,1943}   &  &    \\ 
ACS/WFC   & F814W &   170 &       {\bf NGC\,1718}  &   &    &    ACS/WFC   & F555W &   50  \\ 
{\bf KMHK\,240}    &    & &       WFC3/UVIS & F475W  & 1440 &    ACS/WFC   & F814W &   40  \\ 
ACS/WFC   & F475W &   575 &       WFC3/UVIS & F814W  & 1430 &    {\bf NGC\,1953}   &  &    \\ 
ACS/WFC   & F814W &  1330 &       {\bf NGC\,1751}  &  &     &    WFC3/UVIS & F336W & 2503  \\ 
{\bf KMHK\,250}    &    & &       WFC3/UVIS  & F336W & 3580 &    WFC3/UVIS & F656N & 1440  \\ 
ACS/WFC   & F435W &   720 &       ACS/WFC    & F435W &  770 &    WFC3/UVIS & F814W &  756  \\ 
ACS/WFC   & F555W &   700 &       ACS/WFC    & F555W &  984 &    ACS/WFC   & F555W &  115  \\ 
ACS/WFC   & F814W &   700 &       ACS/WFC    & F814W &  888 &    ACS/WFC   & F814W &   90  \\ 
{\bf KMHK\,291}    &    & &       {\bf NGC\,1755}  &   &    &    {\bf NGC\,1966}   &  &    \\ 
ACS/WFC   & F475W &   575 &       WFC3/UVIS  & F336W & 5688 &    ACS/WFC   & F475W & 2829  \\ 
ACS/WFC   & F814W &  1330 &       WFC3/UVIS  & F814W & 3072 &    ACS/WFC   & F814W & 1403  \\ 
{\bf KMHK\,316}    &    & &       ACS/WFC    & F555W &   50 &    {\bf NGC\,1978}   &  &    \\ 
ACS/WFC   & F475W &   575 &       ACS/WFC    & F814W &   40 &    WFC3/UVIS & F275W &17970  \\ 
ACS/WFC   & F814W &  1330 &       {\bf NGC\,1756}  &  &     &    WFC3/UVIS & F336W & 3720  \\ 
{\bf KMHK\,676}    &    & &       ACS/WFC    & F555W &  170 &    WFC3/UVIS & F343N & 3975  \\ 
ACS/WFC   & F475W &   575 &       ACS/WFC    & F814W &  120 &    WFC3/UVIS & F438W & 2475  \\ 
ACS/WFC   & F814W &  1330 &       {\bf NGC\,1783}  &  &     &    WFC3/UVIS & F555W & 1040  \\ 
{\bf KMHK\,1231}    &   & &       WFC3/UVIS  & F275W & 9045 &    WFC3/UVIS & F814W & 2334  \\ 
ACS/WFC   & F435W &   130 &       WFC3/UVIS  & F336W & 3580 &    ACS/WFC   & F555W &  300  \\ 
ACS/WFC   & F555W &   100 &       WFC3/UVIS  & F343N &27651 &    ACS/WFC   & F814W &  200  \\ 
ACS/WFC   & F656N &   600 &       WFC3/UVIS  & F438W & 5628 &    {\bf NGC\,1983}   &  &    \\ 
ACS/WFC   & F814W &    80 &       ACS/WFC    & F435W &  770 &    ACS/WFC   & F555W &   20  \\ 
{\bf KMK\,8827}    &    & &       ACS/WFC    & F555W &  870 &    ACS/WFC   & F814W &   20  \\        
ACS/WFC   & F475W &   601 &       ACS/WFC    & F814W &  858 &    {\bf NGC\,1987}   &  &    \\             
ACS/WFC   & F814W &  1330 &       {\bf NGC\,1786}  &  &     &    ACS/WFC   & F435W &  770  \\         
{\bf Kron\,1}     &     & &       WFC3/UVIS  & F336W & 2030 &    ACS/WFC   & F555W &  970  \\ 
ACS/WFC   & F555W &   480 &       WFC3/UVIS  & F438W &  400 &    ACS/WFC   & F814W &  858  \\ 
ACS/WFC   & F814W &   290 &       WFC3/UVIS  & F814W &  100 &    {\bf NGC\,2002}   &  &    \\              
{\bf Kron\,3}     &     & &       {\bf NGC\,1793}  &  &     &    ACS/WFC   & F555W &   20  \\ 
ACS/WFC   & F555W &  2024 &       ACS/WFC    & F475W & 2829 &    ACS/WFC   & F814W &   20  \\ 
ACS/WFC   & F814W &  1916 &       ACS/WFC    & F814W & 1403 &    {\bf NGC\,2005}   &  &    \\            
{\bf Kron\,21}    &     & &       {\bf NGC\,1795}  &  &     &    ACS/WFC   & F475W & 1440  \\  
ACS/WFC   & F555W &   480 &       ACS/WFC    & F555W &  300 &    ACS/WFC   & F814W & 1430  \\ 
ACS/WFC   & F814W &   290 &       ACS/WFC    & F814W &  200 &    {\bf NGC\,2010}   &  &    \\ 
{\bf Kron\,29} &    &     &       {\bf NGC\,1801}  &  &     &    ACS/WFC   & F555W &   20  \\ 
ACS/WFC   & F435W &   440 &       WFC3/UVIS  & F336W & 2503 &    ACS/WFC   & F814W &   20  \\ 
ACS/WFC   & F555W &   560 &       WFC3/UVIS  & F656N & 1440 &    {\bf NGC\,2031}   &  &    \\ 
ACS/WFC   & F814W &   560 &       WFC3/UVIS  & F814W &  756 &    ACS/WFC   & F435W &  130  \\ 
{\bf Kron\,34} &    &     &       ACS/WFC    & F555W &  115 &    ACS/WFC   & F555W &  100  \\ 
WFC3/UVIS & F336W &   2517&       ACS/WFC    & F814W &   90 &    ACS/WFC   & F658N &  600  \\ 
WFC3/UVIS & F656N &   1440&       {\bf NGC\,1805}  &  &     &    ACS/WFC   & F814W &   80  \\ 
WFC3/UVIS & F814W &    770&       WFC3/UVIS  & F225W & 4800 &    {\bf NGC\,2056}   &  &    \\ 
ACS/WFC   & F555W &    165&       WFC3/UVIS  & F336W & 3741 &    ACS/WFC   & F555W &  170  \\ 
ACS/WFC   & F814W &    130&       WFC3/UVIS  & F656N & 1440 &    ACS/WFC   & F814W &  120  \\ 
{\bf Lindsay\,1} &  &     &       WFC3/UVIS  & F814W &  756 &    {\bf NGC\,2107}   &  &    \\ 
WFC3/UVIS & F275W & 27341 &       {\bf NGC\,1806}  &  &     &    ACS/WFC   & F555W &  170  \\ 
WFC3/UVIS & F336W &  2900 &       WFC3/UVIS  & F336W & 3580 &    ACS/WFC   & F814W &  120  \\ 
WFC3/UVIS & F343N &  4800 &       WFC3/UVIS  & F343N & 2945 &    {\bf NGC\,2108}   &  &    \\ 
WFC3/UVIS & F438W &  1040 &       ACS/WFC    & F435W &  770 &    ACS/WFC   & F435W &  770  \\ 
ACS/WFC   & F555W &  2504 &       ACS/WFC    & F555W & 1020 &    ACS/WFC   & F555W &  970  \\ 
ACS/WFC   & F814W &  2206 &       ACS/WFC    & F814W &  888 &    ACS/WFC   & F814W &  858  \\ 
{\bf Lindsay\,38} &  &    &       {\bf NGC\,1810}  &  &     &    {\bf NGC\,2121}   &  &    \\ 
WFC3/UVIS & F336W &  1688 &       ACS/WFC    & F475W & 1357 &    WFC3/UVIS & F275W & 18239 \\ 
WFC3/UVIS & F343N &  3630 &       ACS/WFC    & F814W &  572 &    WFC3/UVIS & F336W & 1700  \\ 
WFC3/UVIS & F438W &  1199 &       {\bf NGC\,1818}  &  &     &    WFC3/UVIS & F343N & 2660  \\ 
ACS/WFC   & F555W &  2460 &       WFC3/UVIS  & F225W & 4800 &    WFC3/UVIS & F438W & 1120  \\ 
ACS/WFC   & F814W &  2162 &       WFC3/UVIS  & F275W & 1962 &    WFC3/UVIS & F814W & 2350  \\ 
{\bf Lindsay\,114} &  &   &       WFC3/UVIS  & F336W & 3741 &    {\bf NGC\,2154}   &  &    \\ 
ACS/WFC   & F555W &   480 &       WFC3/UVIS  & F475W &  100 &    WFC3/UVIS & F336W & 2140  \\ 
ACS/WFC   & F814W &   290 &       WFC3/UVIS  & F606W & 2268 &    WFC3/UVIS & F555W & 1040  \\ 
{\bf NGC\,121}    &   &   &       WFC3/UVIS  & F814W & 3156 &    ACS/WFC   & F555W &  300  \\ 
WFC3/UVIS & F336W &  4244 &       {\bf NGC\,1831} &  &      &    ACS/WFC   & F814W &  200  \\ 
WFC3/UVIS & F343N &  2950 &       WFC3/UVIS  & F336W & 4180 &    {\bf NGC\,2155}   &  &    \\ 
WFC3/UVIS & F438W &   800 &       WFC3/UVIS  & F814W & 1480 &    WFC3/UVIS & F336W & 1160  \\ 
WFC3/UVIS & F814W &   200 &       {\bf NGC\,1841} &  &      &    WFC3/UVIS & F343N & 2650  \\ 
ACS/WFC   & F555W &  2024 &       WFC3/UVIS  & F336W & 11668&    WFC3/UVIS & F438W & 1210  \\ 
ACS/WFC   & F814W &  1916 &       ACS/WFC    & F606W & 4336 &    {\bf NGC\,2156}   &  &    \\ 
{\bf NGC\,152}    &   &   &       ACS/WFC    & F814W & 7082 &    ACS/WFC   & F475W & 1357  \\ 
WFC3/UVIS & F438W &  1500 &       {\bf NGC\,1844} &  &      &    ACS/WFC   & F814W &  664  \\ 
WFC3/UVIS & F814W &   700 &       ACS/WFC    & F475W & 6300 &    {\bf NGC\,2164}   &  &    \\ 
{\bf NGC\,265}    &   &   &       ACS/WFC    & F814W & 1686 &    WFC3/UVIS & F225W & 4800  \\ 
ACS/WFC   & F435W &   440 &       {\bf NGC\,1846} &  &      &    WFC3/UVIS & F336W & 3741  \\ 
ACS/WFC   & F555W &   589 &       WFC3/UVIS  & F336W & 9156 &    WFC3/UVIS & F656N &  848  \\ 
ACS/WFC   & F814W &   589 &       WFC3/UVIS  & F343N & 2945 &    WFC3/UVIS & F814W & 1440  \\ 
{\bf NGC\,290}    &   &   &       WFC3/NIR   & F160W & 2844 &    {\bf NGC\,2173}   &  &    \\ 
ACS/WFC   & F435W &   440 &       ACS/WFC    & F435W &  770 &    WFC3/UVIS & F336W & 2200  \\ 
ACS/WFC   & F555W &   560 &       ACS/WFC    & F555W & 1020 &    WFC3/UVIS & F475W & 1520  \\ 
ACS/WFC   & F814W &   560 &       ACS/WFC    & F814W &  888 &    WFC3/UVIS & F814W & 1980  \\ 
{\bf NGC\,294}    &   &   &       {\bf NGC\,1850} &  &      &    {\bf NGC\,2203} & &       \\ 
WFC3/UVIS & F336W &  2517 &       WFC3/UVIS  & F275W & 1720 &    WFC3/UVIS & F336W & 2200  \\ 
WFC3/UVIS & F656N &  1440 &       WFC3/UVIS  & F336W & 2550 &    WFC3/UVIS & F475W & 1520  \\ 
WFC3/UVIS & F814W &   770 &       WFC3/UVIS  & F343N & 4075 &    WFC3/UVIS & F814W & 1980  \\ 
ACS/WFC   & F555W &   165 &       WFC3/UVIS  & F438W & 1045 &    {\bf NGC\,2209} & &       \\ 
ACS/WFC   & F814W &   130 &       WFC3/UVIS  & F467M & 1980 &    WFC3/UVIS & F438W & 1700  \\ 
{\bf NGC\,299}    &   &   &       WFC3/UVIS  & F475W & 1070 &    WFC3/UVIS & F814W & 1030  \\ 
WFC3/UVIS & F656N  & 2394 &       WFC3/UVIS  & F502N & 2051 &    {\bf NGC\,2210} & &       \\ 
ACS/WFC   & F555W  & 1858 &       WFC3/UVIS  & F547M &  782 &    WFC3/UVIS & F336W & 10806 \\ 
ACS/WFC   & F814W  & 1966 &       WFC3/UVIS  & F555W & 1137 &    ACS/WFC   & F606W & 4306  \\ 
{\bf NGC\,330}  &   &     &       WFC3/UVIS  & F657N & 2979 &    ACS/WFC   & F814W & 6950  \\ 
WFC3/UVIS & F225W  & 4860 &       WFC3/UVIS  & F656N & 4225 &    {\bf NGC\,2213} & &       \\ 
WFC3/UVIS & F336W  & 3795 &       WFC3/UVIS  & F673N & 1982 &    WFC3/UVIS & F475W & 1440  \\ 
WFC3/UVIS & F656N  & 1440 &       WFC3/UVIS  & F814W & 1867 &    WFC3/UVIS & F814W & 1430  \\ 
WFC3/UVIS & F814W  &  770 &       WFC3/NIR   & F160W & 9445 &    {\bf NGC\,2249} & &       \\ 
{\bf NGC\,339}  &   &     &       WFC3/NIR   & F164N & 2012 &    WFC3/UVIS & F438W & 1650  \\ 
WFC3/UVIS & F336W  & 3060 &       {\bf NGC\,1852}   &  &    &    WFC3/UVIS & F814W &  910  \\ 
WFC3/UVIS & F343N  & 4220 &       WFC3/UVIS  & F336W & 2140 &    {\bf NGC\,2257} & &       \\ 
WFC3/UVIS & F438W  & 1520 &       WFC3/UVIS  & F555W & 1040 &    WFC3/UVIS & F336W & 10619 \\ 
ACS/WFC   & F555W  & 2024 &       ACS/WFC    & F555W & 330  &    ACS/WFC   & F606W & 4360  \\ 
ACS/WFC   & F814W  & 1926 &       ACS/WFC    & F814W & 200  &    ACS/WFC   & F814W & 6788  \\ 
{\bf NGC\,346}  &   &     &       {\bf NGC\,1854}   &  &    &    {\bf Reticulum} & &       \\ 
WFC3/UVIS & F225W  & 2    &       ACS/WFC    & F555W & 50   &    WFC3/UVIS & F336W & 10978 \\ 
ACS/WFC   & F555W  & 9110 &       ACS/WFC    & F814W & 40   &    WFC3/UVIS & F438W &  400  \\ 
ACS/WFC   & F656N  & 1542 &       {\bf NGC\,1856}   &  &    &    WFC3/UVIS & F814W &  100  \\ 
ACS/WFC   & F814W  & 8632 &       WFC3/UVIS  & F336W & 5688 &    ACS/WFC   & F555W &  330  \\ 
{\bf NGC\,376}  &   &     &       WFC3/UVIS  & F343N & 2750 &    ACS/WFC   & F606W & 4327  \\ 
WFC3/UVIS & F656N  & 2389 &       WFC3/UVIS  & F438W & 1045 &    ACS/WFC   & F814W & 6979  \\ 
ACS/WFC   & F555W  & 1806 &       WFC3/UVIS  & F555W &  700 &    {\bf SL\,862} &      &    \\ 
ACS/WFC   & F814W  & 1966 &       WFC3/UVIS  & F656N & 2615 &    ACS/WFC   & F435W &   735 \\ 
{\bf NGC\,411}  &   &     &       WFC3/UVIS  & F814W & 4037 &    ACS/WFC   & F555W &   705 \\ 
WFC3/UVIS & F336W  & 2200 &       {\bf NGC\,1858}   &  &    &    ACS/WFC   & F814W &   695 \\ 
WFC3/UVIS & F475W  & 1520 &       ACS/WFC    & F555W & 20   &    {\bf Cluster\,1} &   &    \\ 
WFC3/UVIS & F814W  & 1980 &       ACS/WFC    & F814W & 20   &    ACS/WFC   & F475W & 12695 \\ 
{\bf NGC\,416}  &   &     &       {\bf NGC\,1866}   &  &    &    ACS/WFC   & F775W & 28900 \\ 
WFC3/UVIS & F275W  &27349 &       WFC3/UVIS  & F336W & 5688 &    ACS/WFC   & F850LP&  9190 \\ 
WFC3/UVIS & F336W  & 3060 &       WFC3/UVIS  & F343N & 3900 &    {\bf Cluster\,2} &   &    \\ 
WFC3/UVIS & F343N  & 4605 &       WFC3/UVIS  & F438W & 1195 &    WFC3/NIR  & F110W & 12592 \\ 
WFC3/UVIS & F438W  & 1125 &       WFC3/UVIS  & F555W & 2520 &    WFC3/NIR  & F160W & 24784 \\ 
ACS/WFC   & F555W  & 2064 &       WFC3/UVIS  & F814W & 3072 &              &       &       \\ 
\end{longtable} 
\end{center}    
\twocolumn

 \onecolumn

\begin{center}
\begin{longtable}{lcccccccccccccc}

\caption{Cluster-center coordinates, average proper motions, distance modulus, reddening, metallicities, and ages for the studied clusters. }\label{tab:info}\\
\multicolumn{12}{c}%

%{{\bfseries \tablename\ \thetable{} -- continued from previous page}} \\
\endhead

\hline \multicolumn{3}{|r|}{{Continued on next page}} \\ \hline
\endfoot

\hline \hline
\endlastfoot
 
  ID            &  RA         &  error            &  DEC           & error                   &  $\mu_{\alpha}$cos$\delta$          & $\mu_{\delta}$      & (m$-$M)$_{0}$ & E(B$-$V) & [M/H]  & age\footnote{The ages of clusters with the eMSTO are inferred by fitting the isochrone to the lower part of the eMSTO. See Section\,\ref{subsec:CMDs} for details.}  & age\footnote{The ages of clusters with the eMSTO are inferred by fitting the isochrone to the upper part of the eMSTO.  See Section\,\ref{subsec:CMDs} for details.} \\  
                & h m s       &  [arcsec]           & d m s          & [arcsec]                  & [mas yr$^{-1}$]                      & [mas yr$^{-1}$]      &  [mag]          &  [mag]     & [dex]    & [Gyr] & [Gyr] \\
  \hline
  BRHT\,5b      & 05 08 52.65 &    ---           & $-$68 45 18.0  & ---          & 2.09$\pm$0.03      &   0.04$\pm$0.09    &     18.40  &   0.12       &  $-$0.4    &  0.015 & ---  \\
  BS\,90        & 00 59 04.86 & $\pm$0.4         & $-$72 09 10.3  & $\pm$4.7     & 0.67$\pm$0.05      &$-$1.06$\pm$0.06    &     18.91  &   0.03       &  $-$0.7    &  4.2   & ---  \\
  BSDL\,1650    & 05 25 50.01 & ---              & $-$68 49 16.0  & ---          & 1.54$\pm$0.19      &   0.34$\pm$0.15    &     18.38  &   0.18       &  $-$0.4    &  0.30  & ---  \\ 
  ESO\,121-03   & 06 02 02.40 & $\pm$0.8         & $-$60 31 26.0  & $\pm$1.2     & 1.61$\pm$0.04      &   0.92$\pm$0.06    &     18.34  &   0.05       &  $-$0.8    &  6.9   & ---  \\
  Hodge\,02     & 05 17 48.88 & $\pm$0.1         & $-$69 38 43.4  & $\pm$0.3     & 2.29$\pm$0.13      &   0.43$\pm$0.15    &     18.32  &   0.12       &  $-$0.5    &  1.70  & 1.35 \\
  Hodge\,06     & 05 42 17.65 & $\pm$0.8         & $-$71 35 28.2  & $\pm$0.5     & 1.95$\pm$0.06      &   0.76$\pm$0.06    &     18.40  &   0.15       &  $-$0.5    &  2.30  & ---  \\ %(ESO\,057-30) 
  Hodge\,07     & 05 50 02.99 & $\pm$0.4         & $-$67 43 06.6  & $\pm$1.0     & 1.77$\pm$0.05      &   0.75$\pm$0.07    &     18.33  &   0.06       &  $-$0.5    &  2.00  & 1.75 \\ %(ESO\,057-43) 
  Hodge\,11     & 06 14 22.89 & $\pm$0.2         & $-$69 50 50.6  & $\pm$0.2     & 1.56$\pm$0.08      &   0.75$\pm$0.07    &     18.57  &   0.05       &  $-$1.7    & 13.4   & ---  \\
  HW\,57        & 01 07 43.22 & $\pm$0.7         & $-$71 52 37.5  & $\pm$0.8     & 0.77$\pm$0.14      &$-$1.03$\pm$0.06    &     19.18  &   0.12       &  $-$1.3    &  5.3   & ---  \\
  IC\,1641      & 01 09 38.82 & $\pm$0.6         & $-$71 46 04.4  & $\pm$1.3     & 1.04$\pm$0.06      &$-$1.11$\pm$0.09    &     18.93  &   0.02       &  $-$0.6    &  1.20  & 0.80 \\
  IC\,1660      & 01 12 37.59 & $\pm$0.3         & $-$71 45 41.4  & $\pm$0.3     & 1.07$\pm$0.05      &$-$1.30$\pm$0.04    &     19.06  &   0.07       &  $-$0.2    &  0.11  & ---  \\
  IC\,2146      & 05 37 47.37 & $\pm$0.6         & $-$74 47 01.3  & $\pm$0.6     & 2.02$\pm$0.03      &   0.73$\pm$0.03    &     18.42  &   0.05       &  $-$0.4    &  2.20  & 2.00 \\
  KMHK\,240     & 04 54 26.88 & $\pm$0.9         & $-$68 14 55.1  & $\pm$0.6     & 2.26$\pm$0.12      &$-$0.01$\pm$0.11    &     18.52  &   0.12       &  $-$0.5    &  2.10  & 1.90 \\
  KMHK\,250     & 04 54 30.32 & $\pm$0.2         & $-$69 55 15.1  & $\pm$0.4     & 2.01$\pm$0.09      &   0.01$\pm$0.08    &     18.51  &   0.11       &  $-$0.5    &  1.80  & 1.45 \\
  KMHK\,291     & 04 55 45.26 & $\pm$0.2         & $-$68 16 56.2  & $\pm$0.5     & 1.74$\pm$0.12      &$-$0.13$\pm$0.11    &     18.55  &   0.12       &  $-$0.2    &  0.30  & 0.15 \\
  KMHK\,361     & 04 56 37.46 & $\pm$1.4         & $-$68 09 55.8  & $\pm$0.7     & 1.69$\pm$0.16      &   0.06$\pm$0.13    &     18.43  &   0.06       &  $-$0.3    &  1.35  & 1.00 \\
  KMHK\,598     & 05 09 35.72 &    ---           & $-$67 48 31.2  & ---          & 1.76$\pm$0.12      &   0.04$\pm$0.22    &     18.50  &   0.06       &  $-$0.2    &  0.15  & ---  \\
  KMHK\,676     & 05 14 44.37 & $\pm$1.8         & $-$65 20 08.5  & $\pm$0.9     & 1.61$\pm$0.04      &   0.15$\pm$0.08    &     18.49  &   0.07       &  $-$0.2    &  0.14  & ---  \\
  KMHK\,987     & 05 30 32.71 & $\pm$0.8         & $-$66 54 12.4  & $\pm$0.4     & 1.55$\pm$0.03      &   0.47$\pm$0.04    &     18.52  &   0.07       &  $-$0.2    &  0.017 & ---  \\
  KMHK\,1073    & 05 33 10.75 & ---              & $-$71 01 21.0  & ---          & 2.10$\pm$0.07      &   0.54$\pm$0.06    &     18.34  &   0.11       &  $-$0.4    &  0.45  & 0.30 \\   
  KMHK\,1231    & 05 41 09.62 & $\pm$3.1         & $-$69 54 12.4  & $\pm$0.4     & 1.96$\pm$0.09      &   0.63$\pm$0.15    &     18.47  &   0.13       &  $-$0.3    &  0.35  & 0.20 \\
  KMK\,8827     & 05 08 54.14 & ---              & $-$69 00 15.2  &  ---         & 1.81$\pm$0.58      &$-$0.30$\pm$0.61    &     18.40  &   0.11       &  $-$0.2    &  0.20  & ---  \\
  KMK\,8849     & 05 21 10.76 & ---              & $-$69 56 30.3  & ---          & 2.51$\pm$0.41      &   0.34$\pm$0.13    &     18.44  &   0.22       &  $-$0.4    &  0.50  & 0.25 \\ 
  Kron\,1       & 00 21 25.78 & $\pm$1.2         & $-$73 44 55.7  & $\pm$0.5     & 0.35$\pm$0.06      &$-$1.35$\pm$0.06    &     18.90  &   0.03       &  $-$0.9    &  6.8   & ---  \\
  Kron\,3       & 00 24 46.63 & $\pm$0.8         & $-$72 47 37.0  & $\pm$0.2     & 0.53$\pm$0.02      &$-$1.35$\pm$0.03    &     18.93  &   0.02       &  $-$0.9    &  5.6   & ---  \\
  Kron\,21      & 00 41 24.39 & $\pm$0.3         & $-$72 53 23.8  & $\pm$0.2     & 0.62$\pm$0.02      &$-$1.53$\pm$0.03    &     18.84  &   0.06       &  $-$1.0    &  4.4   & ---  \\
  Kron\,29         & 00 51 53.15 & $\pm$0.2         & $-$72 57 12.2  & $\pm$0.3     & 0.65$\pm$0.05      &$-$1.21$\pm$0.03    &     19.10  &   0.07       &  $-$0.4    &  0.25  & 0.12 \\
  Kron\,34      & 00 55 33.44 & $\pm$1.5         & $-$72 49 57.6  & $\pm$1.5     & 0.69$\pm$0.04      &$-$1.25$\pm$0.04    &     18.90  &   0.12       &  $-$0.6    &  0.85  & 0.55 \\
  Lindsay\,1    & 00 03 54.44 & $\pm$2.0         & $-$73 28 18.7  & $\pm$1.3     & 0.54$\pm$0.03      &$-$1.49$\pm$0.03    &     18.86  &   0.04       &  $-$1.2    &  7.2   & ---  \\
  Lindsay\,38   & 00 48 49.80 & $\pm$2.5         & $-$69 52 12.6  & $\pm$1.6     & 0.54$\pm$0.03      &$-$0.86$\pm$0.03    &     19.22  &   0.01       &  $-$1.2    &  5.4   & ---  \\
  Lindsay\,91   & 01 12 51.76 & $\pm$0.2         & $-$73 07 07.4  & $\pm$0.5     & 0.76$\pm$0.07      &$-$1.09$\pm$0.04    &     18.97  &   0.10       &  $-$0.8    &  4.2   & ---  \\
  Lindsay\,113  & 01 49 29.69 & $\pm$4.0         & $-$73 43 40.2  & $\pm$1.3     & 1.30$\pm$0.02      &$-$1.18$\pm$0.03    &     18.76  &   0.03       &  $-$0.8    &  3.6   & ---  \\
  Lindsay\,114  & 01 50 19.27 & $\pm$0.3         & $-$74 21 20.5  & $\pm$0.4     & 1.09$\pm$0.03      &$-$1.14$\pm$0.04    &     18.86  &   0.07       &  $-$0.4    &  0.04  & ---  \\
  NGC\,121      & 00 26 48.94 & $\pm$0.1         & $-$71 32 09.4  & $\pm$0.1     & 0.23$\pm$0.03      &$-$1.23$\pm$0.03    &     19.05  &   0.04       &  $-$1.2    &  9.7   & ---  \\
  NGC\,152      & 00 32 56.47 & $\pm$1.2         & $-$73 06 59.2  & $\pm$2.2     & 0.41$\pm$0.03      &$-$1.26$\pm$0.04    &     19.07  &   0.03       &  $-$0.6    &  1.90  & 1.45 \\
  NGC\,265      & 00 47 11.82 & $\pm$0.6         & $-$73 28 38.2  & $\pm$0.2     & 0.64$\pm$0.03      &$-$1.31$\pm$0.04    &     19.03  &   0.06       &  $-$0.5    &  0.45  & 0.25 \\
  NGC\,290      & 00 51 14.24 & ---              & $-$73 09 42.2  & ---          & 0.67$\pm$0.09      &$-$1.47$\pm$0.07    &     18.95  &   0.05       &  $-$0.5    &  0.30  & 0.20 \\
  NGC\,294      & 00 53 05.58 & $\pm$1.4         & $-$73 22 48.7  & $\pm$3.2     & 0.53$\pm$0.05      &$-$1.27$\pm$0.04    &     18.98  &   0.12       &  $-$0.7    &  0.70  & 0.45 \\
  NGC\,299      & 00 53 24.51 & $\pm$0.2         & $-$72 11 50.6  & $\pm$0.2     & 0.69$\pm$0.04      &$-$1.25$\pm$0.02    &     18.99  &   0.06       &  $-$0.4    &  0.08  & 0.02 \\
  NGC\,330      & 00 56 18.23 & $\pm$0.1         & $-$72 27 32.3  & $\pm$0.1     & 0.75$\pm$0.03      &$-$1.31$\pm$0.03    &     19.04  &   0.04       &  $-$0.4    &  0.09  & 0.04 \\
  NGC\,339      & 00 57 46.56 & $\pm$0.5         & $-$74 28 13.2  & $\pm$0.4     & 0.70$\pm$0.03      &$-$1.25$\pm$0.04    &     18.96  &   0.07       &  $-$1.3    &  5.9   & ---  \\
  NGC\,346      & 00 59 04.93 & $\pm$0.6         & $-$72 10 37.4  & $\pm$0.4     & 0.70$\pm$0.04      &$-$1.23$\pm$0.03    &     18.94  &   0.08       &  $-$0.4    &  0.005 & ---  \\
  NGC\,376      & 01 03 52.75 & $\pm$0.7         & $-$72 49 32.0  & $\pm$0.3     & 0.72$\pm$0.04      &$-$1.31$\pm$0.03    &     18.98  &   0.07       &  $-$0.4    &  0.028 & 0.018 \\
  NGC\,411      & 01 07 55.95 & $\pm$0.4         & $-$71 46 04.1  & $\pm$0.4     & 0.87$\pm$0.08      &$-$1.12$\pm$0.06    &     18.97  &   0.06       &  $-$0.7    &  1.95  & 1.55 \\
  NGC\,416      & 01 07 59.17 & $\pm$0.2         & $-$72 21 19.7  & $\pm$0.1     & 0.88$\pm$0.04      &$-$1.24$\pm$0.03    &     18.96  &   0.11       &  $-$1.2    &  6.0   & ---  \\
  NGC\,419      & 01 08 17.57 & $\pm$0.7         & $-$72 53 03.8  & $\pm$1.0     & 0.77$\pm$0.06      &$-$1.22$\pm$0.04    &     18.85  &   0.07       &  $-$0.7    &  2.00  & 1.55 \\
  NGC\,422      & 01 09 24.48 & $\pm$6.2         & $-$71 45 59.3  & $\pm$0.9     & 0.93$\pm$0.04      &$-$1.27$\pm$0.03    &     18.88  &   0.04       &  $-$0.4    &  0.30  & 0.20 \\
  NGC\,602      & 01 29 31.50 & $\pm$0.8         & $-$73 33 40.8  & $\pm$0.4     & 0.96$\pm$0.02      &$-$1.26$\pm$0.05    &     19.01  &   0.06       &  $-$0.2    &  0.005 & 0.002 \\
  NGC\,1466     & 03 44 32.76 & $\pm$1.3         & $-$71 40 15.5  & $\pm$0.3     & 1.72$\pm$0.06      &$-$0.74$\pm$0.07    &     18.58  &   0.05       &  $-$1.5    &  13.2  & ---  \\
  NGC\,1644     & 04 37 39.85 & $\pm$0.3         & $-$66 11 56.3  & $\pm$0.5     & 1.80$\pm$0.02      &$-$0.29$\pm$0.11    &     18.45  &   0.03       &  $-$0.6    &  1.80  & 1.45 \\
  NGC\,1651     & 04 37 32.23 & $\pm$0.6         & $-$70 35 10.8  & $\pm$0.3     & 2.02$\pm$0.04      &$-$0.30$\pm$0.05    &     18.48  &   0.13       &  $-$0.6    &  2.20  & 2.05 \\
  NGC\,1652     & 04 38 22.77 & $\pm$0.3         & $-$68 40 19.8  & $\pm$0.2     & 1.87$\pm$0.06      &$-$0.37$\pm$0.07    &     18.46  &   0.08       &  $-$0.6    &  2.25  & ---  \\
  NGC\,1718     & 04 52 25.89 & $\pm$0.2         & $-$67 03 06.6  & $\pm$0.4     & 1.85$\pm$0.03      &$-$0.41$\pm$0.04    &     18.43  &   0.22       &  $-$0.5    &  2.05  & 1.85 \\
  NGC\,1749     & 04 54 56.73 &    ---           & $-$68 11 19.1  & ---          & 1.94$\pm$0.08      &$-$0.10$\pm$0.12    &     18.30  &   0.10       &  $-$0.4    &  0.13  & 0.07 \\
  NGC\,1751     & 04 54 11.99 & $\pm$1.1         & $-$69 48 27.1  & $\pm$0.6     & 1.93$\pm$0.07      &$-$0.09$\pm$0.10    &     18.52  &   0.15       &  $-$0.5    &  1.75  & 1.45 \\
  NGC\,1755     & 04 55 15.56 & $\pm$0.2         & $-$68 12 18.8  & $\pm$0.6     & 1.88$\pm$0.04      &$-$0.11$\pm$0.05    &     18.33  &   0.11       &  $-$0.2    &  0.11  & 0.07 \\
  NGC\,1756     & 04 54 49.69 & $\pm$0.7         & $-$69 14 13.2  & $\pm$0.3     & 1.83$\pm$0.04      &   0.10$\pm$0.03    &     18.55  &   0.22       &  $-$0.4    &  0.20  & 0.15 \\
  NGC\,1783     & 04 59 08.97 & $\pm$0.5         & $-$65 59 13.8  & $\pm$0.2     & 1.64$\pm$0.04      &$-$0.06$\pm$0.04    &     18.51  &   0.03       &  $-$0.4    &  1.95  & 1.65 \\
  NGC\,1786     & 04 59 07.99 & $\pm$0.1         & $-$67 44 43.9  & $\pm$0.3     & 1.95$\pm$0.03      &   0.06$\pm$0.03    &     18.42  &   0.09       &  $-$1.5    &  12.9  & ---  \\
  NGC\,1793     & 04 59 38.74 & $\pm$0.1         & $-$69 33 27.8  & $\pm$0.1     & 2.09$\pm$0.08      &$-$0.05$\pm$0.05    &     18.48  &   0.14       &  $-$0.2    &  0.15  & 0.06 \\
  NGC\,1795     & 04 59 47.35 & $\pm$1.1         & $-$69 48 06.5  & $\pm$0.3     & 1.90$\pm$0.05      &   0.23$\pm$0.11    &     18.45  &   0.09       &  $-$0.4    &  1.85  & 1.50 \\
  NGC\,1801     & 05 00 35.41 & $\pm$0.2         & $-$69 36 49.9  & $\pm$0.7     & 1.90$\pm$0.05      &   0.05$\pm$0.04    &     18.39  &   0.12       &  $-$0.3    &  0.45  & 0.30 \\
  NGC\,1805     & 05 02 21.78 & $\pm$0.1         & $-$66 06 41.9  & $\pm$0.1     & 1.56$\pm$0.04      &   0.10$\pm$0.06    &     18.32  &   0.05       &  $-$0.4    &  0.10  & 0.045\\
  NGC\,1806     & 05 02 11.72 & $\pm$0.4         & $-$67 59 08.0  & $\pm$0.5     & 1.85$\pm$0.05      &$-$0.06$\pm$0.07    &     18.52  &   0.04       &  $-$0.4    &  1.90  & 1.60 \\
  NGC\,1810     & 05 03 23.06 & $\pm$0.7         & $-$66 22 56.7  & $\pm$1.4     & 1.72$\pm$0.05      &   0.07$\pm$0.04    &     18.45  &   0.04       &  $-$0.2    &  0.08  & 0.045\\
  NGC\,1818     & 05 04 13.43 & $\pm$0.4         & $-$66 26 01.7  & $\pm$1.2     & 1.64$\pm$0.04      &   0.09$\pm$0.06    &     18.44  &   0.07       &  $-$0.2    &  0.07  & 0.035\\
  NGC\,1831     & 05 06 16.38 & $\pm$0.4         & $-$64 55 06.1  & $\pm$0.8     & 1.69$\pm$0.11      &$-$0.04$\pm$0.10    &     18.41  &   0.05       &  $-$0.3    &  0.90  & 0.70 \\
  NGC\,1841     & 04 45 22.75 & $\pm$0.4         & $-$83 59 55.6  & $\pm$0.6     & 2.05$\pm$0.02      &   0.00$\pm$0.03    &     18.34  &   0.13       &  $-$1.3    &  12.4  & ---  \\
  NGC\,1844     & 05 07 30.38 & $\pm$0.9         & $-$67 19 28.6  & $\pm$1.9     & 1.68$\pm$0.02      &$-$0.03$\pm$0.03    &     18.47  &   0.07       &  $-$0.2    &  0.17  & 0.09 \\
  NGC\,1846     & 05 07 34.15 & $\pm$0.4         & $-$67 27 36.7  & $\pm$0.2     & 1.71$\pm$0.04      &   0.03$\pm$0.04    &     18.52  &   0.05       &  $-$0.4    &  1.95  & 1.60 \\
  NGC\,1850     & 05 08 45.19 & $\pm$1.3         & $-$68 45 42.0  & $\pm$1.5     & 2.02$\pm$0.04      &   0.11$\pm$0.04    &     18.38  &   0.13       &  $-$0.4    &  0.12  & 0.07 \\
  NGC\,1850A\footnote{
  NGC\,1850A is a clump of stars located on the west side of NGC\,1850 and is often considered a separate cluster \citep[e.g.\,][]{caloi1998a}. 
  Based on the results of this table, NGC\,1850 and NGC\,1850A share similar proper motions, distance, metallicity, and age.}   & 05 08 39.44 &    ---           & $-$68 45 44.2  & ---                      & 1.95$\pm$0.04                     &   0.13$\pm$0.01    &     18.36          &   0.11       &  $-$0.4    &  0.020  & 0.10\\
  NGC\,1852     & 05 09 23.95 & $\pm$0.3         & $-$67 46 45.6  & $\pm$0.2     & 1.78$\pm$0.04      &   0.16$\pm$0.06    &     18.52  &   0.07       &  $-$0.4    &  1.75   & 1.40 \\
  NGC\,1854     & 05 09 19.83 & $\pm$0.5         & $-$68 50 52.0  & $\pm$0.8     & 2.09$\pm$0.04      &   0.15$\pm$0.02    &     18.42  &   0.09       &  $-$0.2    &  0.15   & 0.09 \\
  NGC\,1856     & 05 09 30.08 & $\pm$0.1         & $-$69 07 43.9  & $\pm$0.3     & 1.88$\pm$0.05      &   0.20$\pm$0.05    &     18.32  &   0.17       &  $-$0.4    &  0.45   & 0.25 \\
  NGC\,1858     & 05 10 00.07 &  ---             & $-$68 54 15.1  & ---          & 1.87$\pm$0.04      &   0.25$\pm$0.03    &     18.46  &   0.12       &  $-$0.2    &  0.017  & 0.004\\
  NGC\,1866     & 05 13 38.65 & $\pm$0.3         & $-$65 27 52.8  & $\pm$0.4     & 1.55$\pm$0.03      &   0.16$\pm$0.03    &     18.30  &   0.06       &  $-$0.4    &  0.40   & 0.20 \\
  NGC\,1868     & 05 14 35.91 & $\pm$0.4         & $-$63 57 15.1  & $\pm$0.1     & 1.83$\pm$0.04      &   0.05$\pm$0.07    &     18.45  &   0.06       &  $-$0.4    &  1.45   & 1.15 \\
  NGC\,1872     & 05 13 11.29 & $\pm$0.8         & $-$69 18 44.9  & $\pm$0.3     & 1.79$\pm$0.08      &   0.52$\pm$0.05    &     18.31  &   0.18       &  $-$0.4    &  0.60   & 0.40 \\
  NGC\,1898     & 05 16 41.57 & $\pm$0.3         & $-$69 39 24.1  & $\pm$0.1     & 1.98$\pm$0.05      &   0.35$\pm$0.05    &     18.60  &   0.06       &  $-$1.5    &  11.7   & ---  \\
  NGC\,1903     & 05 17 22.62 & $\pm$0.3         & $-$69 20 17.0  & $\pm$0.5     & 2.01$\pm$0.07      &   0.16$\pm$0.05    &     18.40  &   0.06       &  $-$0.2    &  0.15   & 0.10 \\
  NGC\,1917     & 05 19 01.94 & $\pm$0.4         & $-$69 00 05.5  & $\pm$0.5     & 1.65$\pm$0.11      &   0.57$\pm$0.09    &     18.36  &   0.05       &  $-$0.3    &  1.70   & 1.40 \\
  NGC\,1928     & 05 20 57.49 & $\pm$1.4         & $-$69 28 41.6  & $\pm$2.1     & 1.84$\pm$0.10      &   0.13$\pm$0.12    &     18.43  &   0.06       &  $-$1.5    &  13.0   & ---  \\
  NGC\,1938     & 05 21 25.00 & ---              & $-$69 56 21.5  & ---          & 2.04$\pm$0.09      &   0.34$\pm$0.07    &     18.48  &   0.23       &  $-$0.3    &  0.15   & ---  \\ 
  NGC\,1939     & 05 21 26.37 & $\pm$0.3         & $-$69 56 58.4  & $\pm$1.3     & 2.21$\pm$0.07      &   0.44$\pm$0.03    &     18.42  &   0.06       &  $-$1.5    &  13.3   & ---  \\
  NGC\,1943     & 05 22 29.36 & $\pm$0.4         & $-$70 09 18.5  & $\pm$0.8     & 2.04$\pm$0.04      &   0.08$\pm$0.07    &     18.46  &   0.14       &  $-$0.3    &  0.20   & 0.15 \\
  NGC\,1953     & 05 25 27.95 & $\pm$0.2         & $-$68 50 16.1  & $\pm$0.1     & 1.75$\pm$0.06      &   0.41$\pm$0.07    &     18.41  &   0.11       &  $-$0.4    &  0.50   & 0.35 \\
  NGC\,1966     & 05 26 45.53 & ---              & $-$68 49 50.9  & ---          & 1.59$\pm$0.04      &   0.52$\pm$0.21    &     18.40  &   0.06       &  $-$0.2    &  0.005  & 0.003\\
  NGC\,1978     & 05 28 44.71 & $\pm$1.2         & $-$66 14 10.9  & $\pm$1.1     & 1.76$\pm$0.03      &   0.40$\pm$0.04    &     18.53  &   0.07       &  $-$0.5    &  2.50   & ---  \\
  NGC\,1983     & 05 27 44.95 & $\pm$0.8         & $-$68 59 06.5  & $\pm$0.3     & 1.60$\pm$0.05      &   0.49$\pm$0.05    &     18.56  &   0.03       &     0.0    &  0.014  & ---  \\
  NGC\,1987     & 05 27 17.03 & $\pm$0.1         & $-$70 44 11.4  & $\pm$0.2     & 1.94$\pm$0.06      &   0.46$\pm$0.04    &     18.43  &   0.07       &  $-$0.7    &  1.35   & 1.00 \\
  NGC\,2002     & 05 30 20.82 & $\pm$0.4         & $-$66 53 01.1  & $\pm$0.3     & 1.57$\pm$0.04      &   0.47$\pm$0.07    &     18.52  &   0.07       &  $-$0.2    &  0.02   & ---  \\
  NGC\,2005     & 05 30 10.13 & $\pm$0.1         & $-$69 45 10.6  & $\pm$0.2     & 1.88$\pm$0.04      &   0.56$\pm$0.04    &     18.44  &   0.09       &  $-$1.6    &  13.1   & ---  \\
  NGC\,2010     & 05 30 33.93 & $\pm$0.8         & $-$70 49 07.8  & $\pm$1.3     & 2.23$\pm$0.04      &   0.45$\pm$0.04    &     18.54  &   0.09       &  $-$0.2    &  0.12   & ---  \\
  NGC\,2031     & 05 33 40.27 & $\pm$2.9         & $-$70 59 12.6  & $\pm$1.5     & 2.30$\pm$0.07      &   0.61$\pm$0.07    &     18.40  &   0.09       &  $-$0.4    &  0.30   & 0.15 \\
  NGC\,2056     & 05 36 33.95 & $\pm$0.2         & $-$70 40 15.7  & $\pm$0.2     & 2.16$\pm$0.07      &   0.62$\pm$0.08    &     18.38  &   0.08       &  $-$0.4    &  0.45   & 0.30 \\
  NGC\,2107     & 05 43 12.39 & $\pm$0.1         & $-$70 38 24.4  & $\pm$0.2     & 1.96$\pm$0.10      &   0.83$\pm$0.11    &     18.37  &   0.16       &  $-$0.3    &  0.50   & 0.30 \\
  NGC\,2108     & 05 43 56.54 & $\pm$0.1         & $-$69 10 52.9  & $\pm$0.3     & 1.72$\pm$0.06      &   0.84$\pm$0.08    &     18.48  &   0.14       &  $-$0.3    &  1.25   & 1.00 \\
  NGC\,2121     & 05 48 13.22 & $\pm$1.4         & $-$71 28 46.9  & $\pm$0.8     & 1.76$\pm$0.05      &   0.96$\pm$0.04    &     18.48  &   0.09       &  $-$0.5    &  2.9    & ---  \\
  NGC\,2154     & 05 57 38.22 & $\pm$0.3         & $-$67 15 41.7  & $\pm$0.9     & 1.42$\pm$0.04      &   0.78$\pm$0.05    &     18.37  &   0.04       &  $-$0.5    &  2.00   & 1.80 \\
  NGC\,2155     & 05 58 32.24 & $\pm$0.3         & $-$65 28 39.7  & $\pm$0.6     & 1.73$\pm$0.07      &   0.88$\pm$0.05    &     18.39  &   0.05       &  $-$0.4    &  2.8    & ---  \\
  NGC\,2156     & 05 57 49.88 & $\pm$1.0         & $-$68 27 42.5  & $\pm$1.4     & 1.73$\pm$0.09      &   0.91$\pm$0.08    &     18.44  &   0.06       &  $-$0.2    &  0.17   & 0.10 \\
  NGC\,2164     & 05 58 55.83 & $\pm$0.4         & $-$68 30 57.6  & $\pm$0.3     & 1.60$\pm$0.04      &   0.78$\pm$0.04    &     18.43  &   0.07       &  $-$0.3    &  0.20   & 0.10 \\
  NGC\,2173     & 05 57 58.40 & $\pm$0.2         & $-$72 58 43.2  & $\pm$0.2     & 1.97$\pm$0.04      &   0.83$\pm$0.05    &     18.37  &   0.06       &  $-$0.4    &  2.05   & 1.70 \\
  NGC\,2203     & 06 04 42.62 & $\pm$0.7         & $-$75 26 16.1  & $\pm$0.5     & 1.93$\pm$0.03      &   0.88$\pm$0.03    &     18.38  &   0.07       &  $-$0.3    &  1.95   & 1.65 \\
  NGC\,2209     & 06 08 36.19 & $\pm$0.9         & $-$73 50 09.9  & $\pm$0.7     & 1.94$\pm$0.04      &   0.96$\pm$0.05    &     18.39  &   0.10       &  $-$0.4    &  1.45   & 1.15 \\
  NGC\,2210     & 06 11 31.63 & $\pm$0.2         & $-$69 07 18.7  & $\pm$0.2     & 1.44$\pm$0.05      &   1.36$\pm$0.05    &     18.36  &   0.04       &  $-$1.4    &  12.0   & ---  \\
  NGC\,2213     & 06 10 42.13 & $\pm$0.2         & $-$71 31 45.9  & $\pm$0.8     & 1.77$\pm$0.02      &   0.99$\pm$0.04    &     18.36  &   0.09       &  $-$0.4    &  1.85   & 1.60 \\
  NGC\,2249     & 06 25 49.65 & $\pm$0.3         & $-$68 55 14.2  & $\pm$0.2     & 1.55$\pm$0.06      &   1.09$\pm$0.05    &     18.34  &   0.06       &  $-$0.4    &  1.20   & 0.95 \\
  NGC\,2257     & 06 30 12.42 & $\pm$0.3         & $-$64 19 36.6  & $\pm$0.5     & 1.39$\pm$0.05      &   1.00$\pm$0.04    &     18.37  &   0.04       &  $-$1.4    &  11.8   & ---  \\
  OGLEclLMC390& 05 21 18.91 & ---              & $-$69 28 33.8  & ---          & 1.99$\pm$0.21      &   0.60$\pm$0.29    &     18.44  &   0.10       &  $-$0.5    &  1.55   & 1.30 \\ 
  Reticulum    & 04 36 10.99 & $\pm$0.3         & $-$58 51 45.5  & $\pm$0.5     & 1.95$\pm$0.05      &$-$0.27$\pm$0.02    &     18.40  &   0.00       &  $-$1.2    &  11.5   & ---  \\
  SL\,075       & 06 13 27.26 & $\pm$0.9         & $-$70 41 45.0  & $\pm$0.4     & 1.68$\pm$0.04      &   1.07$\pm$0.04    &     18.49  &   0.06       &  $-$0.4    &  1.95   & 1.70 \\  % (ESO\,057-75)
  %                &  & $\pm$         & $-$  & $\pm$                 & $\pm$                     &$-$ $\pm$    &               &          &  $-$    &    \\
\end{longtable} 
\end{center}    
\twocolumn

\begin{table}
  \caption{ The quantities listed in this table are indicative of the precision of values of distance modulus, reddening, age, and metallicity inferred from isochrone fitting. The horizontal lines separate the couples of clusters with similar ages and different photometric qualities. See text for details. }
\begin{tabular}{l c c c c}
\hline \hline
ID         & $\Delta$(m$-$M)$_{0}$ & $\Delta$E(B$-$V) & $\Delta$age & $\Delta$[M/H] \\
           & [mag] & [mag] & [Myr] & [dex] \\
\hline
NGC\,2005  & 0.10 & 0.010 & 500 & 0.10 \\
NGC\,1939  & 0.15 & 0.015 & 1000 & 0.15 \\
\hline
Kron\,3    & 0.10 & 0.010 & 200  & 0.10 \\
Kron\,1    & 0.10 & 0.020 & 350  & 0.15  \\
\hline
NGC\,1846  & 0.10 & 0.010 & 75  & 0.10  \\
Hodge\,7   & 0.15 & 0.020 & 150 &  0.15 \\
\hline
NGC\,1866  & 0.10 & 0.010 & 30  & 0.10 \\
BSDL\,1650 & 0.20 & 0.035 & 100 & 0.20  \\
\hline
%%%%%%%%%%%%%%%%%%%%%%%%%%%%%%%
\end{tabular}
\label{tab:errori}
\end{table}

 \onecolumn
\tiny
\begin{center}
\begin{longtable}{lccccccccccc}

    \caption{Description of the dataset of the thirteen GCs with {\it HST} proper-motion determinations. For completeness, we include information on F275W images, although they are not used for deriving proper motions.} \label{tab:dataPM}\\

\multicolumn{9}{c}%

%{{\bfseries \tablename\ \thetable{} -- continued from previous page}} \\
\endhead

\hline \multicolumn{3}{|r|}{{Continued on next page}} \\ \hline
\endfoot

\hline \hline
\endlastfoot
 ID & CAMERA & FILTER  & DATE & N$\times$EXPTIME & PROGRAM & PI \\
\hline
 BS\,90, NGC\,346 & ACS/WFC   & F555W & Jul 13-18 2004  & 4$\times$3s$+$380s$+$4$\times$456s$+$4$\times$483s      & 10248 & A.\,Nota\\
         & ACS/WFC   & F658N & Jul 15 2004     & 3$\times$514s                                           & 10248 & A.\,Nota\\
         & ACS/WFC   & F814W & Jul 13-18 2004  & 4$\times$2s$+$380s$+$4$\times$450s$+$4$\times$484s      & 10248 & A.\,Nota\\
         & ACS/WFC   & F555W & Jul 15-22 2015  & 4$\times$3s$+$11$\times$450s                            & 13680 & E.\,Sabbi\\
         & ACS/WFC   & F814W & Jul 15-20 2015  & 4$\times$2s$+$10$\times$450s                            & 13680 & E.\,Sabbi\\

 KRON\,34  & UVIS/WFC3 & F336W & Jan 13 2017  & 3$\times$839s       & 14710 & A.\,P.\,Milone\\
           & UVIS/WFC3 & F656N & Jan 13 2017  & 2$\times$720s       & 14710 & A.\,P.\,Milone\\
           & UVIS/WFC3 & F814W & Jan 13 2017  & 90s$+$680s       & 14710 & A.\,P.\,Milone\\
          & ACS/WFC   & F555W & Aug 12 2003  & 165s      & 9891 & G.\,Gilmore\\
          & ACS/WFC   & F814W & Aug 12 2003  & 130s      & 9891 & G.\,Gilmore\\

LINDSAY\,1 & ACS/WFC & F555W & Jul 11 2003  & 480s      & 9891 & G.\,Gilmore\\
 & ACS/WFC & F814W & Jul 11 2003  & 290s      & 9891 & G.\,Gilmore\\
 & ACS/WFC & F555W & Aug 21 2005  & 2$\times$20s$+$4$\times$496s       & 10396 & J.\,Gallagher\\
 & ACS/WFC & F814W & Aug 21 2005  & 2$\times$10s$+$4$\times$474s       & 10396 & J.\,Gallagher\\
 & UVIS/WFC3 & F275W & Jun 12 2019  & 1500s$+$1501s$+$2$\times$1523s$+$2$\times$1525s       & 15630 & N.\,Bastian\\
 & UVIS/WFC3 & F275W & Jun 24-25 2020  & 1500s$+$2$\times$1512s$+$4$\times$1523s$+$2$\times$1524s$+$2$\times$1525s$+$1530s       & 15630 & N.\,Bastian\\
 & UVIS/WFC3 & F336W & Jun 19 2019  & 500s$+$2$\times$1200s      & 14069 & N.\,Bastian\\
 & UVIS/WFC3 & F343N & Jun 19 2019  & 500s$+$800s$+$1650s$+$1850s       & 14069 & N.\,Bastian\\
 & UVIS/WFC3 & F438W & Jun 19 2019  & 120s$+$2$\times$460s      & 14069 & N.\,Bastian\\

NGC\,294  & UVIS/WFC3 & F336W & Feb 17 2017  & 3$\times$839s       & 14710 & A.\,P.\,Milone\\
          & UVIS/WFC3 & F656N & Feb 17 2017  & 2$\times$680s       & 14710 & A.\,P.\,Milone\\
          & UVIS/WFC3 & F814W & Feb 17 2017  & 90s$+$666s       & 14710 & A.\,P.\,Milone\\
          & ACS/WFC   & F555W & Oct 24 2003  & 165s      & 9891 & G.\,Gilmore\\
          & ACS/WFC   & F814W & Oct 24 2003  & 130s      & 9891 & G.\,Gilmore\\

NGC\,339 & UVIS/WFC3 & F336W & Aug  8 2016  & 700s$+$1160s$+$1200s              & 14069 & N.\,Bastian \\
         & UVIS/WFC3 & F343N & Aug  8 2016  & 520s$+$800s$+$1250s$+$1650s       & 14069 & N.\,Bastian \\
         & UVIS/WFC3 & F438W & Aug  8 2016  & 120s$+$180s$+$560s$+$660s         & 14069 & N.\,Bastian \\
         & ACS/WFC   & F555W & Nov 28 2005  & 2$\times$10s$+$4$\times$496s      & 10396 & J.\,Gallagher \\
         & ACS/WFC   & F814W & Nov 28 2005  & 2$\times$10s$+$4$\times$474s      & 10396 & J.\,Gallagher \\

NGC\,416 & UVIS/WFC3 & F275W & Jul 31, Aug 16 2019  & 3$\times$1500s$+$1512s$+$2$\times$1515s  & 15630 & N.\,Bastian \\
         &           &       &                      & $+$1523s$+$1530s$+$2$\times$1533s$+$2$\times$1534  & &  \\
         & UVIS/WFC3 & F336W & Jun 16 2016  & 700s$+$1160s$+$1200s              & 14069 & N.\,Bastian \\
         & UVIS/WFC3 & F343N & Jun 16 2016  & 500s$+$800s$+$1650s$+$1655s       & 14069 & N.\,Bastian \\
         & UVIS/WFC3 & F438W & Jun 16 2016  & 75s$+$150s$+$440s$+$460s          & 14069 & N.\,Bastian \\
         & ACS/WFC   & F555W & Mar 03 2006  & 2$\times$10s$+$4$\times$496s      & 10396 & J.\,Gallagher \\
         & ACS/WFC   & F814W & Nov 22 2005  & 2$\times$10s$+$4$\times$474s      & 10396 & J.\,Gallagher \\
         & ACS/WFC   & F814W & Mar 03 2006  & 2$\times$10s$+$4$\times$474s      & 10396 & J.\,Gallagher \\

NGC\,419 
         & UVIS/WFC3 & F336W & Aug 25 2011  & 400s$+$690s$+$2$\times$700s$+$740s & 12257 & L.\,Girardi  \\
         & UVIS/WFC3 & F343N & Aug 03 2016  & 450s$+$2$\times$1250s$+$1625s     & 14069 & N.\,Bastian \\
         & UVIS/WFC3 & F438W & Aug 03 2016  & 70s$+$150s$+$350s$+$550s          & 14069 & N.\,Bastian \\
         & ACS/WFC   & F555W & Jul 08 2006  & 2$\times$20s$+$4$\times$496s      & 10396 & J.\,Gallagher \\
         & ACS/WFC   & F814W & Jan 01 2006  & 2$\times$10s$+$4$\times$474s      & 10396 & J.\,Gallagher \\
         & ACS/WFC   & F814W & Jul 07 2006  & 2$\times$10s$+$4$\times$474s      & 10396 & J.\,Gallagher \\

NGC\,1755 & UVIS/WFC3 & F336W & Oct 05 2015  & 2$\times$711s       & 14204 & A.\,P.\,Milone\\
          & UVIS/WFC3 & F336W & Dec 28 2015  & 2$\times$711s       & 14204 & A.\,P.\,Milone\\
          & UVIS/WFC3 & F336W & Mar 26 2016  & 2$\times$711s       & 14204 & A.\,P.\,Milone\\
          & UVIS/WFC3 & F336W & Jun 10 2016  & 2$\times$711s       & 14204 & A.\,P.\,Milone\\
          & UVIS/WFC3 & F814W & Oct 05 2015  & 90s$+$678s          & 14204 & A.\,P.\,Milone\\
          & UVIS/WFC3 & F814W & Dec 28 2015  & 90s$+$678s          & 14204 & A.\,P.\,Milone\\
          & UVIS/WFC3 & F814W & Mar 26 2016  & 90s$+$678s          & 14204 & A.\,P.\,Milone\\
          & UVIS/WFC3 & F814W & Jun 10 2016  & 90s$+$678s          & 14204 & A.\,P.\,Milone\\
          & ACS/WFC   & F555W & Aug 08 2003  & 50s      & 9891 & G.\,Gilmore\\
          & ACS/WFC   & F814W & Aug 08 2003  & 40s      & 9891 & G.\,Gilmore\\

NGC\,1783 & UVIS/WFC3 & F275W & Sep 16 2019  & 2$\times$1500s$+$4$\times$1512s       & 15630 & N.\,Bastian\\
 & UVIS/WFC3 & F336W & Oct 12 2011  & 2$\times$1190s$+$1200s       & 12257 & L.\,Girardi\\
 & UVIS/WFC3 & F343N & Sep 14 2016  & 450s$+$845s$+$1650s       & 14069 & N.\,Bastian\\
 & ACS/WFC & F814W & Oct 07 2003  & 170s       & 9891 &  G.\,Gilmore  \\
 & ACS/WFC & F555W & Jan 14 2006  & 40s$+$2$\times$340s  & 10595 & P.\,Goudfrooij  \\
 & ACS/WFC & F555W & Oct 07 2003  & 250s       & 9891 &  G.\,Gilmore  \\
 & ACS/WFC & F814W & Jan 14 2006  & 8s$+$2$\times$340s  & 10595 &  P.\,Goudfrooij \\
 & ACS/WFC & F435W & Jan 14 2006  & 90s$+$2$\times$340s  & 10595 & P.\,Goudfrooij  \\
 
 NGC\,1801 & UVIS/WFC3 & F336W & Feb 26 2017  & 2$\times$834s$+$835s       & 14710 & A.\,P.\,Milone\\
          & UVIS/WFC3 & F656N & Feb 26 2017  & 2$\times$720s       & 14710 & A.\,P.\,Milone\\
          & UVIS/WFC3 & F814W & Feb 26 2017  & 90s$+$666s       & 14710 & A.\,P.\,Milone\\
          & ACS/WFC   & F555W & Oct 08 2003  & 115s      & 9891 & G.\,Gilmore\\
          & ACS/WFC   & F814W & Oct 08 2003  & 90s      & 9891 & G.\,Gilmore\\

 NGC\,1953 & UVIS/WFC3 & F336W & Jul 18-19 2017  & 2$\times$834s$+$835s       & 14710 & A.\,P.\,Milone\\
           & UVIS/WFC3 & F656N & Jul 18 2017  & 2$\times$720s       & 14710 & A.\,P.\,Milone\\
           & UVIS/WFC3 & F814W & Jul 18 2017  & 90s$+$666s       & 14710 & A.\,P.\,Milone\\
          & ACS/WFC   & F555W & Oct 07 2003  & 115s      & 9891 & G.\,Gilmore\\
          & ACS/WFC   & F814W & Oct 07 2003  & 90s      & 9891 & G.\,Gilmore\\

 NGC\,1978 & UVIS/WFC3 & F275W & Sep 17 2019  & 1492s$+$2$\times$1493s$+$1495s$+$2$\times$1498s       & 15630 & N.\,Bastian\\
          &           &       &                      & $+$2$\times$1499s$+$2$\times$1500s$+$1501s$+$1502s  & &  \\
      & UVIS/WFC3 & F336W & 2011 Aug 15 2011 & 380s$+$460s & 12257 & L. Girardi\\
      & UVIS/WFC3 & F336W & 2016 Sep 25 2016 & 660s$+$740s & 14069 & N. Bastian\\
      & UVIS/WFC3 & F343N & 2016 Sep 25 2016 & 425s$+$450s$+$500s$+$2$\times$800s$+$1000s  & 14069 & N. Bastian\\
      & UVIS/WFC3 & F438W & 2016 Sep 25 2016 & 75s$+$120s$+$420s$+$460s$+$650s$+$750s & 14069 & N. Bastian\\
      & UVIS/WFC3 & F814W & Sep 14 2019  &  3$\times$200s$+$348s$+$2$\times$349s$+$688s     & 15630 & N.\,Bastian\\

           & ACS/WFC   & F555W & Oct 07 2003  & 300s      & 9891 & G.\,Gilmore\\
           & ACS/WFC   & F555W & Aug 15 2011 & 60s$+$300s$+$680s & 12257 & L. Girardi\\
           & ACS/WFC   & F814W & Oct 07 2003  & 200s      & 9891 & G.\,Gilmore\\

%  ID            &  RA         &  error            &  DEC           & error                   &  $\mu_{\alpha}$cos$\delta$          & $\mu_{\delta}$      & (m$-$M)$_{0}$ & E(B$-$V)  \\  
%  BRHT\,5b      & 05 08 52.65 &    ---           & $-$68 45 18.0  & ---          & 2.09$\pm$0.03      &   0.04$\pm$0.09    &     18.40  &   0.12       &  $-$0.4    &  0.015 & ---  \\
\end{longtable} 
\end{center}    
\twocolumn

%%%%%%%%%%%%%%%%%%%%%%%%%%%%%%%%%%%%%%%%%%%%%%%%%%%%%%%%%%%%%%%%%%%%%%%%%%%     
\begin{table*}
%\small
  \caption{Proper motions, relative to the main cluster in the FoVs, of field stellar populations represented with red, blue, and aqua colors in Figures\,\ref{fig:PMsSMC} and \ref{fig:PMsLMC}. For each population we provide the ID of the reference cluster, the number of stars, $N$, the median proper motions ($\delta \mu_{\rm \alpha} \cos{\delta}$ and $\delta \mu_{\rm \delta}$, the ellipticity, $\epsilon$, and the position angle, $\theta$, of the best-fitting ellipse.}
\begin{tabular}{l l l c c c c}
\hline \hline
ID & Population & $N$  & $\delta \mu_{\rm \alpha} \cos{\delta} $ &   $\delta \mu_{\rm \delta}$  & $\epsilon$ & $\theta$ \\
   &  &   & [mas yr$^{-1}$] &   [mas yr$^{-1}$]  &  & [deg] \\
% 1-b/a
\hline
KRON\,34 & Red   &539 &    0.001$\pm$0.004 &    0.003$\pm$0.004  & 0.23$\pm$0.05 & 43$\pm$9 \\ 
         & Blue  &193 &    0.031$\pm$0.006 & $-$0.013$\pm$0.004  & 0.34$\pm$0.06 & 28$\pm$5 \\ 
\hline
NGC\,294 & Red   &657 &    0.041$\pm$0.004 & $-$0.026$\pm$0.003  & 0.19$\pm$0.06 & 37$\pm$6 \\ 
         & Blue  &256 &    0.045$\pm$0.004 & $-$0.024$\pm$0.004  & 0.42$\pm$0.07 & 28$\pm$6 \\ 
\hline
NGC\,339 & Red   & 54 &    0.014$\pm$0.014 &    0.013$\pm$0.012  & 0.10$\pm$0.08 & 33$\pm$11 \\ 
         & Blue  &192 &    0.110$\pm$0.011 & $-$0.059$\pm$0.009  & 0.62$\pm$0.03 & 38$\pm$3 \\ 
\hline
NGC\,416 & Red   &330 &    0.017$\pm$0.005 &    0.047$\pm$0.004  & 0.08$\pm$0.07 & 19$\pm$13 \\ 
         & Blue  &688 &    0.109$\pm$0.006 &    0.017$\pm$0.004  & 0.30$\pm$0.05 & 26$\pm$6 \\ 
\hline
NGC\,419 & Red   &211 & $-$0.070$\pm$0.008 &    0.050$\pm$0.008  & 0.31$\pm$0.07 & 33$\pm$6 \\ 
         & Blue  &267 & $-$0.077$\pm$0.009 &    0.054$\pm$0.006  & 0.42$\pm$0.07 & 29$\pm$5 \\ 
\hline
NGC\,1755& Red   &296 &    0.077$\pm$0.006 &    0.169$\pm$0.007  & 0.02$\pm$0.08 & 86$\pm$16 \\ 
         & Blue  &183 & $-$0.019$\pm$0.005 &    0.139$\pm$0.005  & 0.17$\pm$0.07 &314$\pm$14 \\ 
\hline
NGC\,1801& Red   &321 &    0.058$\pm$0.007 & $-$0.036$\pm$0.007  & 0.05$\pm$0.06 & 12$\pm$12 \\ 
         & Blue  &209 &    0.001$\pm$0.005 &    0.009$\pm$0.005  & 0.20$\pm$0.13 & 48$\pm$16 \\ 
         & Aqua  & 44 &    0.023$\pm$0.012 &    0.003$\pm$0.017  & 0.09$\pm$0.12 & 46$\pm$18 \\ 
\hline
NGC\,1953& Red   &441 &    0.095$\pm$0.007 &    0.039$\pm$0.006  & 0.00$\pm$0.05 &360$\pm$13 \\ 
         & Blue  &186 & $-$0.029$\pm$0.006 & $-$0.007$\pm$0.006  & 0.28$\pm$0.06 &317$\pm$16 \\ 
         & Aqua  &53  & $-$0.092$\pm$0.009 & $-$0.022$\pm$0.008  & 0.17$\pm$0.11 &323$\pm$17 \\
\hline
%     \hline\hline
\end{tabular} 
 \label{tab:ellissi}
 \end{table*}
%%%%%%%%%%%%%%%%%%%%%%%%%%%%%%%%%%%%%%%%%%%%%%%%%%%%%%%%%%%%%%%%%%%%%%%%%%%     

\begin{table*}
%\small
  \caption{Proper motions relative to NGC\,346 and proper-motion dispersions  for the clusters NGC\,346 and BS\,90 and for the selected populations of pre-MS, MS, and RGB field stars.}
\begin{tabular}{l c c c c c c}
\hline \hline
ID & $\delta \mu_{\rm \alpha} \cos{\delta}$ & $\delta \mu_{\rm \delta} $ &   $\sigma \mu_{\rm \alpha} \cos{\delta}$  & $\sigma \mu_{\rm \delta} $ & $N$ \\
     & [mas yr$^{-1}$] & [mas yr$^{-1}$]    & [mas yr$^{-1}$] & [mas yr$^{-1}$] & \\

% 1-b/a
\hline
NGC\,346  &   0.000$\pm$0.001  &    0.000$\pm$0.001 & 0.028  & 0.025  &  945\\ 
BS\,90    &$-$0.016$\pm$0.001  &    0.153$\pm$0.001 & 0.036  & 0.036  &  2220\\ 
pre-MS\,I &$-$0.032$\pm$0.005  &    0.019$\pm$0.004 & 0.083  & 0.075  &  345\\ 
pre-MS\,II &   0.004$\pm$0.004  &    0.007$\pm$0.004 & 0.045  & 0.044  & 162 \\
MS\,I    &$-$0.024$\pm$0.002  &    0.005$\pm$0.002 & 0.059  & 0.062  &  2136\\ 
MS\,II   &   0.188$\pm$0.003  & $-$0.034$\pm$0.004 & 0.079  & 0.088  &  582  \\ 
RGB      &$-$0.031$\pm$0.003  &    0.024$\pm$0.008 & 0.084  & 0.089  &  713  \\ 
%         &$\pm$  & $\pm$ &   &   & &   \\ 
%         &$\pm$  & $\pm$ &   &   & &   \\ 
\hline
         \end{tabular} 
 \label{tab:n346}
 \end{table*}
%%%%%%%%%%%%%%%%%%%%%%%%%%%%%%%%%%%%%%%%%%%%%%%%%%%%%%%%%%%%%%%%%%%%%%%%%%%     

\bibliographystyle{mnras}
\bibliography{ms} % your references Yourfile.bib

%\bibliography{ms}
\end{document}